\documentclass[10pt,twocolumn]{article}

\usepackage[T1]{fontenc}
\usepackage{mathpazo}      
\usepackage[scaled=0.9]{helvet}  
\usepackage{courier}       
\usepackage{microtype}
\usepackage{graphicx}
\usepackage{amsmath}
\usepackage{amssymb}
\usepackage{booktabs}
\usepackage{tabularx}
\usepackage{xcolor}
\usepackage{listings}
\usepackage[round,authoryear]{natbib}
\usepackage{xurl}          
\usepackage{hyperref}
\usepackage{cleveref}
\usepackage[title]{appendix}
\usepackage{tikz}
\usetikzlibrary{shapes,arrows,positioning,shadows,calc,backgrounds,decorations.pathreplacing}
\usepackage{pgfplots}
\pgfplotsset{compat=1.18}
\usepackage{subcaption}
\usepackage{multirow}
\usepackage{enumitem}
\usepackage{titlesec}
\titlespacing*{\section}{0pt}{10pt plus 2pt minus 2pt}{4pt plus 1pt minus 1pt}
\titlespacing*{\subsection}{0pt}{8pt plus 2pt minus 2pt}{3pt plus 1pt minus 1pt}
\titlespacing*{\subsubsection}{0pt}{6pt plus 2pt minus 1pt}{2pt plus 1pt minus 1pt}
\usepackage{fancyhdr}
\usepackage{xspace}
\usepackage[section]{placeins}
\usepackage{afterpage}

\newcommand{\mlsysim}{\mbox{\textsc{MLSys}\,\textperiodcentered\,\textsc{im}}\xspace}
\newcommand{\MLSYSIM}{\textbf{\mlsysim}}

\newcommand{\fullmark}{\checkmark}

\newcommand{\emptymark}{--}

\newcommand{\cval}[1]{\pgfmathprintnumber[fixed, precision=1, 1000 sep={\,}]{#1}}
\newcommand{\cvalii}[1]{\pgfmathprintnumber[fixed, precision=2, 1000 sep={\,}]{#1}}
\newcommand{\cvalint}[1]{\pgfmathprintnumber[fixed, precision=0, 1000 sep={\,}]{#1}}

\pgfmathsetmacro{\HhPeakTFLOPS}{989}          
\pgfmathsetmacro{\HhBWTBs}{3.35}              
\pgfmathsetmacro{\AhBWTBs}{2.04}              
\pgfmathsetmacro{\HhHBMGB}{80}                
\pgfmathsetmacro{\NVLinkBWGBs}{900}           
\pgfmathsetmacro{\IBBWGBs}{50}                
\pgfmathsetmacro{\HhCostK}{30}                 
\pgfmathsetmacro{\NodeMTBFhrs}{10000}         

\pgfmathsetmacro{\LLaMAParams}{70}            
\pgfmathsetmacro{\LLaMAWeightsGB}{140}        

\pgfmathsetmacro{\HhRidgePoint}{\HhPeakTFLOPS / \HhBWTBs}  

\pgfmathsetmacro{\ServTPGPUs}{2}                              
\pgfmathsetmacro{\ServSin}{512}                               
\pgfmathsetmacro{\ServEta}{0.5}                               
\pgfmathsetmacro{\ServPrefillMs}{(2 * \LLaMAParams / (\ServTPGPUs * \ServEta)) * (\ServSin / \HhPeakTFLOPS)}
\pgfmathsetmacro{\ServDecodeMs}{\LLaMAWeightsGB / (\ServTPGPUs * \HhBWTBs * 1000) * 1000}
\pgfmathsetmacro{\ServPrefillTps}{\ServSin / (\ServPrefillMs / 1000)}
\pgfmathsetmacro{\ServDecodeTps}{1 / (\ServDecodeMs / 1000)}
\pgfmathsetmacro{\ServTpsRatio}{round(\ServPrefillTps / \ServDecodeTps)}

\pgfmathsetmacro{\BetaFatTree}{1.0}
\pgfmathsetmacro{\BetaTorus}{0.67}
\pgfmathsetmacro{\LinkBWGbps}{400}
\pgfmathsetmacro{\TorusBisectGbps}{\LinkBWGbps * \BetaTorus}
\pgfmathsetmacro{\TopoRatio}{\BetaFatTree / \BetaTorus}

\pgfmathsetmacro{\ChinchillaPstarB}{91}       
\pgfmathsetmacro{\ChinchillaDstarT}{1.8}      

\pgfmathsetmacro{\NVLinkIBgap}{\NVLinkBWGBs / \IBBWGBs}  

\pgfmathsetmacro{\ClusterGPUs}{1024}
\pgfmathsetmacro{\ClusterCapExM}{\ClusterGPUs / 1000 * \HhCostK}  
\pgfmathsetmacro{\TrainDays}{30}
\pgfmathsetmacro{\AmortYears}{3}
\pgfmathsetmacro{\TrainCapExK}{\ClusterCapExM * \TrainDays / (\AmortYears * 365) * 1000}

\pgfmathsetmacro{\BetaOpt}{7}                 
\pgfmathsetmacro{\CkptBytesPerParam}{14}      
\pgfmathsetmacro{\CkptSizeTB}{\LLaMAParams * \CkptBytesPerParam / 1000}

\pgfmathsetmacro{\SynthTargetMs}{50}           
\pgfmathsetmacro{\SynthMsToSInv}{1000 / \SynthTargetMs}  
\pgfmathsetmacro{\SynthBWreqGBs}{\LLaMAWeightsGB * \SynthMsToSInv}  
\pgfmathsetmacro{\SynthAhMultiple}{\SynthBWreqGBs / (\AhBWTBs * 1000)}

\pgfmathsetmacro{\GPTenergyMWh}{1287}
\pgfmathsetmacro{\GPTgridCI}{429}              
\pgfmathsetmacro{\GPTcarbonTonnes}{\GPTenergyMWh / 1000 * \GPTgridCI}

\pgfmathsetmacro{\RthreeGPUs}{512}
\pgfmathsetmacro{\RthreeNodes}{64}
\pgfmathsetmacro{\RthreeTPsize}{8}
\pgfmathsetmacro{\RthreeDPsize}{\RthreeNodes}  
\pgfmathsetmacro{\RthreeEta}{0.40}
\pgfmathsetmacro{\RthreeGradShardGB}{\LLaMAWeightsGB / \RthreeTPsize}  
\pgfmathsetmacro{\RthreeClusterMTBF}{\NodeMTBFhrs / \RthreeGPUs}
\pgfmathsetmacro{\RthreeARms}{2 * (\RthreeDPsize - 1) / \RthreeDPsize * \RthreeGradShardGB / \IBBWGBs * 1000}
\pgfmathsetmacro{\RthreeOverlap}{0.85}
\pgfmathsetmacro{\RthreeExposedMs}{(1 - \RthreeOverlap) * \RthreeARms}

\pgfmathsetmacro{\StwoCPUworkers}{64}
\pgfmathsetmacro{\StwoAugRate}{850}             
\pgfmathsetmacro{\StwoDecodeRate}{1200}          
\newcommand{\StwoEffThroughput}{54{,}400}        
\newcommand{\StwoRawThroughput}{76{,}800}        

\usepackage[
  letterpaper,
  top=0.75in,
  bottom=1in,
  left=0.75in,
  right=0.75in,
  columnsep=0.25in
]{geometry}

\definecolor{codegreen}{rgb}{0,0.5,0}
\definecolor{codegray}{rgb}{0.5,0.5,0.5}
\definecolor{codepurple}{rgb}{0.44,0.0,0.55}
\definecolor{codeblue}{rgb}{0.0,0.25,0.55}
\definecolor{backcolour}{rgb}{0.975,0.975,0.985}
\definecolor{rulecolour}{rgb}{0.78,0.78,0.82}

\hypersetup{
    colorlinks=true,
    linkcolor=blue,
    citecolor=blue,
    urlcolor=blue,
    pdftitle={MLSys·im: First-Principles Infrastructure Modeling for Machine Learning Systems},
    pdfauthor={Vijay Janapa Reddi}
}

\title{
  {\Huge\bfseries \mlsysim}\\[0.4em]
  \Large\normalfont\itshape First-Principles Infrastructure Modeling for Machine Learning Systems
}

\author{
  \fontsize{12}{15}\selectfont
  Vijay Janapa Reddi\\[0.2em]
  \fontsize{11}{14}\selectfont
  Harvard University\\[0.6em]
  \fontsize{10}{12}\selectfont
  \textcolor{gray!60}{\href{https://mlsysbook.ai/mlsysim}{mlsysbook.ai/mlsysim}}
}

\date{}

\begin{document}

\maketitle
\vspace{-1.5em}

\begin{abstract}
As machine learning shifts from laboratory curiosity to critical infrastructure, the systems that sustain it span an extraordinary range, from sub-milliwatt microcontrollers to multi-gigawatt datacenter fleets. Reasoning across this range is hard: empirical profiling requires the target hardware in hand, while cycle-accurate simulation costs hours per configuration, leaving no tool for rapid, full-stack architectural reasoning. We present \MLSYSIM{} (\textbf{M}achine \textbf{L}earning \textbf{Sys}tems \textbf{I}nfrastructure \textbf{M}odeling), a first-principles analytical framework that formalizes the ``physics of systems'' into a dimensionally-strict Python engine. \mlsysim is built on a \emph{demand--supply} abstraction that decouples computational demand from silicon supply and environmental context, and it enforces unit integrity at runtime so the silent conversion errors that plague ad-hoc modeling cannot occur. Every input is drawn from a typed, provenance-tracked registry, so no number enters an analysis without a documented source. On this engine we codify a taxonomy of 22 ``Systems Walls'' resolved by 28 composable models and solvers, enabling sub-second design-space exploration that identifies binding constraints and synthesizes ideal hardware specifications across the entire ML systems lifecycle.
\end{abstract}

\section{Introduction}
\label{sec:intro}

Machine learning has become infrastructure~\citep{sutton2019bitter}. Training a frontier model now requires orchestrating tens of thousands of accelerators across datacenter fabrics where memory ceilings, network bandwidth, power delivery, and regional carbon intensities interact in non-obvious ways~\citep{dean2012large,shoeybi2019megatron}. The pace of this scaling is accelerating: frontier models have grown from billions to trillions of parameters in under five years, and the infrastructure cost of a single training run now rivals that of a small datacenter~\citep{deepseek2025v3}.

Yet the hardware required to develop intuition for these systems is prohibitively scarce: a student cannot requisition a 100{,}000-GPU cluster to explore how topology affects AllReduce latency, and a researcher cannot easily sweep parallelism strategies across hardware generations. This creates a growing \emph{reasoning gap}. The systems are getting more complex, but the tools to think about them have not kept pace.

Consider a concrete example. A team deploying LLaMA-3 70B for interactive serving must answer: \emph{How many H100 GPUs are needed to meet a 50\,ms time-to-first-token SLA at 95th-percentile latency?} The answer depends on at least seven interacting constraints, each a ``wall'' (a hard bound imposed by physics, economics, or algorithmic scaling; see \Cref{tab:walls}).  We say a wall \emph{binds} when it is the tightest constraint, the single bottleneck that limits end-to-end performance.  The \emph{binding constraint} is the wall whose relaxation would yield the largest throughput improvement; all other walls are slack.\footnote{This usage mirrors linear programming: a constraint \emph{binds} (is \emph{active}) when the optimal point lies on it.  Equivalently, in Roofline analysis, the binding ceiling is the one that determines achievable performance.}  For the LLaMA-3 serving example, the model's 70 billion parameters require $\sim$\cvalint{\LLaMAWeightsGB}\,GB in FP16, exceeding a single GPU's \cvalint{\HhHBMGB}\,GB HBM capacity (Wall~2: Memory). Tensor-parallel sharding across two GPUs introduces NVLink synchronization overhead (Wall~14: Communication). Continuous batching with PagedAttention determines KV-cache memory utilization (Wall~5: Batching). The decode phase is memory-bandwidth-bound at \cvalii{\HhBWTBs}\,TB/s per device (Wall~4: Serving). Tail latency under load follows Erlang-C queueing dynamics (Wall~7: Tail Latency). And the fleet's total cost of ownership constrains what is economically viable (Wall~17: Capital). A similar multi-wall analysis applies to training: determining the optimal parallelism strategy for a 512-GPU run involves compute throughput, memory capacity, communication overhead, scaling laws, reliability, and economics simultaneously. No single equation suffices.

Existing tools are constrained along three axes (fidelity, speed, and scope), and no tool occupies the region where all three are adequate (\Cref{tab:comparison}). Profilers require physical silicon; cycle-accurate simulators like ASTRA-sim~2.0~\citep{won2023} require hours per configuration; analytical tools like Calculon~\citep{isaev2023} achieve speed but focus narrowly on LLM training, ignoring data pipelines, reliability, sustainability, and inference. Patterson and Hennessy faced an analogous tradeoff in computer architecture education and chose taxonomic completeness over cycle accuracy: MIPS exposed every architectural concept through a model simple enough to reason about yet faithful enough to develop correct intuition~\citep{hennessy2024architecture}. \mlsysim makes the same choice for ML systems. Beyond coverage gaps, none of these tools enforce dimensional correctness. In ML systems, confusing GB with GiB, FLOP/s with FLOPs, or bandwidth with throughput can silently invalidate capacity-planning spreadsheets, yet no existing tool catches these errors at runtime.

\begin{table*}[!t]
\centering
\caption{\textbf{The 22 ML Systems Walls.} Each wall represents a physical or logical constraint resolved by a dedicated resolver (model or solver). Walls 1--2 (Compute and Memory) share the \texttt{SingleNodeModel}; four companion models deepen individual walls (\texttt{TrainingMemoryModel} for Wall~2, \texttt{ServingCapacityModel} for Wall~4, \texttt{MoERoutingModel} for Wall~14, and \texttt{NetworkRooflineModel}, which lifts the Roofline to fleet scale). Together with 3 optimizers (parallelism, batching, placement), the framework provides 28 resolvers across 22 walls. Domains progress from local node resources through data movement and algorithmic scaling to fleet coordination, operations, and cross-cutting analysis. Each wall is formalized in \Cref{sec:taxonomy}.}
\label{tab:walls}
\small
\renewcommand{\arraystretch}{1.1}
\begin{tabularx}{\textwidth}{@{}r l l l X l@{}}
\toprule
\textbf{\#} & \textbf{Wall} & \textbf{Resolver} & \textbf{Bounded} & \textbf{Core Equation} & \textbf{Ref.} \\
\midrule
\multicolumn{6}{@{}l}{\textit{Node (Single-Accelerator Resources)}} \\
1  & Compute       & SingleNode      & Peak FLOP/s       & $T = \text{OPs} / (\text{Peak} \times \eta)$ & \citealt{williams2009} \\
2  & Memory        & SingleNode      & HBM BW + cap.     & $T = |W| / BW_{\text{HBM}}$ & \citealt{williams2009} \\
3  & Software      & Efficiency      & Achieved MFU      & $\eta = \text{FLOPS}_{\text{ach}} / \text{Peak}$ & \citealt{chowdhery2022palm} \\
4  & Serving       & Serving         & Prefill vs.\ dec. & $T_{\text{pf}} = 2PS/(F\eta);\; T_{\text{dec}} = |W|/BW$ & \citealt{pope2023llm} \\
5  & Batching      & Cont.\ Batch    & KV-cache frag.    & $\text{KV} = 2LHD \lceil S/p \rceil pBb$ & \citealt{kwon2023} \\
6  & Streaming     & WeightStream    & Injection BW      & $T = \max(|W_\ell|/BW, 2P_\ell B/F\eta)$ & \citealt{lie2022cerebras} \\
7  & Tail Latency  & TailLatency     & P99 queueing      & Erlang-C M/M/$c$ & \citealt{dean2013} \\
\midrule
\multicolumn{6}{@{}l}{\textit{Data (Movement \& Pipelines)}} \\
8  & Ingestion     & Data            & Storage I/O       & $\rho = BW_{\text{demand}} / BW_{\text{supply}}$ & \citealt{mohan2021} \\
9  & Transform.    & Transform.      & CPU preproc.      & $T = B / R_{\text{cpu}}$ & \citealt{murray2021tf} \\
10 & Locality      & Topology        & Bisection BW      & $BW_{\text{eff}} = BW_{\text{link}} \cdot \beta / \text{osub}$ & \citealt{leiserson1985} \\
\midrule
\multicolumn{6}{@{}l}{\textit{Algorithm (Scaling \& Compression)}} \\
11 & Complexity    & Scaling         & Scaling laws      & $C = 6PD;\; P^{*} = \sqrt{C/120}$ & \citealt{hoffmann2022chinchilla} \\
12 & Reasoning     & Inf.\ Scaling   & Inf.-time comp.   & $T = K \times T_{\text{step}}$ & \citealt{snell2025scaling} \\
13 & Fidelity      & Compression     & Acc.--efficiency  & $r = b_{\text{base}}/b;\; r = 1/(1{-}s)$ & \citealt{han2016deep} \\
\midrule
\multicolumn{6}{@{}l}{\textit{Fleet (Multi-Node Coordination)}} \\
14 & Communic.     & Distributed     & AllReduce         & $T = 2\tfrac{N{-}1}{N}\tfrac{M}{B_{\text{link}}} + 2(N{-}1)\alpha$ & \citealt{shoeybi2019megatron} \\
15 & Fragility     & Reliability     & Cluster MTBF      & $\text{MTBF}_{\text{cl}} = \text{MTBF}_{\text{node}}/N$ & \citealt{daly2006} \\
16 & Multi-tenant  & Orchestration   & Queue wait        & $T_{\text{wait}} = \rho / [2\mu(1{-}\rho)]$ & \citealt{little1961} \\
\midrule
\multicolumn{6}{@{}l}{\textit{Operations (Economics, Sustainability \& Safety)}} \\
17 & Capital       & Economics       & TCO               & $\text{TCO} = \text{CapEx} + \text{OpEx}$ & \citealt{barroso2019} \\
18 & Sustain.      & Sustainability  & Carbon + water    & $\text{CO}_2 = E \times \text{PUE} \times \text{CI}$ & \citealt{patterson2021carbon} \\
19 & Checkpoint    & Checkpoint      & I/O burst penalty & $\text{penalty} = T_{\text{write}} / T_{\text{interval}}$ & \citealt{eisenman2022checknrun} \\
20 & Safety        & Resp.\ Eng.     & DP-SGD overhead   & $\sigma \propto 1/\varepsilon$ & \citealt{abadi2016} \\
\midrule
\multicolumn{6}{@{}l}{\textit{Analysis (Cross-Cutting Diagnostics)}} \\
21 & Sensitivity   & Sensitivity     & Binding constr.   & $\partial T / \partial x_i$ & \citealt{williams2009} \\
22 & Synthesis     & Synthesis       & Inverse spec      & $BW_{\text{req}} = |W| / T_{\text{target}}$ & \citealt{kwon2023} \\
\midrule
\multicolumn{6}{@{}l}{\textit{Optimizers (Design-Space Search, no dedicated wall)}} \\
   & Parallelism   & Parallelism     & Max MFU           & $\max_{c} \eta(c)$ over TP$\times$PP$\times$DP & \citealt{shoeybi2019megatron} \\
   & Batching      & Batching        & SLA-aware $B^*$   & $\max B$ s.t.\ $P_{99} \le T_{\text{SLA}}$ & \citealt{kwon2023} \\
   & Placement     & Placement       & Cost--perf.       & $\min$ cost s.t.\ $T \le T_{\text{SLA}}$ & --- \\
\bottomrule
\end{tabularx}
\end{table*}

\MLSYSIM{} (Machine Learning Systems Infrastructure Modeling) is an open-source, pure-Python, first-principles analytical modeling framework for ML systems, designed for education and early design-space reasoning before empirical benchmarking. It formalizes back-of-the-envelope ML systems reasoning into a dimensionally strict, composable engine. The framework codifies 22 ``Systems Walls'' into 28 composable resolvers organized across six domains: Node, Data, Algorithm, Fleet, Operations, and Analysis. It separates computational \emph{demand} from silicon \emph{supply} and environmental \emph{context} through a 5-layer demand--supply architecture, enforces SI unit correctness at runtime via the \texttt{pint} library, and produces full-stack analysis in milliseconds on any laptop.

Each of the resolvers computes structural physics from first-principles equations and hardware datasheet constants, then bridges the gap to measured performance through a single, explicit efficiency coefficient~$\eta$ that absorbs second-order effects (kernel launch overhead, straggler variance, framework dispatch costs). This is the same abstraction that makes the Roofline model useful: replacing peak bandwidth with \emph{achievable} bandwidth sacrifices cycle-level precision for the ability to reason correctly about which constraint binds and why~\citep{williams2009}.

\mlsysim serves three communities. \emph{Students} build quantitative intuition through hands-on tutorials where changing a single parameter reveals which wall binds and why. \emph{Instructors} run live classroom demonstrations that produce concrete numbers in real time, replacing hand-waving with Roofline diagrams. \emph{Researchers} perform rapid what-if analysis by comparing parallelism strategies, evaluating procurement decisions, or projecting carbon footprints before committing resources. Designed as the analytical companion to the \emph{Machine Learning Systems} textbook~\citep{mlsysbook2025} (see Appendices~\ref{sec:appendix-mlsysbook} and \ref{sec:appendix-hardware} for pedagogical integration examples and registry snapshots), \mlsysim is fully open-source (Apache-2.0 license) with deterministic, hardware-free execution, ensuring that every result in this paper is independently reproducible. This matters for educational equity. A student at a community college with a Chromebook can run the same quantitative exercises and verify the same binding constraints as one with access to a university GPU cluster. We validate against published empirical anchors spanning five taxonomy domains, matching values within 7\% error while sweeping over 1{,}000 configurations in under one second.

This paper makes the following contributions:

\begin{enumerate}[leftmargin=*,itemsep=3pt,label=\textbf{C\arabic*.}]
    \item \textbf{A Taxonomy of 22 Systems Walls} (\Cref{tab:walls}), each grounded in a published equation and resolved by a dedicated resolver (model or solver). The taxonomy provides a complete, structured vocabulary for reasoning about the constraints that bind ML system performance (\Cref{sec:taxonomy}).

    \item \textbf{Demand--Supply Separation with Dimensional Strictness.} A 5-layer architecture formally decouples computational demand from silicon supply through a single workload-lowering step. Every physical quantity carries SI units at runtime, transforming dimensional analysis from a manual discipline into a machine-checked invariant (\Cref{sec:architecture}).

    \item \textbf{Composable Resolver Algebra.} 28 resolvers (23 models, 2 solvers, and 3 optimizers) compose through chaining: each is a pure function $f(\text{config}) \to \text{metrics}$, producing a three-level evaluation (Feasibility, Performance, Macro) that identifies binding constraints. The algebra includes an inverse-Roofline \emph{synthesis} solver that derives minimum hardware specifications from SLA requirements (\Cref{sec:solver-formalism,sec:usage}).

    \item \textbf{Accessible Full-Stack Reasoning without Hardware.} \mlsysim runs on any laptop without GPUs, clusters, or cloud credits. It powers deterministic labs and browser-friendly examples, providing an accessible systems engineering environment for students and instructors (\Cref{sec:usage}).
\end{enumerate}

The paper builds the framework in layers. We first survey the modeling landscape and identify the void (\Cref{sec:related}). \Cref{sec:architecture} presents the architecture, including four design principles, a 5-layer input stack, the dimensionally strict type system, and the extensibility model. \Cref{sec:taxonomy} formalizes the 22-wall taxonomy, and \Cref{sec:solver-formalism} defines the 3-tier resolver algebra (Models, Solvers, Optimizers) that composes over it. We validate against published benchmarks spanning five domains (\Cref{sec:validation}), demonstrate the framework through student, instructor, and researcher use cases (\Cref{sec:usage}), surface common misconceptions the framework is designed to expose (\Cref{sec:fallacies}), discuss limitations and future work (\Cref{sec:discussion}), and conclude (\Cref{sec:conclusion}).

\section{Related Work}
\label{sec:related}

Tools for modeling and evaluating ML systems span a wide spectrum of fidelity, scope, and intended audience. We organize prior work into four categories and position \mlsysim relative to each: (1)~cycle-level simulators that model hardware at the microarchitectural level, (2)~accelerator design tools that evaluate individual chip architectures, (3)~analytical and co-design tools that trade fidelity for speed, and (4)~pedagogical simulators that prioritize conceptual clarity. \Cref{tab:comparison} provides a quantitative summary across these categories.

\subsection{Cycle-Level Simulators}

Cycle-accurate simulators provide the highest fidelity by modeling hardware behavior at the microarchitectural level. gem5~\citep{binkert2011} is the canonical general-purpose architecture simulator, capable of modeling CPUs and GPUs down to individual pipeline stages. While invaluable for processor design, gem5 lacks ML-specific abstractions (it has no notion of a transformer layer, a training step, or a parallelism strategy), and simulating even a single forward pass of a modern model can require hours of wall-clock time.

ASTRA-sim 2.0~\citep{won2023} addresses the ML gap by providing a hierarchical network simulator purpose-built for distributed training. It models collective communication patterns such as AllReduce across realistic network topologies, producing high-fidelity estimates of communication overhead. SimAI~\citep{wang2025simai} extends this approach with a full-stack training simulator that integrates NS3-based network modeling with kernel computation traces, achieving 98\% alignment with real-world results on 1024-node A100 clusters. Both inherit the fundamental cost of high fidelity. Simulating one training step of a large model at cluster scale can take minutes to hours, making iterative design-space exploration impractical. Their scope is also limited to communication and compute; neither models economics, sustainability, data pipelines, or reliability.

\mlsysim occupies a different point on the fidelity--speed spectrum. Where ASTRA-sim answers ``how many microseconds does this AllReduce take on this exact topology,'' \mlsysim answers ``which of 22 possible bottlenecks binds this system, and how does changing hardware shift the binding constraint?'' The two classes are complementary: \mlsysim narrows the design space, and high-fidelity simulators validate specific points within it.

\begin{table*}[!t]
    \centering
    \caption{\textbf{Comparison of ML systems modeling tools.} \emph{Scope}: which system aspects a tool models (e.g., ``compute \& comm.''\ means compute throughput and collective communication only, while ``full-stack'' spans all six taxonomy domains). \emph{Speed}: wall-clock time for a single evaluation. \emph{Walls}: how many of the 22 Systems Walls (\Cref{tab:walls}) each tool addresses. \emph{Phase}: whether training, inference, or both are modeled. \emph{Dist.}: whether multi-node (multi-accelerator) distributed analysis is supported.}
    \label{tab:comparison}
    \small
    \renewcommand{\arraystretch}{1.1}
    \begin{tabularx}{\textwidth}{@{}l l X c c c c@{}}
    \toprule
    \textbf{Tool} & \textbf{Approach} & \textbf{Scope} & \textbf{Speed} & \textbf{Walls} & \textbf{Phase} & \textbf{Dist.} \\
    \midrule
    \multicolumn{7}{@{}l}{\textit{Cycle-level}} \\
    gem5 & Cycle-accurate & CPU/GPU microarchitecture & Hours & 1--2 & Both & \emptymark \\
    ASTRA-sim 2.0 & Cycle-accurate & Network collectives, topology & Hours & 1--2 & Train & \fullmark \\
    SimAI & Trace-driven & Full-stack distributed training & Minutes & 2--3 & Train & \fullmark \\
    \midrule
    \multicolumn{7}{@{}l}{\textit{Accelerator design}} \\
    Timeloop + Accelergy & Analytical & Accelerator dataflow \& energy & Minutes & 1--2 & Infer & \emptymark \\
    LLMCompass & Analytical & LLM inference HW design space & Minutes & 2--3 & Infer & \emptymark \\
    \midrule
    \multicolumn{7}{@{}l}{\textit{Analytical \& co-design}} \\
    Paleo & Analytical & DNN training compute \& comm. & Seconds & 2 & Train & \fullmark \\
    Calculon & Analytical & LLM training performance & Seconds & 2--3 & Train & \fullmark \\
    Lumos & Trace-driven & LLM training perf.\ modeling & Seconds & 2--3 & Train & \fullmark \\
    Vidur & Empirical & LLM inference scheduling & Seconds & 3--4 & Infer & \emptymark \\
    GenZ & Analytical & LLM inference platform design & Seconds & 2--3 & Infer & \emptymark \\
    LLM-Viewer & Analytical & LLM inference memory/latency & Seconds & 1--2 & Infer & \emptymark \\
    \midrule
    \multicolumn{7}{@{}l}{\textit{Sustainability}} \\
    LLMCarbon & Analytical & LLM carbon footprint (op.\ + embodied) & Seconds & 1 & Both & \emptymark \\
    CodeCarbon & Empirical & Runtime energy \& carbon tracking & Seconds & 1 & Both & \emptymark \\
    \midrule
    \MLSYSIM & \textbf{Analytical} & \textbf{Full-stack: compute, memory, network, data,} & \textbf{Sub-sec.} & \textbf{22} & \textbf{Both} & \fullmark \\
     & & \textbf{scaling, reliability, econ., sustainability, safety} & & & & \\
    \bottomrule
    \end{tabularx}
    \end{table*}

\subsection{Accelerator Design Tools}

A second class of tools targets the design and evaluation of individual accelerator architectures. Timeloop~\citep{parashar2019} provides a systematic methodology for evaluating DNN accelerator dataflows, modeling how data tiles map onto spatial architectures and estimating latency and energy for each mapping. Accelergy~\citep{wu2019}, its companion framework, supplies the energy estimation primitives that Timeloop consumes. Together, they form a powerful toolkit for accelerator architects exploring the design space of novel silicon. LLMCompass~\citep{zhang2024} brings this approach to LLM inference, combining an automated mapper with an area-based cost model to explore compute, memory bandwidth, and buffer configurations, achieving 4\% error for end-to-end LLM inference on A100 nodes within minutes.

These tools operate at the operator and tile level, modeling how a single convolution or matrix multiplication executes on a specific microarchitecture. They do not reason about system-level concerns such as how multiple accelerators communicate across a network fabric, how the data pipeline feeds those accelerators, or what the total cost of ownership looks like at fleet scale. \mlsysim operates one abstraction level higher. It consumes the \emph{outputs} of accelerator-level analysis (peak FLOP/s, memory bandwidth, TDP) as inputs to its hardware registry and reasons about how those specifications interact with workload demands, network topologies, and infrastructure constraints.

\subsection{Analytical and Co-Design Tools}

Closest in spirit to \mlsysim are analytical tools that sacrifice microarchitectural detail for speed. Calculon~\citep{isaev2023} is an analytical co-design tool for large language model training. It models training time as a function of hardware specifications, parallelism strategies, and model architecture, achieving execution speeds comparable to \mlsysim. However, Calculon's scope is narrow by design, targeting transformer-based LLM training exclusively, with no support for CNNs, mixture-of-experts architectures, or inference workloads. It does not model data pipelines, reliability, sustainability, economics, or safety considerations, and it lacks dimensional enforcement. Lumos~\citep{liang2025lumos} takes a trace-driven approach, using profiled kernel traces to predict LLM training performance at scale with 3.3\% average error on up to 512 H100 GPUs. While Lumos achieves higher single-point accuracy than \mlsysim, it requires empirical traces from the target hardware, limiting its use in what-if exploration across hypothetical configurations. The DeepSeek-V3 systems paper~\citep{deepseek2025v3} exemplifies the kind of hardware-aware co-design analysis that \mlsysim targets: it demonstrates how FP8 mixed-precision training, MoE sparsity, and multi-plane network topology interact to achieve frontier-model training at a fraction of conventional cost, a multi-wall optimization that spans Walls 1, 3, 13, 14, and 17 in our taxonomy.

Paleo~\citep{qi2017paleo} pioneered the analytical approach, decomposing DNN training time into computation and communication components across data- and model-parallel configurations. FlexFlow~\citep{jia2019flexflow} optimizes parallelism strategies through simulation-guided search, and Habitat~\citep{yu2021habitat} provides cross-hardware extrapolation of training performance using execution-time scaling curves. While these tools advance specific aspects of analytical modeling, they predate or do not address the full scope of modern concerns (inference serving, fleet economics, sustainability). Vidur~\citep{agrawal2024vidur} extends analytical modeling to LLM \emph{inference}, using operator-level profiling to build a fine-grained runtime estimator validated at less than 5\% error across multiple LLMs and scheduling policies. GenZ~\citep{bambhaniya2024genz} provides an analytical framework for LLM inference platform design that models multi-dimensional network topologies and serving optimizations. DistServe~\citep{zhong2024distserve} and Sarathi-Serve~\citep{agrawal2025} advance LLM serving through prefill-decode disaggregation and chunked-prefill scheduling respectively, demonstrating that the two-phase inference model (\Cref{eq:serving}) requires increasingly sophisticated scheduling to achieve high goodput. All of these tools focus on inference performance in isolation; they do not model training, data pipelines, economics, or sustainability.

LLM-Viewer~\citep{yuan2024llmviewer} and llm-analysis~\citep{kim2023llmanalysis} provide lightweight memory and latency estimation for transformer inference. These tools are useful for single-model profiling but do not extend to fleet-level reasoning, multi-tenant scheduling, or cross-domain constraint analysis.

A parallel line of work targets sustainability and fleet efficiency. LLMCarbon~\citep{faiz2024llmcarbon} projects end-to-end carbon footprints (operational and embodied) for dense and MoE LLMs, validated within 8\% of Google's published figures. CodeCarbon~\citep{lottick2019codecarbon} provides empirical energy tracking at runtime via hardware power monitors. \citet{wongpanich2025fleet} introduce \emph{ML Productivity Goodput} (MPG) as a fleet-level efficiency metric for warehouse-scale TPU clusters, demonstrating that traditional utilization metrics are insufficient for characterizing ML fleet performance across model, data, framework, compiler, and scheduling layers. These tools each address one or two domains and do not model the full cross-stack interactions that determine \emph{why} a workload consumes so much energy.

\mlsysim generalizes the analytical approach to the full ML systems stack. Where each tool above models one or two domains, \mlsysim composes resolvers spanning all six (\Cref{tab:walls}): compute, memory, serving, data pipelines, scaling laws, fleet coordination, economics, sustainability, and responsible engineering. Runtime dimensional strictness (SI units, no bare physical floats) is the invariant of \Cref{sec:dimensional}.

\subsection{Pedagogical Precedents}

\mlsysim draws direct inspiration from the tradition of pedagogical simulators in systems education. Patterson and Hennessy's MIPS/SPIM simulator~\citep{patterson2014organization} taught generations of students computer architecture not by replicating a production processor, but by providing a simplified model that made architectural concepts tangible through rapid experimentation. Similarly, xv6~\citep{cox2011xv6} and MINIX~\citep{tanenbaum2006minix} teach operating systems by stripping away production complexity to reveal core abstractions.

\mlsysim follows this pedagogical philosophy for the ML systems domain. Production ML infrastructure (spanning millions of lines of code across frameworks, compilers, schedulers, and orchestrators) is too complex for students to reason about directly. \mlsysim provides a controlled environment in which students can sweep hardware configurations, vary parallelism strategies, and observe how binding constraints shift, all in rapid iteration cycles. Its integration with the companion textbook~\citep{mlsysbook2025} provides structured laboratory exercises with autogradable assessments, a capability absent from every research-oriented tool surveyed in \Cref{sec:related}.

\section{Architecture}
\label{sec:architecture}

\begin{figure*}[!t]
\centering
\includegraphics[width=\textwidth]{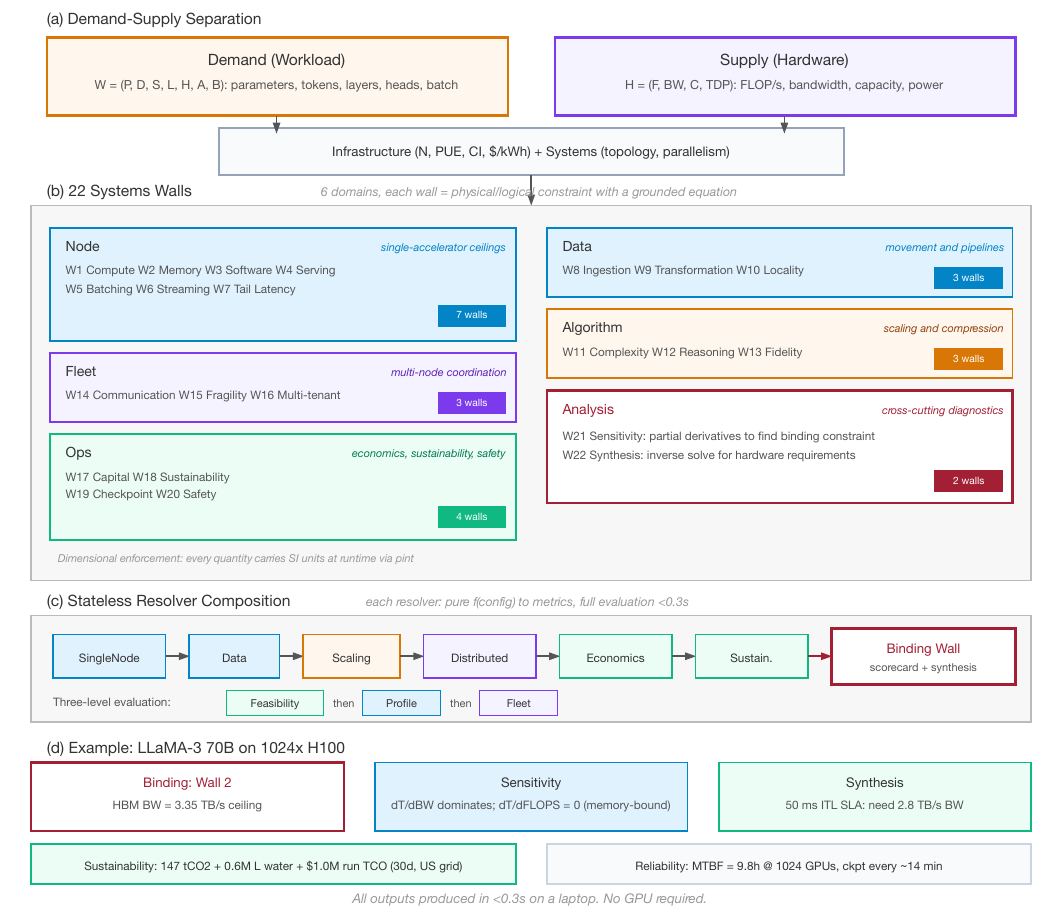}
\caption{\textbf{\mlsysim Framework Overview.} (a)~Demand--supply separation decouples workload specifications from hardware capabilities and environmental context. (b)~All 22 Systems Walls organized into six domains (Node, Data, Algorithm, Fleet, Operations, Analysis), each grounded in a published equation. (c)~Stateless resolver composition chains 28 resolvers to identify binding constraints through a three-level evaluation. (d)~Example outputs for LLaMA-3 70B on 1024$\times$ H100, produced in $<$0.3\,s on a laptop.}
\label{fig:overview}
\end{figure*}

\Cref{fig:overview} presents the end-to-end framework. This section unpacks the design in five parts. \Cref{sec:philosophy} introduces the four principles (analytical speed, dimensional strictness, taxonomic completeness, demand--supply separation) that constrain every subsequent decision. \Cref{sec:stack} translates those principles into a five-layer input stack of workloads, hardware, infrastructure, topology, and resolvers. \Cref{sec:types} describes the dimensionally strict type system that enforces correctness at every API boundary, \Cref{sec:extensibility} explains how new workloads, hardware, and resolvers compose with the existing pipeline without modification, and \Cref{sec:api} maps the public import surface and the presentation layer that sits outside the analytical engine.

\subsection{Design Philosophy}
\label{sec:philosophy}

The central question \mlsysim addresses is: \emph{Where does the complexity of an ML system come from?} We argue that complexity arises from the non-linear interaction of constraints across six distinct domains (node resources, data movement, algorithms, fleet coordination, operations, and cross-cutting analysis) and that an effective modeling tool must formalize \emph{all six} simultaneously. Four design principles govern \mlsysim's approach.

\subsubsection{Analytical Speed over Cycle Accuracy}
\label{sec:speed}

Analytical models that execute in sub-second time enable iterative design-space exploration that cycle-accurate simulators cannot support at comparable speed. By absorbing microarchitectural detail (cache hit rates, warp scheduling) into a single efficiency parameter~$\eta$, \mlsysim achieves sub-second execution per solve. This enables sweeps over thousands of hardware--model--topology combinations in seconds, the kind of rapid ``what-if'' exploration that \citet{hennessy2024architecture} identify as necessary for quantitative architectural reasoning.

To maintain rigor despite analytical simplification, \mlsysim enforces a \textbf{``No Magic Numbers'' invariant}. Every registry value carries a typed \texttt{Provenance} record whose \emph{kind} (datasheet, literature, industry report, convention, estimate, derived, illustrative, or heuristic) determines what evidence it must supply: evidence-backed kinds require a source URL and a verification date, while estimates and derived values require explanatory notes, and all records require a dated verification. The H100's peak FP16 throughput is not a bare \texttt{989} floating-point literal; it is \texttt{989\,*\,TFLOPs\,/\,second}, sourced from NVIDIA's published datasheet~\citep{nvidia2023h100}. Shared sources live once in a central catalog of 126 provenance records (\texttt{core.provenance\_catalog}) that registry entries reference by key, and an audit tool (\texttt{python -m mlsysim.tools.audit\_provenance}) walks every registry and fails on any value with missing or weak provenance. This discipline ensures that analytical speed does not come at the cost of reproducibility.

\subsubsection{Dimensional Strictness as an Invariant}
\label{sec:dimensional}

Dimensional consistency is a pervasive challenge in systems modeling. Mixing gigabits with gigabytes, or omitting the refractive index of fiber in latency calculations, are canonical failure modes that silently corrupt results. \mlsysim treats dimensional correctness not as a feature but as a \textbf{runtime invariant}, analogous to memory safety in Rust. The framework wraps every physical quantity using the \texttt{pint} unit library, and gives compute work its own base dimension: a FLOP is not a dimensionless count, so compute throughput can never silently add to, or convert into, a memory bandwidth --- the category error at the heart of many back-of-the-envelope mistakes. Mixing any two of \{compute rate, bandwidth, latency\} raises a deterministic \texttt{DimensionalityError} before any computation proceeds, while the ratios that \emph{should} compose still do: arithmetic intensity (FLOP/byte) divides cleanly, and utilization (achieved over peak FLOP rate) reduces to a dimensionless fraction:

\begin{lstlisting}[caption={\textbf{Dimensional Strictness.} Prevents silent unit errors at the API level.},label={lst:units}]
from mlsysim.core.units import Q_
rate = Q_("989 TFLOPs/s")    # Compute throughput
bw   = Q_("3.35 TB/s")       # Memory bandwidth
(rate / bw).to("flop/byte")  # -> 295 flop/byte (ridge point)
rate + bw  # -> raises DimensionalityError
\end{lstlisting}

This design eliminates the single most common class of bugs in back-of-the-envelope systems analysis, namely silent unit conversion errors.\footnote{The most infamous unit-conversion failure is the Mars Climate Orbiter, lost in 1999 because ground software produced thrust data in pound-force seconds while the spacecraft expected newton seconds~\citep{stephenson1999mco}.} Hardware and model specifications live in typed registries (e.g., \texttt{Hardware.Cloud.H100.memory.bandwidth = 3.35\,*\,TB/s}); the unit vocabulary lives in \texttt{core.units}, and physical constants and formulas live in \texttt{mlsysim.physics}, preventing unit mismatches at the API level. The physics formulas additionally validate the \emph{dimensionality} of their arguments, raising \texttt{DimensionalityError} when, for example, a byte count is passed where a bandwidth is expected.

\subsubsection{Taxonomic Completeness}
\label{sec:taxonomic}

We define a modeling framework as ``complete'' only when every fundamental bottleneck to scaling has a mathematical resolver. \mlsysim codifies 22 such bottlenecks, which we call \emph{Systems Walls}, organized into six domains: Node (single-accelerator resources), Data (movement and pipelines), Algorithm (scaling and compression), Fleet (multi-node coordination), Operations (economics, sustainability, and safety), and Analysis (cross-cutting diagnostics). Each wall maps to a dedicated resolver with a formal equation grounded in the systems literature. \Cref{sec:taxonomy} presents the full taxonomy with formal definitions, and \Cref{sec:solver-formalism} describes how solvers compose.

\subsubsection{Demand--Supply Separation}
\label{sec:demand-supply}

\mlsysim enforces a strict separation between \emph{what} a model computes and \emph{where} it runs. A \texttt{TransformerWorkload} describes computational demand (parameters, layers, FLOPs, arithmetic intensity) without reference to any specific accelerator. A \texttt{HardwareNode} describes physical supply (peak throughput, memory bandwidth, TDP) without reference to any specific model. This decoupling, inspired by the compiler IR philosophy of separating representation levels~\citep{hennessy2024architecture}, enables hardware--software co-design. The same GPT-3 workload can be evaluated against an H100, a TPU~v5p, or a hypothetical future accelerator in a single parametric sweep, with all dimensional conversions handled automatically.

\subsection{The Layered Input Stack}
\label{sec:stack}

\begin{figure*}[!t]
\centering
\includegraphics[width=\textwidth]{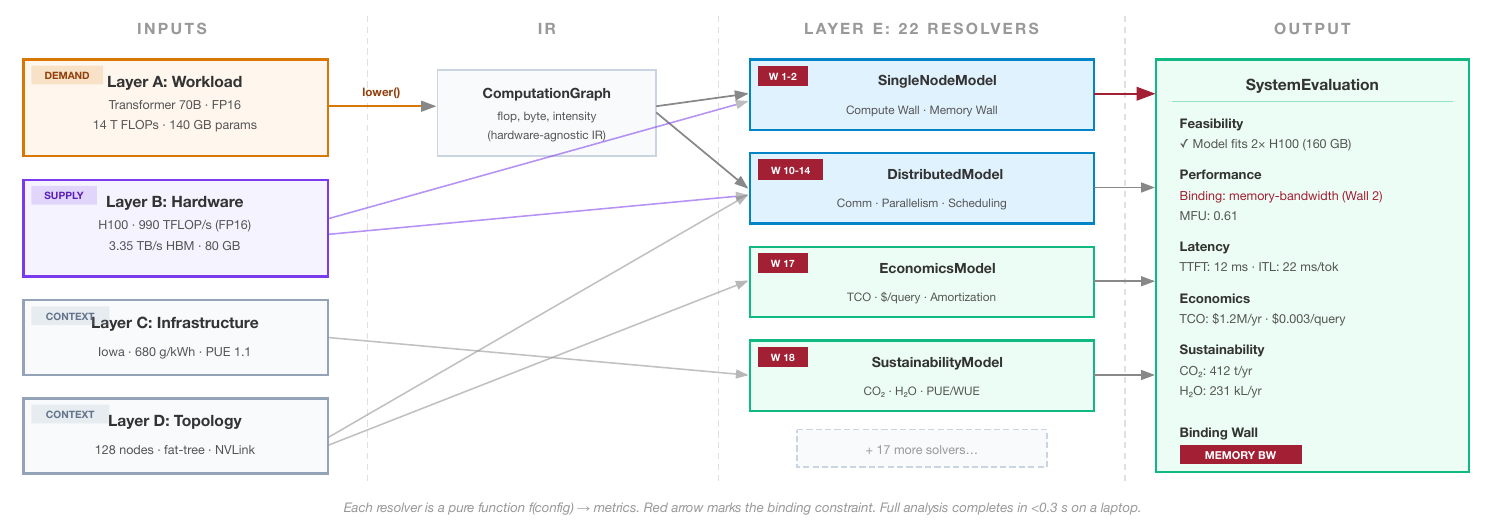}
\caption{\textbf{The \mlsysim 5-Layer Architecture.} Layers A--D provide typed inputs: workload demand, hardware supply, infrastructure context, and network topology. The single lowering step occurs in Layer~A, where \texttt{lower()} produces a hardware-agnostic Computation Graph (an intermediate representation of total FLOPs, weight bytes, and arithmetic intensity). Layer~E's 28 stateless resolvers consume these inputs and produce a three-level SystemEvaluation scorecard. The red arrow marks the binding constraint identified by the solver chain.}
\label{fig:stack}
\end{figure*}

\mlsysim implements these design principles through a five-layer architecture (\Cref{fig:stack}). Layers A--D are independent \emph{input layers} that describe demand, supply, context, and topology respectively; they do not depend on one another.  The single lowering step occurs in Layer~A, where a workload's \texttt{lower()} method produces a hardware-agnostic \texttt{ComputationGraph} intermediate representation.  Layer~E (Resolvers) then consumes any combination of layers A--D as needed; a single-node analysis requires only A+B, while a fleet-wide carbon estimate draws on A+B+C+D.

\textbf{Layer A: Workloads (Demand).} A workload is a hardware-agnostic description of computational demand. \mlsysim provides five concrete workload types (\Cref{tab:workloads}), each exposing a \texttt{lower()} method that produces a \texttt{ComputationGraph} (an intermediate representation containing total operations, weight bytes, and arithmetic intensity in \texttt{flop/byte}). This IR is the contract between demand and supply, capturing \emph{what} must be computed without prescribing \emph{how}.

\begin{table}[!t]
\centering
\caption{\textbf{Supported Workload Types.} Each workload lowers to a \texttt{ComputationGraph} with total FLOPs, weight bytes, and arithmetic intensity.}
\label{tab:workloads}
\small
\renewcommand{\arraystretch}{1.1}
\begin{tabularx}{\columnwidth}{@{}l X l@{}}
\toprule
\textbf{Workload} & \textbf{Key Parameters} & \textbf{Scaling} \\
\midrule
Transformer     & $P$, $L$, $H$, $D$, seq.\ length & $2P$ FLOPs/token \\
CNN             & $P$, inference FLOPs             & Fixed per image \\
Sparse (MoE)    & Total vs.\ active $P$, experts   & Active $P$ for FLOPs \\
SSM (Mamba)     & $P$, state dim, $D$              & $O(1)$ state cache \\
Diffusion       & $P$, denoising steps $T$         & $T \times$ FLOPs/step \\
\bottomrule
\end{tabularx}
\end{table}

\textbf{Layer B: Hardware (Supply).} A \texttt{HardwareNode} composes four subsystems: \texttt{ComputeCore} (peak FLOP/s with a precision-keyed dictionary for FP16, TF32, FP8, INT8), \texttt{MemoryHierarchy} (capacity and bandwidth), optional \texttt{StorageHierarchy}, and optional \texttt{IOInterconnect}. Each node also carries TDP, unit cost, and a kernel dispatch tax. The \texttt{ridge\_point()} method computes the Roofline inflection $R = F_{\text{peak}} / BW_{\text{mem}}$ in \texttt{flop/byte}~\citep{williams2009}, enabling immediate classification of any lowered workload as compute-bound or memory-bound.

\textbf{Layer C: Infrastructure (Context).} \texttt{GridProfile} objects encode regional environmental parameters: carbon intensity (gCO$_2$/kWh), Power Usage Effectiveness (PUE), and Water Usage Effectiveness (WUE). A \texttt{Datacenter} composes a grid profile with rack-level power density constraints. This layer converts raw energy consumption into carbon footprint and water usage, following the methodology of \citet{patterson2021carbon}.

\textbf{Layer D: Systems (Topology).} A \texttt{Fleet} composes \texttt{Node}s (each a physical compute server containing one or more accelerators on PCIe or NVLink) with a \texttt{NetworkFabric} (topology, inter-node bandwidth, latency, oversubscription ratio). This layer enables distributed analysis. The \texttt{DistributedModel} decomposes workloads using 4D parallelism (data-parallel~DP $\times$ tensor-parallel~TP $\times$ pipeline-parallel~PP $\times$ expert-parallel~EP) and calculates hierarchical AllReduce costs, pipeline bubble fractions, and scaling efficiency.

\textbf{Layer E: Resolvers (Analysis).} Twenty-eight stateless resolvers (23 models, 2 solvers, and 3 optimizers) consume demand, supply, context, and topology to produce dimensioned performance metrics. The solver formalism (stateless composition, chaining semantics, and three-level evaluation) is detailed in \Cref{sec:solver-formalism}.

The five layers are useful only if their inputs are reproducible. \mlsysim therefore organizes vetted specifications into eight curated registries collectively called the \emph{MLSys Zoo}: \textbf{Hardware} (accelerators and boards, plus technology classes under \texttt{Hardware.Tech}), \textbf{Models} (workload architectures across six family files: language, vision, generative vision, recommendation, state-space, and TinyML), \textbf{Datasets} (benchmark catalog entries), \textbf{Platforms} (deployment envelopes), \textbf{Infrastructure} (grids, datacenters, racks, cooling, pricing, and capacity), \textbf{Systems} (nodes, racks, fabrics, clusters, pods, and storage paths), \textbf{Ops} (operational policies and overhead profiles such as training-run goodput budgets and monitoring thresholds), and executable \textbf{Scenarios} that compose existing model, hardware, infrastructure, and system entries with local constraints such as latency, power, run length, or region without redefining the underlying facts. Two auxiliary surfaces complete the picture: non-executable real-world anchors, such as Waymo data-rate ranges or mobile power envelopes, live in \texttt{ReferenceStats.*}, and cited field figures (MFU bands, scaling laws) live in \texttt{Literature.*}. Solver fallbacks live in \texttt{mlsysim.engine.calibration}; architecture formulas (Roofline, AllReduce, Young--Daly) live in \texttt{mlsysim.physics}; the machine-readable wall taxonomy of \Cref{tab:walls} ships as \texttt{mlsysim.engine.walls}; and domain-reviewed empirical anchors in \texttt{mlsysim.engine.empirical} bind registry entries to published benchmark envelopes for regression testing.

\begin{figure*}[!t]
\centering
\includegraphics[width=\textwidth]{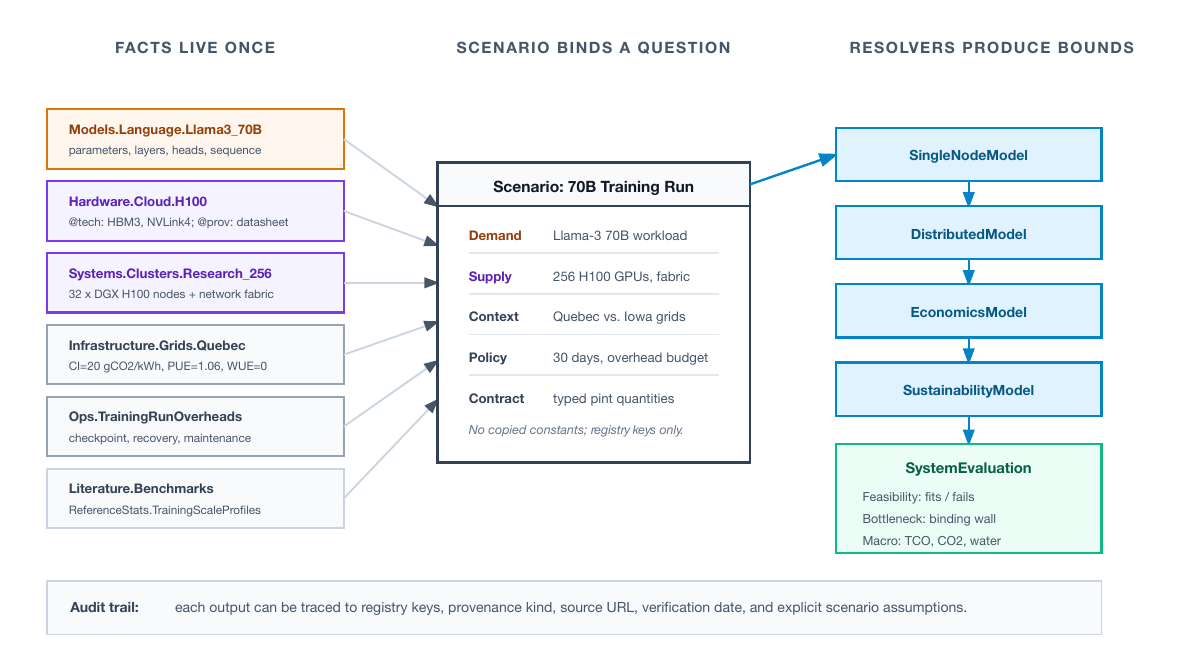}
\caption{\textbf{Registry-Backed System Composition.} \mlsysim stores workload, hardware, systems, infrastructure, operational, and literature facts once, with typed schemas and provenance records. A scenario composes those reusable facts into a local infrastructure question---for example, training a 70B model on a 256-GPU H100 fleet in Qu\'ebec or Iowa---and hands dimensioned objects to the resolver chain. The resulting \texttt{SystemEvaluation} can be traced back to the registry keys, source records, and scenario assumptions that produced it.}
\label{fig:registry-composition}
\end{figure*}

Registry entries are authored as YAML data files and validated against \texttt{pydantic} schemas at import time, so stating a fact is a data edit while computing or composing remains Python. The loader rejects duplicate keys outright and resolves two reference forms: \texttt{@tech:} pointers that let an instance inherit a technology-class fact, and \texttt{@prov:} pointers into the shared provenance catalog (\Cref{sec:speed}). The \texttt{Hardware.Tech} sub-registry encodes the load-bearing distinction between \emph{instances} and \emph{technology classes}: per-part facts that vary between products (capacity, bandwidth, TDP, price) live on instances such as \texttt{Hardware.Cloud.H100}, while facts that are properties of a technology generation (HBM access latency, energy per operation, interconnect latencies) live once on tech-class entries such as \texttt{Hardware.Tech.Memory.HBM3}, with instance lookups falling back to the tech class when unset. Each registry entry is a fully typed object; for example, \texttt{Hardware.Cloud.H100} returns a \texttt{HardwareNode} with all physical quantities dimensioned via \texttt{pint}. \Cref{fig:registry-composition} illustrates how these reusable facts become a concrete infrastructure model: a scenario binds a workload, accelerator fleet, regional grid, and operational policy to a local question, then passes the resulting typed objects to the resolver chain.

Once a scenario has assembled typed inputs, the \texttt{physics} package decomposes the framework's analytical equations into nine domain-aligned modules (\Cref{tab:physics-modules}), each owning one conceptual domain of the systems stack, plus two supporting modules: \texttt{physics.constants} holds universal physical constants (e.g., the speed of light in fiber), and \texttt{physics.quantities} provides generic dimensioned helpers (transfer time, compute time, energy from power, carbon from energy). A shared \texttt{\_ensure\_unit()} helper coerces raw floats into \texttt{pint} quantities at module entry points, guaranteeing dimensional correctness without burdening callers. Chapter and notebook code import from \texttt{mlsysim.physics} directly. The appendix (\Cref{tab:physics-api}) lists the complete function signatures.

\begin{table}[!t]
\centering
\caption{\textbf{Physics Domain Modules.} Each module owns the analytical formulas for one taxonomy domain. All functions accept and return \texttt{pint} quantities.}
\label{tab:physics-modules}
\small
\renewcommand{\arraystretch}{1.08}
\begin{tabularx}{\columnwidth}{@{}l l X@{}}
\toprule
\textbf{Module} & \textbf{Domain} & \textbf{Key Formulas} \\
\midrule
\texttt{performance}    & Node        & Roofline bottleneck, Amdahl's speedup, pipeline bubble \\
\texttt{memory}         & Node        & Model memory, activation memory, KV-cache, checkpoint size \\
\texttt{serving}        & Node        & Erlang-C M/M/$c$ queue latency \\
\texttt{communication}  & Fleet       & Ring/tree/hierarchical AllReduce, All-to-All \\
\texttt{reliability}    & Fleet       & Young--Daly, cluster MTBF, failure probability \\
\texttt{transformer}    & Algorithm   & Training/decode FLOP counts \\
\texttt{economics}      & Operations  & Fleet TCO, egress cost \\
\texttt{networking}     & Data        & Speed-of-light latency \\
\texttt{statistics}     & Analysis    & PSI drift detection, sample sizing \\
\bottomrule
\end{tabularx}
\end{table}

\begin{figure*}[!t]
\centering
\includegraphics[width=\textwidth]{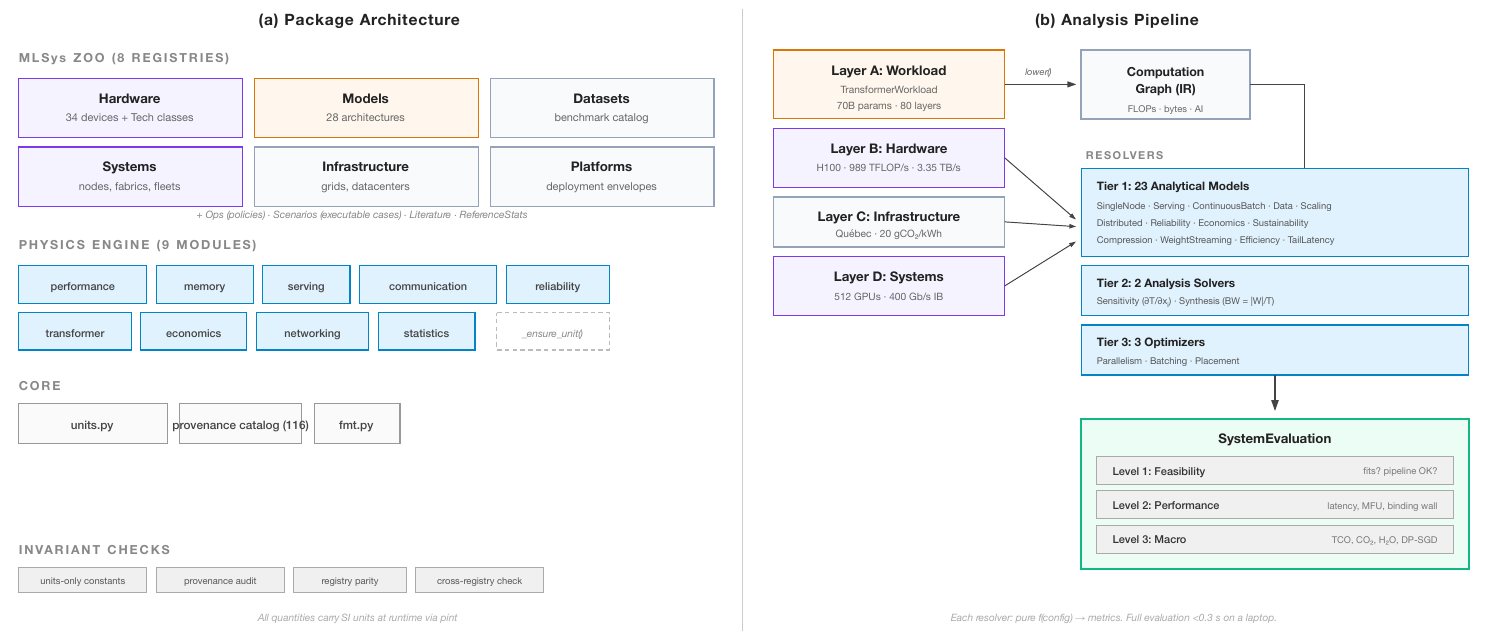}
\caption{\textbf{System Anatomy.} (a)~Package architecture: eight registries (six shown) provide typed, provenance-tracked specifications, nine physics modules supply domain-aligned formulas, and five invariant checks enforce design decisions as the codebase evolves. (b)~Analysis pipeline: a workload lowers to a hardware-agnostic intermediate representation; four input layers feed 28 stateless resolvers organized in three tiers; the output is a three-level \texttt{SystemEvaluation} scorecard (Feasibility, Performance, Macro).}
\label{fig:anatomy}
\end{figure*}

The \textbf{Hardware} registry holds 37 devices across five deployment tiers (Cloud, Workstation, Edge, Mobile, Tiny), spanning nine orders of magnitude in peak throughput from milliwatt-class microcontrollers (\texttt{Hardware.Tiny.ESP32\_S3}, 520\,KiB on-chip SRAM) to wafer-scale engines (\texttt{Hardware.Cloud.Cerebras\_CS3}, 44\,GB on-wafer SRAM, 125\,PFLOP/s~\citep{lie2022cerebras}). Beyond NVIDIA GPUs (T4 and V100 through B200 and the rack-scale GB200~NVL72), the registry includes AMD MI250X and MI300X, Intel Gaudi~2 and Gaudi~3, AWS Trainium~2, and the full Google TPU lineage (v1 through v6/Trillium), enabling cross-vendor design-space exploration, plus workstation (Apple M3 Max, DGX Spark), edge (Jetson, Coral), mobile (Apple A17 Pro, Snapdragon 8 Gen~3, Tensor G3), and TinyML (ESP32-S3, nRF52840, Himax WE-1) devices. The \texttt{list(sort\_by=)} class method enables programmatic comparison, and users extend any zoo by instantiating new typed objects (\Cref{lst:hwnode}) or adding YAML data files, ensuring custom entries participate in the same dimensionally strict pipeline as vetted ones. \Cref{fig:anatomy} maps the implementation structure behind this composition path: panel~(a) shows package architecture (registries, physics modules, core utilities, and invariant checks), while panel~(b) traces a concrete analysis from typed inputs through the resolver chain to the three-level \texttt{SystemEvaluation} scorecard. \Cref{sec:appendix-hardware} gives a representative snapshot of the current registry surface.

\subsection{The Type System}
\label{sec:types}

\mlsysim's type system is built on Pydantic \texttt{BaseModel} classes with \texttt{pint} \texttt{Quantity} fields, providing both schema validation and dimensional enforcement at construction time. The same schemas validate the YAML registry data: every \texttt{*.yaml} entry is instantiated through its \texttt{pydantic} type at import, so a malformed unit string or a missing provenance record fails the import rather than surfacing later as a wrong number. The composition hierarchy is deliberately shallow: \texttt{HardwareNode} aggregates \texttt{ComputeCore} and \texttt{MemoryHierarchy} as direct fields, not through deep inheritance. This design makes the relationship between a hardware specification and its physical quantities immediately legible:

\begin{lstlisting}[caption={\textbf{Custom Hardware Node.} Composing a hardware specification with dimensional types.},label={lst:hwnode}]
from mlsysim.hardware.types import *
from mlsysim.core.units import Q_
node = HardwareNode(
    name="Custom Accelerator",
    release_year=2025,
    compute=ComputeCore(
        peak_flops=Q_("500 TFLOPs/s"),
        precision_flops={"fp8": Q_("1000 TFLOPs/s")}),
    memory=MemoryHierarchy(
        capacity=Q_("96 GiB"),
        bandwidth=Q_("4 TB/s")),
    tdp=Q_("500 W"))
print(node.ridge_point())  # -> 125 flop/byte
\end{lstlisting}

The \texttt{ComputationGraph} IR bridges the demand--supply gap. When a solver calls \texttt{workload.lower()}, the workload computes its total operations, weight bytes, and arithmetic intensity, all in dimensioned quantities. For Mixture-of-Experts models, \texttt{SparseTransformerWorkload.lower()} uses \emph{active} parameters for FLOPs but \emph{total} parameters for memory footprint, correctly modeling the fundamental decoupling between compute cost and capacity requirements in sparse architectures~\citep{shazeer2017outrageously}.

The complete evaluation produces a \texttt{SystemEvaluation} scorecard, a single object containing every metric from every resolver, cross-referenced by wall number. Students can inspect any individual wall or view the aggregate to understand how constraints interact across the full stack.

\subsection{Extensibility}
\label{sec:extensibility}

The layered architecture is designed for extension at every level. New workload types (e.g., a \texttt{RetrievalAugmentedWorkload} for RAG pipelines) require only implementing the \texttt{lower()} method to produce a \texttt{ComputationGraph}; all existing resolvers then apply without modification. New hardware entries are added to the \textbf{Hardware} registry as declarative \texttt{HardwareNode} specifications (\Cref{lst:hwnode}), with no resolver changes needed. New resolvers can be introduced for emerging constraints by implementing the appropriate tier interface: a Tier 1 Model (e.g., a \texttt{PrivacyModel} for federated learning overhead), a Tier 2 Solver, or a Tier 3 Optimizer.

By accepting typed inputs and returning dimensioned outputs, the type system enforces correctness at every boundary, ensuring that custom extensions compose safely with existing components. This design ensures that \mlsysim can track the rapidly evolving ML systems landscape without requiring architectural changes to the core framework.

\subsection{Public API and Presentation Layer}
\label{sec:api}

The package separates analysis from presentation. \texttt{from mlsysim import *} exposes the registries, the physics formulas, the unit vocabulary, and the formatting helpers; solver classes are imported through the stable \texttt{mlsysim.solvers} path, which mechanically re-exports the canonical implementations in \texttt{mlsysim.engine.solvers} so the two surfaces cannot drift. The \texttt{engine} package is the infrastructure-modeling core: the Roofline core (\texttt{engine.engine}), the 28 resolvers, the three-level evaluation, pipeline composition, design-space exploration, executable scenarios, the machine-readable wall taxonomy, calibration defaults, and empirical anchors. Presentation lives outside the engine: \texttt{mlsysim.fmt} provides a typed \texttt{fmt\_*} formatter family (50+ helpers for quantities, counts, currencies, percentages, and scientific notation) whose outputs are Markdown-safe strings and whose precision-safety guards refuse to render non-finite values or to silently round a non-zero value to the literal ``0''; \texttt{mlsysim.viz} supplies matplotlib plotters with a consistent style; and a command-line interface (\texttt{python -m mlsysim}) exposes the zoo browser, scenario evaluation, serving analysis, design-space optimization, a local-hardware Iron Law audit, and schema export.

\section{Taxonomy of ML Systems Walls}
\label{sec:taxonomy}

\begin{table}[!t]
\centering
\caption{\textbf{Notation.} Symbols used throughout; all quantities carry SI units at runtime via \texttt{pint}.}
\label{tab:notation}
\footnotesize
\setlength{\tabcolsep}{4pt}
\renewcommand{\arraystretch}{1.05}
\begin{tabularx}{\columnwidth}{@{}l l X@{}}
\toprule
\textbf{Symbol} & \textbf{Unit} & \textbf{Description} \\
\midrule
\multicolumn{3}{@{}l}{\textit{Model \& Workload}} \\
$P$, $P_{\ell}$ & params & Total / per-layer parameter count \\
$|W|$, $|W_{\ell}|$ & bytes & Bytes read per step / per layer \\
$b_{\text{prec}}$ & B/param & Precision (e.g., 2 for FP16) \\
$L$, $H$, $D$ & -- & Layers, attention heads, head dim \\
$S$ & tokens & Sequence length \\
$B$ & samples & Batch size \\
$K$ & -- & Reasoning steps \\
$C$ & FLOPs & Training compute ($6PD$) \\
$I$ & FLOP/B & Arithmetic intensity \\
\midrule
\multicolumn{3}{@{}l}{\textit{Hardware \& Infrastructure}} \\
$\text{Peak}_{\text{FLOPS}}$ & FLOP/s & Peak accelerator throughput \\
$BW_{\text{HBM}}$ & B/s & HBM bandwidth \\
$BW_{\text{inject}}$ & B/s & Injection BW (wafer-scale) \\
$BW_{\text{link}}$ & B/s & Per-link network bandwidth \\
$N$, $G$ & -- & Nodes in fleet, GPUs per node \\
$\alpha$ & s & Per-hop network latency \\
\midrule
\multicolumn{3}{@{}l}{\textit{Efficiency \& Utilization}} \\
$\eta$ & -- & HW utilization ($\approx$ MFU) \\
$\eta_{\text{overlap}}$ & -- & Compute--comm overlap \\
$\rho$ & -- & Utilization ratio (queue/data) \\
$\beta$, $\beta_{\text{opt}}$ & -- & Bisection BW frac, optimizer multiplier \\
\midrule
\multicolumn{3}{@{}l}{\textit{Parallelism}} \\
TP, PP, DP, EP & -- & Tensor, pipeline, data, expert parallel \\
$V$, $M_{\text{micro}}$ & -- & Virtual stages, microbatches \\
\midrule
\multicolumn{3}{@{}l}{\textit{Sustainability}} \\
PUE, WUE & --, L/kWh & Power / Water Usage Effectiveness \\
$\text{CI}$ & gCO$_2$/kWh & Regional carbon intensity \\
\midrule
\multicolumn{3}{@{}l}{\textit{Key Derived Quantities}} \\
$I^{*}$ & FLOP/B & Ridge point ($\text{Peak}/BW_{\text{HBM}}$) \\
$B^{*}$, $P^{*}$, $D^{*}$ & varies & Optimal batch, model, dataset size \\
$\tau_{\text{opt}}$ & s & Optimal checkpoint interval \\
\bottomrule
\end{tabularx}
\end{table}

The ML systems literature is rich with specialized models for specific bottlenecks, from the original Roofline model~\citep{williams2009} and Chinchilla scaling laws~\citep{hoffmann2022chinchilla} to PagedAttention batching limits~\citep{kwon2023} and datacenter sustainability accounting~\citep{patterson2021carbon}. However, these constraints are typically studied in isolation. We synthesize this disjointed literature into a unified taxonomy of 20 distinct ``Walls'' plus 2 cross-cutting diagnostic tools (Sensitivity and Synthesis). We borrow the term from the computer architecture tradition (the ``memory wall''~\citep{williams2009}, the ``power wall''~\citep{hennessy2024architecture}), where a \emph{wall} denotes a hard physical or logical constraint that bounds system performance and cannot be circumvented by software optimization alone. Codifying these previously disparate equations into a single, composable framework, each wall is resolved by a dedicated resolver (model or solver) that accepts typed inputs and produces dimensionally correct bounds. This integration of theoretical and empirical constraints into a single executable engine has not been done before. The complete taxonomy is summarized in \Cref{tab:walls}.

The taxonomy proceeds outward from the silicon. \Cref{sec:walls-node} covers the seven Node walls that bound a single accelerator (compute, memory, software efficiency, serving, batching, streaming, tail latency). \Cref{sec:walls-data} adds the three Data walls that govern how samples reach the accelerator (ingestion, transformation, locality). \Cref{sec:walls-algorithm} covers the three Algorithm walls that originate in the mathematics of learning (complexity, reasoning, fidelity). \Cref{sec:walls-fleet} adds the three Fleet walls that arise once a job spans multiple nodes (communication, fragility, multi-tenant scheduling). \Cref{sec:walls-operations} covers the four Operations walls that determine whether a system should run at all (capital, sustainability, checkpoint, safety). Finally, \Cref{sec:walls-analysis} introduces the two cross-cutting diagnostic tools (Sensitivity and Synthesis) that operate across the entire taxonomy.

\subsection{Node (Single-Accelerator Resources)}
\label{sec:walls-node}

The Node walls define what a single accelerator (one physical processing unit such as a GPU, TPU, or wafer-scale engine, used interchangeably with ``individual accelerator'') can achieve in isolation, before any distributed coordination is considered. Multi-node analysis, where a server hosts several such accelerators on intra-node links such as NVLink, falls under the Fleet domain (\Cref{sec:walls-fleet}). Node walls are the innermost constraints and the first a practitioner should evaluate.

\textbf{Wall~1: The Compute Wall.} Every accelerator has a hard throughput ceiling determined by the number of arithmetic units and the clock frequency. An H100, for example, provides a peak of \cvalint{\HhPeakTFLOPS}\,TFLOP/s at FP16 with Tensor Cores, establishing an upper bound that no software optimization can exceed. The \texttt{SingleNodeModel} evaluates the classical Roofline model bounds~\citep{williams2009}. If execution is entirely compute-bound, the minimum time to process the workload is dictated by the \emph{Iron Law of ML Systems}:
\begin{equation}
\label{eq:tcompute}
T_{\text{compute}} = \frac{\text{OPs}}{\text{Peak}_{\text{FLOPS}} \times \eta}
\end{equation}
where $\eta \in (0,1]$ is the hardware utilization efficiency and OPs is the total operation count. Throughout this paper, $\eta$ denotes the ratio of sustained to peak throughput; the related metric \emph{Model FLOPS Utilization} (MFU) measures only model-useful FLOPs and excludes overhead such as activation recomputation. For first-order analysis, we treat $\eta \approx \text{MFU}$; the distinction matters only when recomputation or non-model compute is significant. When this wall binds, the only remedy is faster silicon or fewer operations. \textbf{Assumptions:} Peak FLOPS is a hard ceiling; $\eta$ is workload-dependent and must be specified or estimated from benchmark data.

\textbf{Wall~2: The Memory Wall.} High-bandwidth memory (HBM) imposes two ceilings: capacity (the model must fit) and bandwidth (weights must stream to compute units fast enough). An H100 reads HBM at \cvalii{\HhBWTBs}\,TB/s, yet its \cvalint{\HhPeakTFLOPS}\,TFLOP/s demand data at a rate that exceeds this bandwidth for any workload below ${\sim}\cvalint{\HhRidgePoint}$\,flop/byte, making most LLM inference memory-bound, not compute-bound. During training, techniques like \textbf{Low-Rank Adaptation (LoRA)} and \textbf{Activation Recomputation} fundamentally alter the capacity constraint by trading compute or parameter trainability for drastically reduced memory footprints. The same \texttt{SingleNodeModel} computes~\citep{williams2009}:
\begin{equation}
\label{eq:tmemory}
T_{\text{memory}} = \frac{|W|}{BW_{\text{HBM}}}
\end{equation}
where $|W|$ is the total bytes read per inference step. The realized execution time is the maximum of the two bounds:
\begin{equation}
\label{eq:bottleneck}
T = \max(T_{\text{compute}},\; T_{\text{memory}})
\end{equation}
The crossover between compute-bound and memory-bound regimes occurs at the \emph{ridge point}, the arithmetic intensity at which the two ceilings intersect:
\begin{equation}
\label{eq:ridge}
I^{*} = \frac{\text{Peak}_{\text{FLOPS}}}{BW_{\text{HBM}}} \quad \text{(flop/byte)}
\end{equation}
Workloads with arithmetic intensity $I < I^{*}$ are memory-bound; those with $I > I^{*}$ are compute-bound. \textbf{Assumptions:} Peak FLOPS and HBM bandwidth are hard ceilings; MFU accounts for software inefficiency via $\eta$.

For training, capacity pressure is larger than inference weight residency alone. The \texttt{TrainingMemoryModel} exposes the first-order per-accelerator accounting directly:
\begin{equation}
\label{eq:training-memory}
M_{\text{train}} =
M_{\text{weights}} + M_{\text{grad}} + M_{\text{opt}} +
M_{\text{act}} + M_{\text{comm}}.
\end{equation}
Mixed-precision Adam defaults to FP16/BF16 weights and gradients plus FP32 master weights and optimizer moments, while tensor, pipeline, expert, and ZeRO sharding reduce different terms of the sum~\citep{shoeybi2019megatron,rajbhandari2020}. Activation checkpointing reduces $M_{\text{act}}$ by recomputing intermediate tensors during backward pass; the activation term implements the per-layer analytical bounds of \citet{korthikanti2023} ($34\,sbh$ bytes per layer with selective recomputation; $34\,sbh + 5as^2b$ without, where $s$, $b$, $h$, $a$ are sequence length, microbatch size, hidden dimension, and attention heads). The model intentionally reports each term separately so students can see whether a configuration fails because of optimizer state, activations, or communication buffers rather than a generic ``out of memory'' label.

\textbf{Wall~3: The Software Wall.} The gap between peak and achieved FLOP/s is typically larger than the gap between hardware generations. Most na\"ive implementations achieve only 30\% of peak throughput; the remaining 70\% is lost to redundant memory traffic, low warp occupancy, and unfused operations. The \texttt{EfficiencyModel} models this as a multiplicative efficiency factor~\citep{chowdhery2022palm}:
\begin{equation}
\label{eq:efficiency}
\eta = \frac{\text{FLOPS}_{\text{achieved}}}{\text{Peak}_{\text{FLOPS}}}
\end{equation}
where $\eta \in (0,1]$ modulates the Roofline ceiling, reducing the effective peak from $\text{Peak}_{\text{FLOPS}}$ to $\eta \times \text{Peak}_{\text{FLOPS}}$. FlashAttention~\citep{dao2022}, for example, achieves a $2.5\times$ speedup over standard attention by fusing memory-bound operations into a single kernel pass, effectively raising $\eta$ from ${\sim}0.3$ to ${\sim}0.75$ for attention layers. When this wall binds, better kernels, not bigger chips, are the remedy. \textbf{Assumption:} $\eta$ is a single scalar that aggregates all software inefficiencies; in practice, different operations (GEMM vs.\ attention vs.\ normalization) achieve different utilization on the same silicon. To avoid circularity when $\eta$ is unknown, the \texttt{EfficiencyModel} provides default ranges derived from published benchmarks: $\eta \approx 0.30$--$0.45$ for large-scale training~\citep{chowdhery2022palm,llama3team2024}, $\eta \approx 0.50$--$0.60$ for highly optimized GEMM-heavy workloads, and $\eta < 0.10$ for memory-bound inference decode. Students can use these defaults as starting points and refine as they gather profiling data.

\textbf{Wall~4: The Serving Wall.} Autoregressive LLM inference exhibits two distinct phases with fundamentally different Roofline characteristics~\citep{pope2023llm}. For a 70B model on two tensor-parallel H100s, a \cvalint{\ServSin}-token prefill takes ${\sim}\cvalint{\ServPrefillMs}$\,ms (compute-bound) while each decode token costs ${\sim}\cvalint{\ServDecodeMs}$\,ms (memory-bound). Prefill processes all input tokens in parallel (${\sim}\cvalint{\ServPrefillTps}$ tokens/s), while decode generates one token per memory-bandwidth pass (${\sim}\cvalint{\ServDecodeTps}$ tokens/s), a ${\sim}\cvalint{\ServTpsRatio}{\times}$ throughput gap. The \texttt{ServingModel} decomposes end-to-end inference latency as:
\begin{align}
\label{eq:serving}
T_{\text{prefill}} &= \frac{2P \cdot S_{\text{in}}}{\text{Peak}_{\text{FLOPS}} \times \eta} \quad \text{(compute-bound)} \\
T_{\text{decode}} &= \frac{|W|}{BW_{\text{HBM}}} \quad \text{(memory-bound)}
\end{align}
where $S_{\text{in}}$ is the input sequence length, $P$ is the parameter count, and $|W|$ includes both model weights and KV-cache reads (which grow with batch size and context length: $|W| = |W_{\text{model}}| + |W_{\text{KV}}|$). The $2PS$ term accounts for the linear projection FLOPs. For long contexts ($S > 4096$), the self-attention computation adds $\mathcal{O}(S^2)$ FLOPs per layer, which can constitute 20--40\% of total prefill cost. The \texttt{ServingModel} includes this attention term explicitly. When serving multiple concurrent requests ($B > 1$), model weights are loaded from HBM once per step while KV-cache reads grow linearly with the batch: $T_{\text{step}} = (|W| + B \cdot |KV_1|) / BW$, where $|KV_1|$ is the per-request KV-cache size. Because each step produces $B$ tokens simultaneously, the effective per-token cost is $T_{\text{step}}/B$, amortizing the weight read across requests. At large batch sizes the per-request KV cache dominates and throughput saturates.

The model also exposes an optional chunked-prefill estimate inspired by Sarathi-Serve~\citep{agrawal2025}. When \texttt{prefill\_chunk\_tokens} is set, only the new, uncached prefill work is partitioned into chunks:
\begin{align}
\label{eq:chunked-prefill}
C_{\text{prefill}} &=
\max\!\left(1,\left\lceil\frac{S_{\text{in}} - S_{\text{cache}}}{S_{\text{chunk}}}\right\rceil\right), \\
T_{\text{stall}} &= \max_i\left(T_{\text{chunk},i}\right)
\end{align}
where $T_{\text{chunk},i}$ is the compute time for chunk $i$ plus one dispatch tax. The resulting \texttt{ServingResult} reports \texttt{prefill\_chunks} ($C_{\text{prefill}}$), \texttt{prefill\_chunk\_time} ($T_{\text{stall}}$), and \texttt{decode\_stall\_bound}, which is set to $T_{\text{stall}}$ as a first-order bound on how long the slowest prefill chunk can interfere with ongoing decode iterations. This is a coarse analytical bound, not a full scheduler: total TTFT still includes the same prefill compute work plus one dispatch tax per chunk, and the solver does not claim Sarathi-Serve's stall-free packing behavior.

For deployment sizing, the serving-capacity resolver combines serving latency, continuous-batching capacity, and tail-queueing estimates. Given average generated length $G$ and per-replica token throughput $R_{\text{tok}}$, it estimates request capacity as
\begin{align}
\label{eq:serving-capacity}
Q_{\text{replica}} &= \frac{R_{\text{tok}}}{G}, \\
P99_{\text{request}} &\approx T_{\text{base}} + P99_{\text{queue}} \le T_{\text{target}}.
\end{align}
This converts a common serving question---``how many replicas do I need for a QPS and P99 target?''---into a transparent composition of the latency, batching, and queueing walls. The compute efficiency $\eta$ remains an exposed parameter with the same interpretation as in the Roofline model; the default is a starting point, not a claim of universal serving efficiency.

The prefill phase processes all input tokens in parallel and is compute-bound; the decode phase generates one token at a time and is memory-bandwidth-bound. The solver incorporates modern paradigms including \textbf{Prompt Caching} (prefix caching~\citep{zheng2024sglang}, which reduces TTFT by skipping prefill for previously computed KV-cache entries), \textbf{Speculative Decoding} (probability-weighted verification using a smaller draft model~\citep{leviathan2023fast}), \textbf{Disaggregated Serving} (phase splitting onto different hardware with KV-cache network transfer~\citep{patel2024}), and \textbf{Chunked Prefill} (optional chunk-level stall bounds for interleaving prefill work with decode~\citep{agrawal2025}). This duality explains why batching strategies that improve prefill throughput may have no effect on decode latency, because the two phases are bound by different resources. \textbf{Assumptions:} Prefill is compute-bound for sequence lengths $S_{\text{in}} \gg 1$; decode is memory-bandwidth-bound at batch size 1. Chunked prefill partitions the same upper-bound prefill work by token count with a fixed dispatch tax per chunk, so \texttt{decode\_stall\_bound} is a scheduling proxy rather than a measured tail-latency guarantee. At large batch sizes, decode transitions toward compute-bound; the solver models this crossover via the Roofline.

\textbf{Wall~5: The Batching Wall.} Serving throughput depends on how many requests share the accelerator simultaneously, but memory-bound decode means that each additional request in the batch adds KV-cache pressure without reducing per-token latency. Static batching wastes memory through external fragmentation. Each request reserves a contiguous KV-cache block sized for maximum sequence length, even if most requests finish early. The \texttt{ContinuousBatchingModel} models iteration-level scheduling with non-contiguous allocation via PagedAttention~\citep{kwon2023}:
\begin{equation}
\label{eq:pagedkv}
\text{KV}_{\text{paged}} = 2 \times L \times H \times D \times \lceil S / p \rceil \times p \times B \times b
\end{equation}
where $L$ is layers, $H$ is KV heads, $D$ is head dimension, $S$ is sequence length, $p$ is page size in tokens, $B$ is batch size, and $b$ is bytes per element. Internal fragmentation is bounded by the last page, eliminating the 40--50\% external fragmentation of contiguous allocation. \textbf{Assumptions:} Decode is memory-bandwidth-bound for batch $\geq 1$; static batching baseline assumes 50\% fragmentation waste.

\textbf{Wall~6: The Streaming Wall.} Wafer-scale architectures~\citep{lie2022cerebras} (e.g., Cerebras CS-3) invert the conventional memory hierarchy. Activations reside on-wafer in SRAM while model weights stream from external MemoryX nodes, shifting the bottleneck from HBM bandwidth to injection interconnect bandwidth. The \texttt{WeightStreamingModel} models this as:
\begin{equation}
\label{eq:weightstream}
T_{\text{layer}} = \max\!\left(\frac{|W_{\ell}|}{BW_{\text{inject}}},\; \frac{2P_{\ell} \times B}{\text{Peak} \times \eta}\right)
\end{equation}
where $|W_{\ell}|$ is the layer weight size in bytes, $P_{\ell} = |W_{\ell}| / b_{\text{prec}}$ is the parameter count (with $b_{\text{prec}}$ bytes per element), and the factor of 2 accounts for the multiply-accumulate FLOPs per parameter. The two terms inside the $\max$ represent the injection time (weight delivery) and the compute time (matrix arithmetic) for one layer. When $B$ is small, weight injection dominates and the compute engine sits idle; when $B$ is large, compute dominates and the injection link sits idle. Setting the two terms equal and substituting $|W_{\ell}| = P_{\ell} \times b_{\text{prec}}$ yields the optimal batch size $B^{*} = (b_{\text{prec}} \times \text{Peak} \times \eta) / (2 \times BW_{\text{inject}})$. This result depends only on numerical precision, peak compute, and injection bandwidth, independent of layer size. This is the unique operating point where injection and compute perfectly overlap, maximizing utilization of both resources. \textbf{Assumptions:} Layer weights dominate the injection payload; 10\% overhead is reserved for working memory; perfect within-layer pipelining is assumed.

\textbf{Wall~7: The Tail Latency Wall.} At scale, P99 latency governs user experience, not the median. A single slow replica in a fan-out of 100 services dominates end-to-end response time. The \texttt{TailLatencyModel} models inference replicas as an M/M/$c$ queue using the Erlang-C formula~\citep{dean2013}:
\begin{equation}
\label{eq:erlangc}
\mathbb{P}[\text{wait}] = \frac{(c\rho)^c / c! \cdot (1-\rho)^{-1}}{\sum_{k=0}^{c-1}(c\rho)^k/k! + (c\rho)^c/c! \cdot (1-\rho)^{-1}}
\end{equation}
where $c$ is the number of replicas, $\rho = \lambda / (c\mu)$ is per-server utilization, and $\lambda$, $\mu$ are arrival and service rates. The implementation evaluates the Erlang-C sum in log space (a log-sum-exp over log-gamma terms), so deep replica pools do not overflow the factorials in \Cref{eq:erlangc}. P99 latency grows non-linearly as $\rho \to 1$: for a 10-replica deployment, P99 at 80\% utilization is ${\sim}3{\times}$ the service time; at 95\%, it rises to ${\sim}10{\times}$, making the distinction between the two the difference between meeting and violating a latency SLA. \textbf{Assumption:} The M/M/$c$ model is intentionally optimistic; real traffic is bursty and service times are heavy-tailed, so M/M/$c$ provides a lower bound on tail latency. If the system is unstable under M/M/$c$, it will certainly be unstable under real traffic. To bridge toward more realistic service-time distributions, the solver accepts a \texttt{service\_time\_cv} parameter (coefficient of variation); when $\text{cv} \neq 1$, a Kingman correction factor $(c_v^2 + 1)/2$ scales the wait time, capturing the heavier tails of non-exponential service times.

\subsection{Data (Movement \& Pipelines)}
\label{sec:walls-data}

The Data walls govern how data moves \emph{to} the accelerator. A node that is locally unconstrained can still starve if the data pipeline cannot keep pace.

\textbf{Wall~8: The Ingestion Wall.} Storage I/O must supply training samples at the rate the accelerator consumes them. This wall binds most often in vision tasks, where raw images are large. For example, an 8-GPU DGX A100 training ResNet-50 at batch 2{,}048 demands ${\sim}2$\,GB/s of compressed image reads per step, easily saturating a single NVMe SSD and requiring striped or cached storage. The \texttt{DataModel} computes the demand--supply ratio~\citep{mohan2021}:
\begin{equation}
\label{eq:ingestion}
\rho_{\text{data}} = \frac{BW_{\text{demand}}}{BW_{\text{supply}}}
\end{equation}
When $\rho_{\text{data}} > 1$, the accelerator stalls waiting for data. $BW_{\text{demand}}$ is the product of batch size, sample size, and step rate; $BW_{\text{supply}}$ is the effective throughput of the storage subsystem after accounting for read amplification and caching. For LLM training on tokenized text, $BW_{\text{demand}}$ is typically negligible ($<$1\,MB/s); this wall is primarily relevant for vision, audio, and video workloads. \textbf{Assumption:} Storage bandwidth is the bottleneck, not network I/O (single-node training).

\textbf{Wall~9: The Transformation Wall.} JPEG decoding, tokenization, and augmentation execute on CPU cores, not on the accelerator. When the CPU preprocessing pipeline cannot keep pace, the accelerator stalls even if storage bandwidth is abundant. On ImageNet, a single CPU worker typically decodes and augments ${\sim}\cvalint{\StwoAugRate}$ images/s; \cvalint{\StwoCPUworkers} workers therefore cap the pipeline at ${\sim}\StwoEffThroughput$ images/s, below what an 8-GPU DGX A100's compute can consume (see Case~S2, \Cref{sec:usage}). The \texttt{TransformationModel} quantifies this bottleneck~\citep{murray2021tf}:
\begin{equation}
\label{eq:transform}
T_{\text{transform}} = \frac{B}{R_{\text{cpu}}}
\end{equation}
where $B$ is the batch size (in samples) and $R_{\text{cpu}}$ is the aggregate CPU preprocessing rate (in samples/s), computed as $R_{\text{cpu}} = N_{\text{workers}} \times r_{\text{worker}}$ where $r_{\text{worker}}$ is the per-worker throughput after all transformations (decode, augment, normalize). \textbf{Assumption:} Preprocessing is CPU-bound and scales linearly with worker count up to core saturation.

\textbf{Wall~10: The Locality Wall.} Network topology determines the effective bandwidth available between any two nodes in the cluster. The \texttt{TopologyModel} models this through the \emph{bisection bandwidth fraction} $\beta$~\citep{leiserson1985}, which varies by topology: Fat-Tree provides full bisection ($\beta = 1.0$), Dragonfly achieves $\beta \approx 0.85$, and 3D~Torus yields $\beta \approx 0.67$. The effective inter-node bandwidth is:
\begin{equation}
\label{eq:locality}
BW_{\text{eff}} = \frac{BW_{\text{link}} \times \beta}{\text{oversubscription}}
\end{equation}
This wall becomes binding when collective communication patterns demand bandwidth that the topology cannot supply at scale. For example, a 3D~Torus cluster with \cvalint{\LinkBWGbps}\,Gb/s links delivers an effective bisection bandwidth of $\cvalint{\LinkBWGbps} \times \cvalii{\BetaTorus} = \cvalint{\TorusBisectGbps}$\,Gb/s per node; a Fat-Tree with the same links provides the full $\cvalint{\LinkBWGbps}$\,Gb/s ($\beta = \cvalii{\BetaFatTree}$), a $\cval{\TopoRatio}{\times}$ advantage that compounds across multi-hop AllReduce patterns. We place the Locality Wall in the Data domain rather than Fleet because it models the \emph{physical topology constraint} on data movement (bisection bandwidth, oversubscription), whereas the Communication Wall (Wall~14, Fleet) models the \emph{algorithmic cost} of specific collectives (AllReduce, All-to-All) that run atop that topology. The two walls interact but address distinct levels of abstraction. \textbf{Assumption:} $\beta$ values are topology-specific constants; real networks may exhibit dynamic congestion not captured by this static model.

\subsection{Algorithm (Scaling \& Compression)}
\label{sec:walls-algorithm}

The Algorithm walls arise not from hardware but from the mathematics of learning itself. They determine how much computation a workload \emph{requires}, independent of the silicon that executes it.

\textbf{Wall~11: The Complexity Wall.} Chinchilla scaling laws~\citep{hoffmann2022chinchilla} establish that training compute scales jointly with model size $P$ and dataset size $D$: doubling the parameters requires approximately doubling the tokens to remain compute-optimal. The \texttt{ScalingModel} implements:
\begin{align}
\label{eq:chinchilla}
C &= 6PD \quad \text{(total training FLOPs)} \\
D^{*} &\approx 20P \quad \text{(compute-optimal tokens)} \\
P^{*} &= \sqrt{\frac{C}{120}} \quad \text{(optimal model size for budget } C\text{)}
\end{align}
These relations allow students to reason backward from a compute budget to the largest model that can be trained optimally, or forward from a model size to the minimum viable training cluster. For example, a budget of $10^{24}$~FLOPs yields $P^{*} \approx \cvalint{\ChinchillaPstarB}$B parameters with $D^{*} \approx \cval{\ChinchillaDstarT}$T tokens; Chinchilla (70B, 1.4T tokens) was trained near this optimal frontier~\citep{hoffmann2022chinchilla}. \textbf{Assumption:} Scaling law coefficients are fitted to published training runs; extrapolation beyond the fitted range is flagged.

\textbf{Wall~12: The Reasoning Wall.} Inference-time compute scaling introduces a cost that grows linearly with the number of reasoning steps $K$. The \texttt{InferenceScalingModel} models this as~\citep{snell2025scaling}:
\begin{equation}
\label{eq:reasoning}
T_{\text{reason}} = K \times T_{\text{step}}(P, S_{\text{context}})
\end{equation}
where $T_{\text{step}}$ is the per-step latency, itself a function of model size $P$ and context length $S_{\text{context}}$. In practice, $K$ varies dramatically by strategy: simple chain-of-thought uses $K \approx 3$--$8$, best-of-$N$ sampling uses $K = N$ (typically 8--64), and tree-search methods like Monte Carlo Tree Search can require $K > 100$~\citep{snell2025scaling}. A $K{=}8$ chain-of-thought query costs ${\sim}8{\times}$ more compute than a single-pass answer, fundamentally altering serving economics (see Case~S1, \Cref{sec:usage}). \textbf{Assumption:} Each reasoning step is an independent decode sequence; KV-cache is not shared across steps.

\textbf{Wall~13: The Fidelity Wall.} Compression trades model fidelity for efficiency. Quantization reduces precision while pruning removes weights entirely. The accuracy--efficiency frontier is task- and architecture-dependent. The \texttt{CompressionModel} quantifies the two primary mechanisms~\citep{han2016deep, gholami2022}:
\begin{align}
\label{eq:compression}
r_{\text{quant}} &= \frac{b_{\text{base}}}{b_{\text{target}}} \quad \text{(quantization ratio, e.g., } 16/4 = 4{\times}\text{)} \\
r_{\text{prune}} &= \frac{1}{1 - s} \quad \text{(memory reduction at sparsity } s\text{)}
\end{align}
Here $b_{\text{base}}$ is the baseline precision (typically 16 for FP16/BF16 models, not 32) and $b_{\text{target}}$ is the quantized precision. Quantization reduces memory reads but not FLOPs, shifting the arithmetic intensity rightward on the Roofline by a factor of $r_{\text{quant}}$ and potentially crossing the ridge point from memory-bound to compute-bound. For pruning, $r_{\text{prune}}$ gives the \emph{memory reduction} ratio; actual compute speedup depends on sparsity structure. Unstructured sparsity yields no acceleration on current GPUs, while 2:4 structured sparsity~\citep{nvidia2023h100} provides a $2{\times}$ throughput gain via Sparse Tensor Cores. Post-training quantization methods such as GPTQ~\citep{frantar2023gptq} and AWQ~\citep{lin2025} demonstrate that 4-bit quantization can preserve most accuracy for large language models, making $r_{\text{quant}} = 4{\times}$ a practical operating point. However, compression ratio alone does not determine inference speedup. Unstructured pruning reduces storage but yields no compute speedup on GPU GEMM kernels; only structured pruning and $N{:}M$ sparsity (e.g., 2:4 on NVIDIA Ampere+) accelerate hardware execution. The \texttt{CompressionModel} reports both \texttt{memory\_savings\_pct} and \texttt{inference\_speedup} to capture this distinction. The accuracy impact $\Delta_{\text{acc}}$ is modeled as a configurable function, since the fidelity--compression frontier varies by architecture and task. \textbf{Assumptions:} Accuracy degradation follows empirical curves from~\citet{gholami2022}; pruning compute speedup requires structured sparsity with hardware support.

\subsection{Fleet (Multi-Node Coordination)}
\label{sec:walls-fleet}

The Fleet walls arise when systems scale beyond a single node, requiring coordination across multiple accelerators connected by network fabric.

\textbf{Wall~14: The Communication Wall.} Distributed training requires gradient synchronization across $N$ nodes, and the cost of that synchronization grows with both message size and node count. The \texttt{DistributedModel} models the dominant collective operations. For Ring AllReduce~\citep{shoeybi2019megatron} and its \textbf{ZeRO/FSDP} partitioned equivalents (Reduce-Scatter and All-Gather):
\begin{equation}
\label{eq:allreduce}
T_{\text{ring}} = \frac{2(N-1)}{N} \cdot \frac{M}{B_{\text{link}}} + 2(N-1) \cdot \alpha
\end{equation}
where $M$ is the message size, $B_{\text{link}}$ is the per-link bandwidth, and $\alpha$ is the per-hop latency. Modern clusters use \textbf{hierarchical AllReduce} to exploit the bandwidth asymmetry between intra-node interconnect (e.g., NVLink at \cvalint{\NVLinkBWGBs}\,GB/s) and inter-node fabric (e.g., InfiniBand at \cvalint{\IBBWGBs}\,GB/s per port, an $\cvalint{\NVLinkIBgap}{\times}$ gap). The \texttt{DistributedModel} implements a two-level model:
\begin{equation}
\label{eq:hierarchical}
T_{\text{hier}} = T_{\text{intra}}(BW_{\text{NVLink}},\, G) + T_{\text{inter}}(BW_{\text{IB}},\, N)
\end{equation}
where $G$ is GPUs per node and $N$ is the node count. Each level applies the ring formula (\Cref{eq:allreduce}) at its respective bandwidth, making the inter-node phase the dominant cost at scale. After the intra-node reduce-scatter, each lead GPU holds $M/G$ of the reduced result; the inter-node phase therefore operates on $M/G$, not the full $M$, making this the bandwidth-critical optimization of hierarchical collective design. In practice, modern frameworks also use \textbf{Compute/Communication Overlap}, hiding network latency behind backward pass computation. The \texttt{DistributedModel} models this with an overlap efficiency parameter $\eta_{\text{overlap}} \in [0,1]$ (default 0.85, reflecting typical Megatron-LM behavior), yielding an exposed communication cost of $(1 - \eta_{\text{overlap}}) \cdot T_{\text{comm}}$. \textbf{Gradient Accumulation} further reduces effective communication cost. When accumulating gradients over $K_{\text{acc}}$ microsteps before synchronizing, the DP AllReduce cost is amortized by $1/K_{\text{acc}}$, trading latency for communication efficiency at large global batch sizes. Tensor parallelism partitions model weights across devices and requires two AllReduce operations per transformer layer on \emph{activations} (not weights). Each AllReduce exchanges a tensor of size $B \times S \times H \times b$ bytes, where $B$ is batch size, $S$ is sequence length, $H$ is hidden dimension, and $b$ is bytes per element. Weights are partitioned once and remain stationary. Pipeline parallelism~\citep{narayanan2021} introduces a bubble overhead:
\begin{equation}
\label{eq:bubble}
B_{\text{pipeline}} = \frac{P_{\text{stages}} - 1}{V \cdot M_{\text{micro}} + P_{\text{stages}} - 1}
\end{equation}
where $V$ is the number of virtual pipeline stages and $M_{\text{micro}}$ is the number of microbatches. For Mixture-of-Experts All-to-All dispatch:
\begin{equation}
\label{eq:alltoall}
T_{\text{a2a}} = \frac{(N-1)}{N} \cdot \frac{M}{B_{\text{link}}} + (N-1) \cdot \alpha
\end{equation}
where $M$ is the routed activation payload. The \texttt{MoERoutingModel} maintains a first-order MoE abstraction where the total parameter count determines memory residency, the active parameter count determines the compute workload, and any hot-expert imbalance inflates the routed payload. With routing imbalance factor $\gamma \ge 1$:
\begin{align}
\label{eq:moe-imbalance}
E_{\text{active,eff}} &= \min(E_{\text{total}}, \gamma \cdot E_{\text{top-k}}), \\
P_{\text{active,eff}} &= P_{\text{active}} \cdot
\frac{E_{\text{active,eff}}}{E_{\text{top-k}}}.
\end{align}
The same factor is available in \texttt{DistributedModel} as \texttt{moe\_routing\_imbalance\_factor}, where it inflates expert-parallel traffic without introducing a router simulation. This follows the load-balancing motivation of sparsely-gated MoE and Switch-style routing systems~\citep{shazeer2017outrageously,fedus2022switch,lepikhin2020gshard}.

\textbf{Wall~15: The Fragility Wall.} Component failures are inevitable at scale. If each node has a mean time between failures of $\text{MTBF}_{\text{node}}$, then a cluster of $N$ nodes has~\citep{daly2006}:
\begin{equation}
\label{eq:mtbf}
\text{MTBF}_{\text{cluster}} = \frac{\text{MTBF}_{\text{node}}}{N}
\end{equation}
The probability of at least one failure during a training run of duration $T$ is:
\begin{equation}
\label{eq:pfail}
P(\geq 1 \text{ failure}) = 1 - e^{-T / \text{MTBF}_{\text{cluster}}}
\end{equation}
The Young-Daly formula~\citep{young1974, daly2006} gives the optimal checkpoint interval:
\begin{equation}
\label{eq:youngdaly}
\tau_{\text{opt}} = \sqrt{2 \delta \cdot \text{MTBF}_{\text{cluster}}}
\end{equation}
where $\delta$ is the time to write one checkpoint. The \texttt{ReliabilityModel} uses these relations to estimate the fraction of compute lost to checkpointing and recovery, reporting a \texttt{goodput\_ratio} that captures the effective fraction of wall-clock time spent on useful forward/backward computation after subtracting checkpoint writes and failure recovery. \textbf{Assumption:} Failures are independent and exponentially distributed (memoryless).

\textbf{Wall~16: The Multi-tenant Wall.} Production GPU clusters are rarely dedicated to a single job. Shared clusters introduce queueing delays that grow hyperbolically as utilization approaches 1.0, creating a tension between maximizing hardware utilization and minimizing researcher wait times. The \texttt{OrchestrationModel} models job wait times using an M/D/1 queue~\citep{little1961}:
\begin{equation}
\label{eq:queue}
T_{\text{wait}} = \frac{\rho}{2\mu(1 - \rho)}
\end{equation}
where $\rho = \lambda / \mu$ is the cluster utilization, $\lambda$ is the job arrival rate, and $\mu$ is the service rate. As $\rho \to 1$, wait times diverge hyperbolically. A cluster at 90\% utilization has $5{\times}$ the queue depth of one at 50\%. \textbf{Assumption:} Job durations are approximately deterministic, which is reasonable for large training runs with predictable step times.

\subsection{Operations (Cost, Carbon \& Safety)}
\label{sec:walls-operations}

The Operations walls capture constraints that are not about \emph{how fast} a system runs but \emph{whether it should run at all}: economic viability, environmental impact, checkpoint overhead, and responsible deployment.

\textbf{Wall~17: The Capital Wall.} Performance analysis is incomplete without economic constraints. A \cvalint{\ClusterGPUs}-GPU H100 cluster costs ${\sim}$\$\cval{\ClusterCapExM}M in CapEx alone; amortized over three years, the hardware cost of a single 30-day training run exceeds \$800K before adding electricity. The \texttt{EconomicsModel} computes total cost of ownership~\citep{barroso2019}:
\begin{equation}
\label{eq:tco}
\text{TCO} = \text{CapEx} + \text{OpEx}_{\text{energy}} + \text{OpEx}_{\text{maint}}
\end{equation}
where $\text{OpEx}_{\text{energy}} = E_{\text{total}} \times P_{\text{kWh}}$ converts total energy consumption to dollar cost at the regional electricity price. At scale, CapEx dominates. Electricity typically accounts for less than 10\% of total run cost, making hardware utilization (MFU) the primary lever for cost efficiency. \textbf{Assumption:} Linear amortization over a 3--5 year hardware lifetime.

\textbf{Wall~18: The Sustainability Wall.} Regional grid carbon intensity varies by more than an order of magnitude, making geography a first-order systems design variable. The \texttt{SustainabilityModel} converts energy into environmental impact~\citep{patterson2021carbon}:
\begin{align}
\label{eq:sustainability}
E_{\text{total}} &= E_{\text{IT}} \times \text{PUE} \\
\text{CO}_2 &= E_{\text{total}} \times \text{CI}_{\text{region}} \quad \text{(gCO}_2\text{/kWh)} \\
\text{H}_2\text{O} &= E_{\text{total}} \times \text{WUE} \quad \text{(L/kWh)}
\end{align}
where PUE is the power usage effectiveness of the datacenter, $\text{CI}_{\text{region}}$ is the carbon intensity of the local grid, and WUE is the water usage effectiveness. This model captures \emph{operational} carbon only. For edge and IoT deployments at scale ($>10^6$ devices), \emph{embodied} carbon from manufacturing can dominate operational carbon by 10--100$\times$~\citep{gupta2022chasing}. The \texttt{SustainabilityModel} accepts an optional \texttt{embodied\_carbon\_per\_device} parameter for lifecycle analysis. \textbf{Assumption:} Grid carbon intensity is a static regional constant; temporal variation (e.g., renewable intermittency) is not modeled. Energy-proportional power follows~\citet{barroso2007}: idle power is 30\% of TDP, with the remaining 70\% scaling linearly with MFU.

\textbf{Wall~19: The Checkpoint Wall.} Long-running training jobs must periodically save model state (weights and optimizer states) to persistent storage, incurring an I/O penalty that directly reduces effective MFU. This is a classic manifestation of Amdahl's Law: as fleet size scales to accelerate the parallel forward/backward passes, the serial time spent writing to persistent storage begins to dominate. The \texttt{CheckpointModel} models the I/O burst penalty~\citep{eisenman2022checknrun}:
\begin{equation}
\label{eq:checkpoint2}
\text{MFU}_{\text{penalty}} = \frac{T_{\text{write}}}{T_{\text{interval}}} = \frac{|W| \times \beta_{\text{opt}} / BW_{\text{storage}}}{T_{\text{ckpt\_interval}}}
\end{equation}
where $|W|$ is the model weight size in bytes, and $\beta_{\text{opt}}$ is the optimizer state multiplier, the ratio of total checkpoint bytes to model weight bytes. For mixed-precision Adam with FP16 weights, $\beta_{\text{opt}} \approx 7$ (FP32 master weights at 4\,bytes + FP32 momentum at 4\,bytes + FP32 variance at 4\,bytes, plus FP16 model weights at 2\,bytes, totaling 14\,bytes per parameter vs.\ 2\,bytes for the FP16 model alone). Gradients are ephemeral and not checkpointed. For a 70B-parameter model, the checkpoint size is $\cvalint{\LLaMAParams}\text{B} \times \cvalint{\CkptBytesPerParam}\text{\,B/param} \approx \cval{\CkptSizeTB}$\,TB, making storage bandwidth the binding constraint during I/O bursts.

\textbf{Wall~20: The Safety Wall.} Privacy and fairness guarantees impose quantifiable computational overhead. The \texttt{ResponsibleEngineeringModel} models the cost of differential privacy via DP-SGD~\citep{abadi2016}, where the noise multiplier scales inversely with the privacy budget:
\begin{equation}
\label{eq:dpsgd}
\sigma \propto \frac{1}{\varepsilon}
\end{equation}
At moderate privacy ($\varepsilon = 1$), DP-SGD typically incurs a ${\sim}3{\times}$ training slowdown due to per-sample gradient clipping and noise addition; at strong privacy ($\varepsilon = 0.1$), the overhead can exceed $10{\times}$. The solver models this as a compute multiplier: $T_{\text{DP}} = T_{\text{base}} \times f(\varepsilon)$, where $f(\varepsilon)$ is empirically calibrated from published DP-SGD benchmarks. Fairness constraints impose a complementary data requirement. Sufficient representation of minority subgroups demands additional training data proportional to $O(1/p_{\min})$, where $p_{\min}$ is the smallest subgroup prevalence. Unlike the other walls, Wall~20 is more qualitative than quantitative, with overhead depending heavily on task, model architecture, and privacy mechanism. The solver reports the compute multiplier as a range rather than a point estimate. \textbf{Assumption:} Privacy budget $\varepsilon$ is a hard constraint; the solver reports the compute multiplier, not the privacy guarantee.

\subsection{Analysis (Cross-Cutting Diagnostics)}
\label{sec:walls-analysis}

The preceding 20 walls each model a specific physical or logical constraint. The final two entries are \emph{diagnostic tools} rather than walls in the strict sense. They operate \emph{across} the taxonomy rather than within a single domain, providing analysis capabilities that span all walls. The Sensitivity tool identifies which wall is binding, and the Synthesis tool derives minimum hardware from SLA requirements.

\textbf{Wall~21: The Sensitivity Wall.} Optimization is effective only when directed at the binding constraint; improving a non-bottleneck parameter yields no measurable gain. The \texttt{SensitivitySolver} identifies the binding constraint by computing partial derivatives of end-to-end latency with respect to each hardware parameter~\citep{williams2009}:
\begin{equation}
\label{eq:sensitivity}
\frac{\partial T}{\partial BW_{\text{mem}}}, \quad \frac{\partial T}{\partial \text{Peak}_{\text{FLOPS}}}, \quad \frac{\partial T}{\partial BW_{\text{net}}}, \quad \ldots
\end{equation}
The parameter with the largest sensitivity is the binding constraint, that is, the single upgrade that would yield the greatest performance improvement. Each reported sensitivity is the normalized latency response $(T_{\text{perturbed}} - T_{\text{base}}) / T_{\text{base}}$ to a 10\% parameter upgrade. For Llama-3 8B decode on an A100, the solver reports a memory-bandwidth sensitivity of $-0.088$ (a 10\% bandwidth upgrade cuts latency by 8.8\%) versus exactly zero for peak FLOPS, because under the hard $\max$ of \Cref{eq:bottleneck} a strictly memory-bound workload does not respond to a compute upgrade at all. This transforms ``where should I invest?'' from intuition into calculation. \textbf{Assumption:} Finite-difference approximation with 10\% perturbation; second-order effects are ignored. The binding constraint is identified as the parameter with the largest absolute sensitivity.

\textbf{Wall~22: The Synthesis Wall.} The \texttt{SynthesisSolver} addresses the inverse problem: given a service-level objective (e.g., 50\,ms inter-token latency), it derives the minimum hardware specifications required to satisfy it~\citep{kwon2023}:
\begin{align}
\label{eq:synthesis}
BW_{\text{required}} &= \frac{|W|}{T_{\text{target}}} \\
\text{FLOPS}_{\text{required}} &= \frac{\text{OPs}}{T_{\text{target}} \times \eta}
\end{align}
This enables hardware-software co-design. Engineers specify an SLA and the solver derives the minimum hardware that satisfies it. For LLaMA-70B at a \cvalint{\SynthTargetMs}\,ms inter-token latency target, the solver yields $BW_{\text{required}} = \cvalint{\LLaMAWeightsGB}\;\text{GB} / \cvalint{\SynthTargetMs}\;\text{ms} = \cvalint{\SynthBWreqGBs}$\,GB/s, $\cval{\SynthAhMultiple}{\times}$ the A100's \cval{\AhBWTBs}\,TB/s, confirming that this SLA requires at least two tensor-parallel A100s. \textbf{Assumption:} Hardware parameters are independently adjustable; co-design coupling between FLOPS and bandwidth is not modeled.

\section{The 3-Tier Resolver Architecture}
\label{sec:solver-formalism}

The walls define \emph{what} constrains a system. This section formalizes \emph{how} resolvers compose to produce end-to-end system evaluations. To clarify the mathematical intent of each component, \mlsysim organizes its analytical tools into a strict 3-tier taxonomy, summarized below before \Cref{sec:solvers-compose} shows how the tiers chain into multi-stage analyses through a shared, dimensionally typed contract, and \Cref{sec:solvers-scorecard} describes the three-level \texttt{SystemEvaluation} scorecard (Feasibility, Performance, Macro) that the framework returns.

\paragraph{Tier 1: Analytical Models (the ``physics engine'').} Models perform forward evaluation $Y = f(X)$ to determine the physical and logical consequences of a specific system configuration. For example, the \texttt{ServingModel} calculates the exact time-to-first-token for a given LLM and GPU pair. Models are purely deterministic and make no decisions; they comprise the first 23 resolvers in our taxonomy.

\paragraph{Tier 2: Analysis Solvers (the ``math engine'').} Solvers perform algebraic inversion or calculus, $X = f^{-1}(Y)$ or $\nabla f$, to find the exact parameter required to hit a specific target. For example, the \texttt{SynthesisSolver} takes a target latency SLA and works backward to derive the minimum memory bandwidth required.

\paragraph{Tier 3: Optimizers (the ``engineering engine'').} Optimizers perform constrained design-space search, $\max f(X)$ subject to $g(X) \le c$, to find the best configuration among many valid options. Unlike Models and Solvers, which map directly to individual walls, Optimizers operate across the entire taxonomy. The \texttt{ParallelismOptimizer} sweeps all valid 3D tensor/pipeline/data parallel splits to maximize Model FLOPs Utilization (MFU) on a given cluster; the \texttt{BatchingOptimizer} searches for the maximum batch size that satisfies a P99 queueing latency SLA.

\subsection{Stateless Composition and Chaining}
\label{sec:solvers-compose}

\begin{figure*}[!t]
\centering
\includegraphics[width=\textwidth]{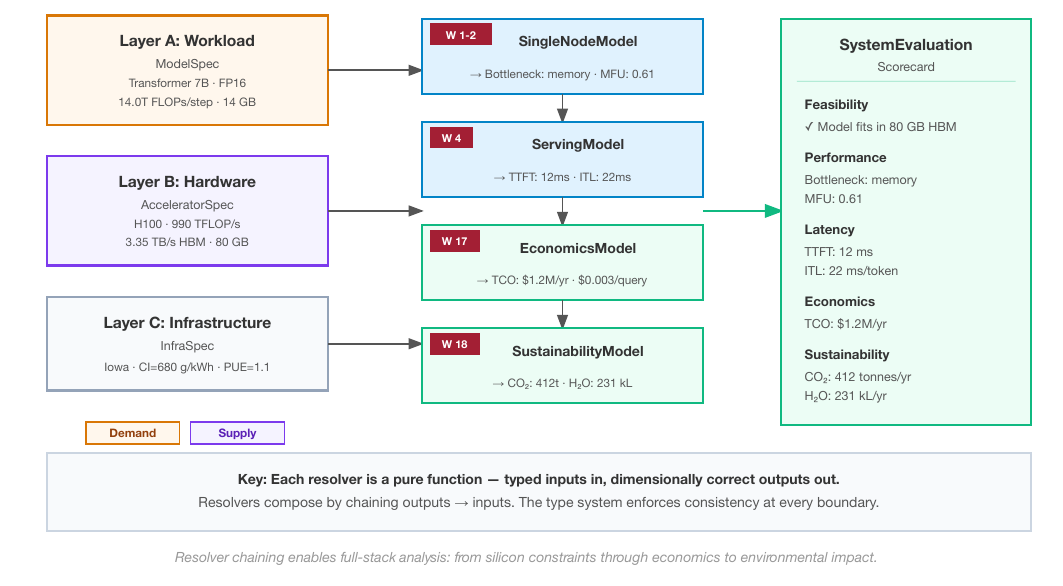}
\caption{\textbf{Resolver Composition.} Three input layers feed four resolvers. Each resolver is a pure function: typed inputs in, dimensionally correct outputs out. The scorecard aggregates three evaluation levels: Feasibility, Performance, and Macro (economics, sustainability, safety).}
\label{fig:solver-chaining}
\end{figure*}

Every resolver in \mlsysim is a pure function: it accepts a typed configuration, performs analytical computation, and returns a typed result object (a Pydantic \texttt{BaseModel} with dimensioned fields). Resolvers maintain no hidden state between invocations. Because they share a common type system, the output of one tier feeds naturally into the next, as illustrated in \Cref{fig:solver-chaining}. The reader should trace, in that figure, how the demand, supply, and topology layers feed a resolver chain that culminates in the three-level scorecard described in \Cref{sec:solvers-scorecard}; the typed contract at each arrow is what enables the chain to compose without runtime checks. A full-stack analysis composes resolvers in sequence to resolve complex design questions. For example, determining the financial cost of training an optimally-sized model on a frontier cluster requires chaining algorithmic scaling, distributed execution, and macro-economics:

\begin{equation}
\label{eq:chain}
\resizebox{0.88\columnwidth}{!}{$\displaystyle
\mathsf{Scaling} \xrightarrow{\;\mathcal{R}_1\;} \mathsf{Distributed} \xrightarrow{\;\mathcal{R}_2\;} \mathsf{Economics} \xrightarrow{\;\mathcal{R}_3\;} \mathsf{Sustainability}
$}
\end{equation}

\Cref{lst:composability} demonstrates this exact chain in \mlsysim. The \texttt{ScalingModel} calculates the optimal model size for a given compute budget ($\mathcal{R}_1$). The \texttt{DistributedModel} takes that workload and computes the real-world execution time on an 8,192-GPU fleet, factoring in 3D parallelism overhead ($\mathcal{R}_2$). Finally, the \texttt{EconomicsModel} converts that execution time into a Total Cost of Ownership ($\mathcal{R}_3$).

\begin{lstlisting}[caption={\textbf{Solver Composition.} Bridging algorithmic scaling, distributed execution, and fleet economics in a single executable chain.},label={lst:composability},float=t]
import mlsysim
from mlsysim.solvers import ScalingModel, DistributedModel, EconomicsModel
from mlsysim.models.types import TransformerWorkload

# 1. Algorithm: Find optimal parameters for a fixed compute budget
budget = mlsysim.Q_("4e24 flop")  # ~100K H100-days at 50% MFU
optimal = ScalingModel().solve(compute_budget=budget)  # P* ~ 183B params

# Instantiate the demand (Layer A: Workload)
model = TransformerWorkload(
    name="Frontier-Model", architecture="transformer",
    parameters=optimal.optimal_parameters,
    layers=80, hidden_dim=8192, heads=64
)

# 2. Fleet: Evaluate on a massive 8K GPU cluster (Layer D: Supply/Topology)
fleet = mlsysim.Systems.Clusters.Frontier_8K
seq_len = 4096
perf = DistributedModel().solve(
    model, fleet,
    batch_size=4096, tp_size=8, pp_size=4, seq_len=seq_len
)

# 3. The Capital: Calculate TCO for the resulting training time
n_steps = optimal.optimal_tokens.to("count").magnitude / (4096 * seq_len)
duration = perf.step_latency_total * n_steps
tco = EconomicsModel().solve(fleet, duration_days=duration.to("day").magnitude)

print(f"Scaling Efficiency: {perf.scaling_efficiency:.1%}")   # 54.9%
print(f"Training Duration:  {duration.to('day'):.1f}")        # 68.1 day
print(f"Total Job Cost:     ${tco.tco_usd:,.0f}")             # $15,905,569
\end{lstlisting}

Each link in the chain preserves dimensional correctness. Units propagate through the computation, and any mismatch raises an immediate error rather than producing a silently wrong result.

\subsection{The SystemEvaluation Scorecard}
\label{sec:solvers-scorecard}

\mlsysim provides a \texttt{Scenario.evaluate()} entry point that orchestrates solver composition automatically through a three-level evaluation. A scenario is a concrete executable case, such as a doorbell workload on a microcontroller or a frontier training workload on a fleet; sourced statistics used only as reference anchors remain outside the executable path in \texttt{ReferenceStats.*}.

\textbf{Level~1: Feasibility.} Does the model fit? Can the data pipeline keep pace? The framework checks memory capacity against model size, ingestion bandwidth against training throughput, and reports any wall where demand exceeds supply.

\textbf{Level~2: Performance.} What are the achievable latency, throughput, and utilization? The Roofline analysis (\Cref{eq:bottleneck}), communication modeling (\Cref{eq:allreduce}), and pipeline bubble (\Cref{eq:bubble}) combine to produce end-to-end training step time.

\textbf{Level~3: Macro.} What does it cost, and what does it emit? TCO (\Cref{eq:tco}), carbon (\Cref{eq:sustainability}), and responsibility overhead (\Cref{eq:dpsgd}) are computed from the performance results.

The three levels are evaluated in order; a feasibility failure at Level~1 short-circuits the evaluation and reports the binding constraint. This ordering reflects the dependency structure, since communication optimization is irrelevant if the model exceeds available memory. The complete implementation details and key assumptions for each solver are documented alongside their respective walls in \Cref{sec:taxonomy}.

\section{Validation}
\label{sec:validation}

An analytical framework earns trust through transparent confrontation with empirical ground truth. We validate \mlsysim along two axes: \emph{accuracy} against published benchmarks, and \emph{speed} relative to alternative modeling tools. Our goal in validation is not to ``curve-fit'' the model to perfectly match empirical results by introducing arbitrary magic numbers. Rather, it is to demonstrate that a first-principles application of physics and generalized efficiency constants places the prediction squarely in the correct ballpark. As we argue in \Cref{sec:discussion}, a model that is off by 10\% because it structurally ignores L2 cache misses is functioning exactly as intended, trading cycle-level precision for architectural intuition. The anchors below illustrate this philosophy.

The remainder of this section presents the validation in four parts. \Cref{sec:validation-anchors} anchors \mlsysim predictions against seven published benchmarks spanning single-node training, distributed training, inference, scaling laws, sustainability, and automated parallelism search. \Cref{sec:validation-speed} reports the design-space exploration speed relative to cycle-accurate simulators. \Cref{sec:accuracy} discusses the accuracy scope and the empirical role of the efficiency parameter $\eta$. Finally, \Cref{sec:validation-dim} argues that runtime dimensional strictness is itself a structural form of validation that numerical accuracy alone cannot supply.

\subsection{Empirical Anchors}
\label{sec:validation-anchors}

We anchor \mlsysim predictions against seven published benchmarks spanning single-node training, distributed training, inference, scaling laws, sustainability, and automated design-space optimization.

\textbf{Anchor~1: MLPerf ResNet-50 on DGX A100 (Single-Node Training).}
For ResNet-50 training on a DGX A100 node (8$\times$ A100 GPUs with NVLink) at batch size 2048, \mlsysim predicts a throughput of approximately 37{,}500 samples/s using the \texttt{SingleNodeModel} with end-to-end hardware utilization $\eta = 0.19$ and 8-way data parallelism within the node. The low $\eta$ reflects that ResNet-50's small convolution kernels, data pipeline, and framework overhead cannot saturate the A100's tensor cores the way large GEMM-dominated Transformer layers can. NVIDIA's MLPerf Training closed-division submissions for the 8-GPU DGX A100 configuration~\citep{mattson2020} report aggregate training rates in the same regime as our prediction; at the order of magnitude that determines whether the system is compute-bound or memory-bound, this validates the Roofline-based throughput model~\citep{williams2009} at the core of \mlsysim's single-node solver. Per-GPU throughput is $\sim$4{,}700 samples/s, consistent with the A100's Roofline ceiling for ResNet-50's arithmetic intensity.

\textbf{Anchor~2: vLLM Llama-2-70B on H100 (Inference).}
For autoregressive decoding of Llama-2-70B (FP16, batch size 1), the model weights total \cvalint{\LLaMAWeightsGB}\,GB, requiring at minimum two tensor-parallel H100s (each with \cvalint{\HhHBMGB}\,GB HBM3). \mlsysim's first-order estimate divides total weights by aggregate bandwidth: $\cvalint{\LLaMAWeightsGB}\;\text{GB} / (2 \times \cvalii{\HhBWTBs}\;\text{TB/s}) = 20.9$\,ms for the weight-read phase alone. Adding KV-cache reads, attention computation, NVLink synchronization, and framework scheduling overhead, the predicted end-to-end ITL is approximately 43\,ms~\citep{nvidia2023h100}. Published vLLM benchmarks for this configuration report ITL values in the 40--50\,ms range~\citep{kwon2023}, confirming that decode-phase LLM inference is memory-bandwidth-bound and that the overhead multiplier ($\sim$2$\times$ over the pure bandwidth floor) is consistent across deployments.

\textbf{Anchor~3: Llama~3 Training at 16K H100s (Distributed Training).}
Meta's Llama~3 training report~\citep{llama3team2024} documents achieving 38--43\% MFU on 16{,}384 H100 GPUs with 4D parallelism (TP$\times$CP$\times$PP$\times$DP). We configure \mlsysim's \texttt{DistributedModel} with the published fleet (2{,}048 nodes $\times$ 8 H100s, 400\,Gb/s InfiniBand per node) training the 405B-parameter model with intra-node tensor parallelism (TP=8), pipeline parallelism over 64 microbatches, and system-level efficiency $\eta = 0.42$. This $\eta$ captures kernel utilization, stragglers, load imbalance, checkpointing pauses, and thermal throttling, effects that the analytical communication model does not represent. After accounting for pipeline bubble overhead (\Cref{eq:bubble}), hierarchical AllReduce cost (\Cref{eq:hierarchical}), and compute--communication overlap ($\eta_{\text{overlap}} = 0.85$), \mlsysim predicts a scaling efficiency of ${\sim}93\%$, yielding an aggregate MFU of $0.93 \times 0.42 \approx 39.1\%$, within the reported 38--43\% range. This validates the distributed training model at production scale.

\textbf{Anchor~4: PaLM Hardware vs.\ Model FLOPs Utilization (Communication Overhead).}
Google's PaLM-540B~\citep{chowdhery2022palm} was trained on 6{,}144 TPU~v4 chips (two pods of 3{,}072 connected over the data-center network), achieving 57.8\% hardware FLOPs utilization but only 46.2\% \emph{model} FLOPs utilization (MFU). The gap reflects activation rematerialization and the cross-pod gradient reduction that the raw hardware-utilization figure hides. Using \mlsysim's \texttt{DistributedModel} with TPU~v4 specifications (275\,TFLOP/s BF16, ICI bandwidth at 24\,GB/s with 2$\times$ oversubscription) and the PaLM-540B workload at system-level efficiency $\eta = 0.47$, the predicted aggregate MFU is ${\sim}43\%$, within ${\sim}3$ percentage points of the reported 46.2\%. The $\eta$ absorbs ICI fabric congestion, straggler effects, and scheduling overhead. \mlsysim correctly identifies the intra-pod to inter-pod bandwidth transition, the cross-pod data-parallel gradient exchange over the slower data-center network, as the dominant communication bottleneck.

\textbf{Anchor~5: Chinchilla Scaling Laws (Algorithmic Scaling).}
The Chinchilla paper~\citep{hoffmann2022chinchilla} establishes that compute-optimal training requires $D \approx 20P$ tokens. \mlsysim's \texttt{ScalingModel} implements the parametric scaling law $C = 6PD$ and derives the optimal allocation $P^{*} = \sqrt{C/120}$. For $C = 10^{24}$ FLOPs, the solver predicts $P^{*} \approx \cvalint{\ChinchillaPstarB}$B parameters with $D^{*} \approx \cval{\ChinchillaDstarT}$T tokens. Chinchilla itself was trained at $C = 6 \times 70\text{B} \times 1.4\text{T} \approx 5.88 \times 10^{23}$ FLOPs, for which the solver predicts $P^{*} \approx 70.0$B, recovering the published 70B model size to better than 1\%. This validates the scaling law implementation against its original calibration data.

\textbf{Anchor~6: Training Carbon Footprint (Sustainability).}
\citet{patterson2021carbon} report that training GPT-3 (175B parameters) on V100 GPUs consumed approximately 1{,}287\,MWh and emitted 552 tonnes CO$_2$. Using the reported energy and US average grid intensity (429~gCO$_2$/kWh), \mlsysim's carbon-accounting layer computes $\cvalint{\GPTenergyMWh} \times \cvalint{\GPTgridCI} / 1{,}000 = \cvalint{\GPTcarbonTonnes}$ tonnes CO$_2$, matching the reported footprint to rounding. This validates the dimensional carbon calculation separately from any hardware power model.

\textbf{Anchor~7: Llama~3 Parallelism Strategy (Optimizer Convergence).}
To validate the Tier 3 design-space search, we configure the \texttt{ParallelismOptimizer} with the Meta Llama~3 405B model and its 16{,}384 H100 cluster constraints~\citep{llama3team2024}. The optimizer sweeps power-of-two TP and PP degrees with $\text{TP} \times \text{PP} \times \text{DP} = 16{,}384$, pre-screening each candidate so that the per-GPU shard of weights, gradients, and Adam optimizer state (the dominant training-memory term) fits in 90\% of the 80\,GB HBM capacity, then ranking survivors by MFU. It recovers Meta's binding decisions on both parallelism axes: $\text{TP}{=}8$, pinning tensor parallelism to a single node so AllReduce traffic stays on intra-node NVLink rather than crossing InfiniBand, and $\text{PP}{=}16$, the shallowest pipeline whose per-GPU memory screen passes once optimizer state is counted. This matches the $\text{TP}{=}8$, $\text{PP}{=}16$ parallelism backbone Meta reports; the optimizer allocates the remaining degrees to data parallelism ($\text{DP}{=}128$), whereas Meta additionally folds in context parallelism ($\text{CP}{=}16$) for batch-size and activation-memory reasons beyond the optimizer's weight-and-optimizer-state screen.

\begin{table*}[t]
\centering
\caption{\textbf{Validation Summary.} Predicted vs.\ reported values across seven empirical anchors. Error is $|(\text{pred.} - \text{rep.}) / \text{rep.}|$.}
\label{tab:validation}
\small
\renewcommand{\arraystretch}{1.15}
\begin{tabularx}{\textwidth}{@{}X X X r@{}}
\toprule
\textbf{Anchor} & \textbf{Predicted} & \textbf{Reported} & \textbf{Error} \\
\midrule
1: ResNet-50 DGX A100 & 37{,}500 s/s & v0.7 same order & order-of-mag.\ \\
2: Llama-2 70B ITL   & 43\,ms       & 40--50\,ms      & in range \\
3: Llama~3 MFU       & 39.1\%       & 38--43\%        & in range \\
4: PaLM MFU          & 43\% MFU     & 46.2\% MFU      & 6.9\% \\
5: Chinchilla $P^*$  & 70.0B        & 70B             & $<$1\% \\
6: GPT-3 CO$_2$      & 552\,t       & 552\,t          & 0.0\% \\
7: Llama~3 Parallelism & TP=8, PP=16, DP=128 & TP=8, PP=16, CP=16 & TP, PP exact \\
\bottomrule
\end{tabularx}
\end{table*}

These seven anchors span five of the six taxonomy domains (Node, Data is validated indirectly via the ResNet pipeline-bound case in \Cref{sec:usage}, Algorithm, Fleet, Operations) and cover both Roofline regimes (compute-bound and memory-bound). \Cref{tab:validation} summarizes the results.

Every entry across all eight MLSys Zoo registries records mandatory \texttt{metadata.provenance} (\Cref{sec:speed}). This enforces strict traceability: datasheet, literature, and industry-report numbers require a verified URL and date, while estimates and derived values require documented justification, and the provenance audit tool (\texttt{mlsysim.tools.audit\_provenance}) fails on any registry value whose provenance is missing or weak. This prevents the silent accumulation of ``magic numbers'' as the framework evolves.

\subsection{Design-Space Exploration Speed}
\label{sec:validation-speed}

\begin{figure}[!t]
\centering
\includegraphics[width=\columnwidth]{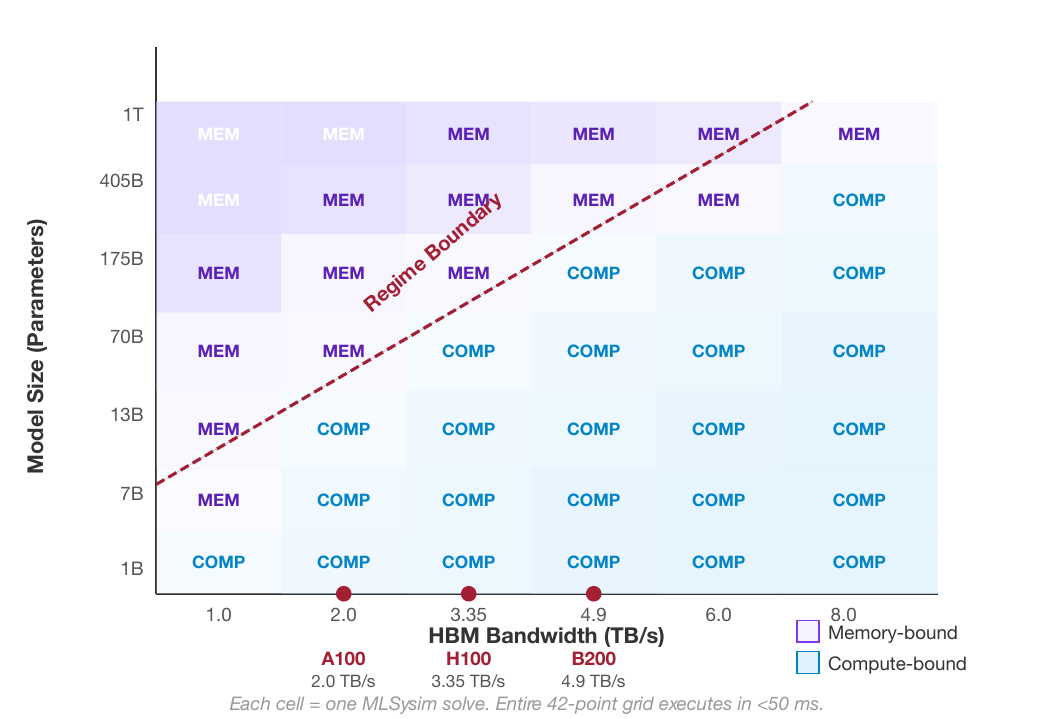}
\caption{\textbf{Design-Space Exploration: Bottleneck Regime Map.} Each cell shows the binding constraint (memory or compute) for a given model size and HBM bandwidth combination under FP16 training at batch size 256. Larger models with lower bandwidth are memory-bound; smaller models with higher bandwidth are compute-bound. The diagonal regime boundary shifts as hardware generations increase bandwidth (A100 $\to$ H100 $\to$ B200). The entire 42-point grid executes in $<$50\,ms.}
\label{fig:heatmap}
\end{figure}

\mlsysim's analytical engine sweeps over 1,000 hardware--model--precision configurations in under one second on a standard laptop. In contrast, ASTRA-sim~2.0 requires hours to simulate a single distributed training configuration at cycle-level fidelity~\citep{won2023}. This three-order-of-magnitude speedup is the design objective. Sub-second execution enables interactive parametric sweeps (e.g., varying HBM bandwidth, substituting fat-tree for torus topology, or relocating the datacenter from Iowa to Singapore) that would be impractical with cycle-accurate simulation. \Cref{fig:heatmap} illustrates one such sweep, a 42-point grid of model size versus HBM bandwidth, where each cell represents a single solver invocation and the entire map executes in under 50\,ms on a standard laptop.

\subsection{Accuracy Scope and Limitations}
\label{sec:accuracy}

\mlsysim provides first-order estimates, not cycle-accurate predictions. We do not re-derive the resolver abstraction here. It is the same first-principles structural bounds plus a single efficiency~$\eta$ for unmodeled second-order effects, as in \Cref{sec:intro,sec:philosophy}. \textbf{Calibrating $\eta$ for validation.} The efficiency parameter $\eta$ must be selected for the workload class and measurement scope: $\eta \approx 0.19$ for end-to-end ResNet-50 training throughput on A100, reflecting convolution, framework, and data-pipeline overhead~\citep{mattson2020}; $\eta \approx 0.42$ for large-scale Transformer training, consistent with Meta's reported system-level efficiency~\citep{llama3team2024}. The Roofline parallel (peak vs.\ achievable ceilings; which constraint binds) is already stated in \Cref{sec:intro}; see \citet{williams2009}. Two of the seven anchors above require no efficiency parameter at all, because the Chinchilla scaling law is pure mathematics and the carbon footprint is a direct measurement, demonstrating that the physics layer is independently sound. \Cref{sec:discussion} details the specific phenomena this abstraction cannot capture and the resulting accuracy boundaries.

For \mlsysim's intended use cases (architectural reasoning, lab exercises, capacity planning, and design-space exploration), first-order accuracy is sufficient and often preferable. A student who understands \emph{why} a system is memory-bound has learned more than one who can predict its throughput to three decimal places.

To safeguard against model drift, the test suite includes empirical anchor tests (\texttt{engine.empirical}) that fail automatically if predictions leave their reviewed tolerance envelopes around published values.

\subsection{Dimensional Correctness as Validation}
\label{sec:validation-dim}

Beyond numerical accuracy, dimensional strictness (\Cref{sec:dimensional}) provides a structural form of validation. Unit mismatches raise immediate errors rather than producing silently wrong results, eliminating a category of bugs that numerical validation alone cannot catch.

\subsection{Design Invariant Enforcement}
\label{sec:validation-ratchets}

Dimensional strictness (\Cref{sec:dimensional}) catches unit errors at runtime. A complementary question is whether the framework's \emph{architectural} invariants hold as the codebase evolves: do all hardware specifications flow through typed registries? Does every formula carry \texttt{pint} units end-to-end? Can a contributor accidentally reintroduce a bare constant that bypasses the registry? \mlsysim answers these with five machine-checked structural invariants, enforced by an 862-test suite, each encoding a design decision that would otherwise erode through incremental changes:

\begin{enumerate}[leftmargin=*,itemsep=2pt]
\item \textbf{Specification separation.} There is no constants module at all: the legacy \texttt{core/constants.py} was deleted outright once its last value migrated, and a CI gate now pins its \emph{absence} (any attempt to reintroduce the module, or a compatibility shim for it, fails), alongside name-pattern checks for hardware, model, and network constants. Every domain figure lives in its category registry; measurement units live in \texttt{core.units}.
\item \textbf{Loader contract.} The YAML registry loader rejects duplicate keys outright and requires every \texttt{@tech:} and \texttt{@prov:} reference to resolve, so a typo in a data file fails at import rather than silently shadowing or orphaning a specification.
\item \textbf{Provenance coverage.} A dedicated audit (\texttt{mlsysim.tools.audit\_provenance}, also run as a test) walks every registry leaf and fails on missing or weak provenance, enforcing the kind-specific evidence rules of \Cref{sec:speed} across all eight zoos.
\item \textbf{Cross-registry consistency.} When the same physical quantity appears in multiple registries (e.g., NVLink bandwidth in both \textbf{Hardware} and \textbf{Systems} node definitions), a dedicated check verifies that the values agree. This prevents silent divergence between overlapping specification surfaces.
\item \textbf{Registry immutability.} Every registry entry is a frozen \texttt{pydantic} model: attribute assignment raises at the call site. Entries are shared singletons read by every notebook cell in a rendering process, so without this guarantee a single stray write (\texttt{H100.tdp = ...}) would silently corrupt every downstream calculation; with it, the mistake fails loudly where it happens. Code that needs a perturbed variant builds an explicit copy (\texttt{model\_copy}).
\end{enumerate}

These checks are additive: each new registry entry or physics module automatically participates without manual test authoring. Empirical-anchor tests (\texttt{engine.empirical}) complete the picture by binding representative solver outputs to sourced benchmark envelopes, so a formula regression or a registry edit that moves a prediction outside its reviewed range fails CI. The effect is that the zoo architecture described in \Cref{sec:stack} is not merely documented but \emph{enforced} — a structural guarantee that the framework's design decisions survive community contributions and long-term evolution.

\section{Usage \& Case Studies}
\label{sec:usage}

\mlsysim is designed for three audiences: students developing quantitative reasoning skills, instructors preparing demonstrations, and researchers evaluating design trade-offs. We present two representative use cases per persona, each illustrating how solvers compose to answer questions that span multiple walls. For a "Cookbook" of exciting, runnable code examples demonstrating how users define hardware, perform algebraic design-space sweeps, and invoke SLA-driven hardware synthesis programmatically in Python, see the Appendix (Listings~\ref{lst:speculative},~\ref{lst:appendix-sweep},~\ref{lst:appendix-compose},~\ref{lst:appendix-data},~\ref{lst:appendix-synthesis},~\ref{lst:appendix-speculative},~\ref{lst:appendix-scorecard},~\ref{lst:appendix-optimizer}, and~\ref{lst:appendix-edge}).

\subsection{Student Use Cases}

Students interact with \mlsysim primarily through single-solver queries, short resolver chains, and browser-friendly exercises in the companion textbook. Students can manipulate hardware parameters (e.g., batch size, SLA targets, carbon taxes) and observe how binding constraints shift without needing a backend server or physical hardware.

We present two examples chosen to illustrate complementary aspects of the framework. The first (S1) demonstrates \emph{vertical} resolver composition, chaining inference and economics models to connect an algorithmic decision (chain-of-thought reasoning) to its infrastructure cost. The second (S2) demonstrates \emph{horizontal} composition, combining data-pipeline and compute models to diagnose a bottleneck that shifts between walls as batch size increases.

\subsubsection{S1: Chain-of-Thought Cost}
A student investigates the inference economics of chain-of-thought (CoT) prompting. Using Llama-3 70B on H100 hardware, they configure the \texttt{InferenceScalingModel} with $K{=}8$ reasoning steps. Each step generates $\sim$50 tokens (the calibrated default) at the memory-bound decode rate, so the total reasoning time is $T_{\text{reason}} = \text{TTFT} + K \cdot 50 \cdot \text{ITL}$. The solver reports a total reasoning time of ${\sim}17.9$\,s, a $6.5{\times}$ per-query \emph{latency} multiplier relative to a single-step answer (the one-time prefill TTFT amortizes across the $K$ steps). The \emph{cost} multiplier is larger and tracks total work rather than latency: CoT generates $8{\times}$ as many decode tokens ($K \cdot 50$ vs.\ 50), and because decode is memory-bandwidth-bound the serving throughput is set by token volume. Feeding this into the \texttt{EconomicsModel}, a deployment sized for 100~QPS of single-pass queries must grow its H100 fleet, and hence its TCO, by ${\sim}8{\times}$ to sustain the same query rate under CoT. The two multipliers are distinct: \mlsysim shows chain-of-thought taxes both latency and infrastructure cost, but by different factors.

\subsubsection{S2: CPU Pipeline Bottleneck}
A student configures ResNet-50 training on a DGX A100 (8 GPUs) with a batch size of 2{,}048. At a kernel-level efficiency of $\eta = 0.30$, the \texttt{SingleNodeModel} predicts a per-step compute time of ${\sim}35$\,ms, a demand rate of ${\sim}58{,}000$ images/s. Adding the \texttt{DataModel} rules out storage I/O. With \cvalint{\StwoCPUworkers} CPU workers (8 per GPU) decoding ImageNet JPEGs at \cvalint{\StwoDecodeRate} images/s each, the raw decode pipeline delivers \StwoRawThroughput images/s. This appears sufficient, but the \texttt{TransformationModel} accounts for the full augmentation pipeline (random crop, color jitter, normalization) at \cvalint{\StwoAugRate} images/s per worker, reducing effective throughput to $\cvalint{\StwoCPUworkers} \times \cvalint{\StwoAugRate} = \StwoEffThroughput$ images/s, below the GPUs' $58{,}000$ images/s appetite. The student discovers that the binding constraint shifts from Wall~1 (Compute) to Wall~9 (Transformation): the GPUs have spare cycles, but the CPUs cannot feed them fast enough, and the achieved \emph{end-to-end} efficiency degrades below the kernel-level $\eta$, consistent with the gap between kernel-level and end-to-end efficiency in Anchor~1 ($\eta = 0.19$, \Cref{sec:validation-anchors}).

\subsection{Instructor Use Cases}

Instructors need demonstrations that run in real time, produce concrete numbers, and connect cleanly to lecture narratives. The following cases show how \mlsysim turns abstract concepts into live, interactive classroom demonstrations.

\subsubsection{I1: Live Roofline Demo, Batch Size Sweep}
An instructor demonstrates the Roofline model by sweeping batch size from 1 to 256 on an H100 for ResNet-50 inference at $\eta = 0.5$. At each batch size, the \texttt{SingleNodeModel} returns the bottleneck label, the arithmetic intensity, and the MFU:

\begin{center}
\small
\begin{tabularx}{\columnwidth}{@{}rXlr@{}}
\toprule
Batch & Bottleneck & AI (FLOP/B) & MFU \\
\midrule
1   & Memory   & 73   & 0.01 \\
8   & Compute  & 356  & 0.06 \\
32  & Compute  & 610  & 0.17 \\
128 & Compute  & 743  & 0.34 \\
256 & Compute  & 771  & 0.40 \\
\bottomrule
\end{tabularx}
\end{center}

\begin{figure}[!t]
\centering
\includegraphics[width=\columnwidth]{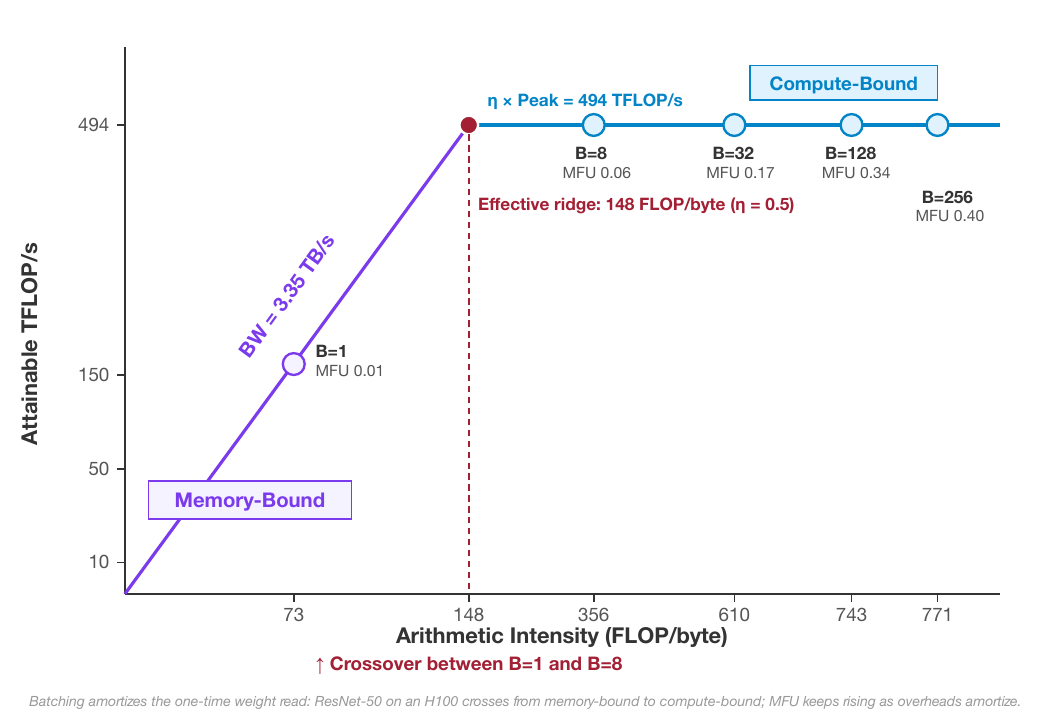}
\caption{\textbf{Roofline Crossover: ResNet-50 Batch Size Sweep on H100.} Increasing batch size moves the operating point rightward along the Roofline, transitioning from memory-bound (purple) to compute-bound (cyan) between batch 1 and batch 8. The peak ridge point is $\cvalint{\HhPeakTFLOPS}/\cvalii{\HhBWTBs} \approx \cvalint{\HhRidgePoint}$\,FLOP/byte; at $\eta = 0.5$ the effective ridge sits at ${\approx}148$\,FLOP/byte. MFU continues rising past the crossover as dispatch and framework overheads amortize.}
\label{fig:roofline}
\end{figure}

The crossover from memory-bound to compute-bound occurs between batch 1 and batch 8, where the arithmetic intensity crosses the effective ridge point $\eta \times F_{\text{peak}} / \text{BW}_{\text{HBM}} \approx 148$\,FLOP/byte: batching amortizes the one-time weight read across samples, while per-sample activation traffic bounds the gain. MFU keeps climbing past the crossover (0.06 $\to$ 0.40) as the fixed dispatch and per-layer framework taxes amortize. \Cref{fig:roofline} visualizes this transition on the Roofline diagram. The contrast case makes the deeper point: repeating the sweep for a 7B-parameter LLM \emph{decode} workload shows the arithmetic intensity saturating near 10\,FLOP/byte, far left of the ridge, because KV-cache and activation reads grow with the batch; decode remains memory-bound at every batch size (Walls~4--5). The entire sweep executes in $<$50\,ms, enabling real-time interaction during lecture.

\subsubsection{I2: Iowa vs.\ Qu\'ebec Carbon}
An instructor poses a policy question: \emph{does geography matter for carbon footprint?} Using the \texttt{DistributedModel}, they configure a 256-GPU cluster (\texttt{Systems.Clusters.Research\_256}) training a 70B model for 30 days at MFU 0.42. The \texttt{SustainabilityModel} then computes emissions under two registered datacenter profiles: Iowa (680~gCO$_2$/kWh, PUE 1.12, circa 2020 coal/gas grid) and Qu\'ebec (20~gCO$_2$/kWh, PUE 1.06, hydroelectric). The configuration is identical in hardware, model, and MFU, yet the run emits 58.4 tonnes CO$_2$ in Iowa versus 1.6 tonnes in Qu\'ebec, a $36{\times}$ gap, demonstrating that grid carbon intensity is a first-order systems design variable (Wall~18). \Cref{fig:carbon} visualizes this contrast.
\afterpage{%
\begin{figure}[!t]
\centering
\includegraphics[width=\columnwidth]{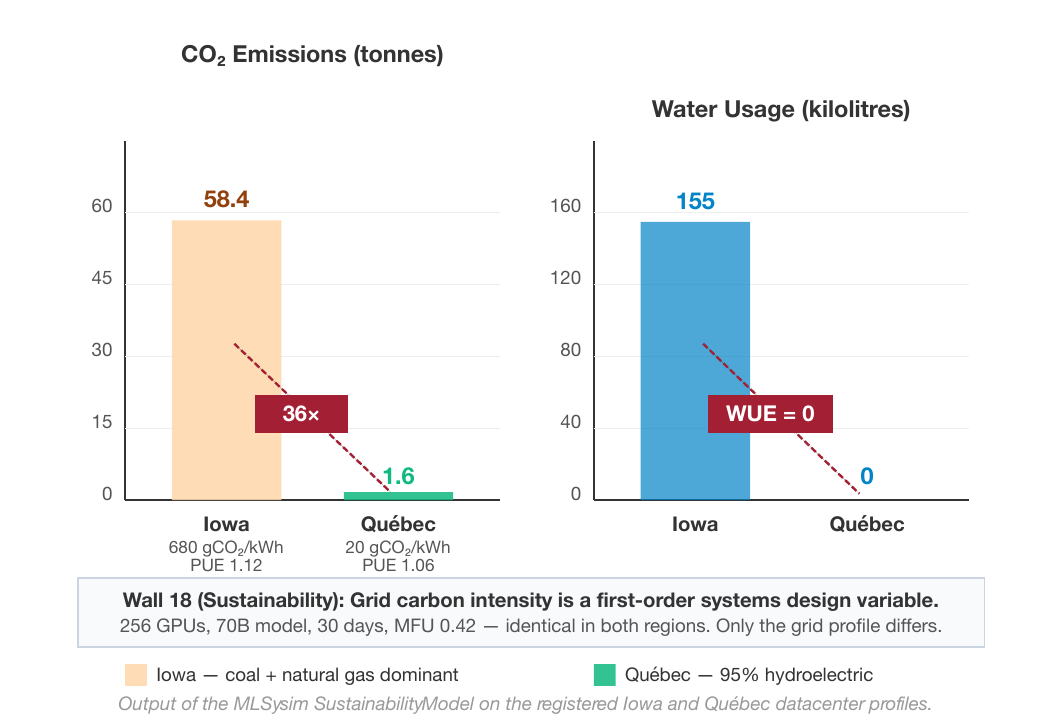}
\caption{\textbf{Geography as a Systems Variable: Iowa vs.\ Qu\'ebec.} An identical 256-GPU cluster training a 70B model for 30 days produces $36{\times}$ more CO$_2$ in Iowa (coal/gas grid circa 2020, 680~gCO$_2$/kWh, PUE 1.12) than in Qu\'ebec (95\% hydroelectric, 20~gCO$_2$/kWh, PUE 1.06): 58.4 vs.\ 1.6 tonnes. Water usage diverges even more sharply because Qu\'ebec's registered WUE is zero. Hardware, model, parallelism strategy, and MFU are identical; only the regional grid profile differs between the two sites.}
\label{fig:carbon}
\end{figure}%
}

\subsection{Researcher Use Cases}

Researchers need to evaluate architectural alternatives and justify procurement decisions with quantitative evidence. The following cases show how \mlsysim enables rapid what-if analysis across hardware generations and parallelism strategies.

\subsubsection{R1: GPU vs.\ Cerebras Crossover}
A researcher evaluates whether wafer-scale silicon changes the economics of large-model inference. For a 180B-parameter model on H100s (the 360\,GB FP16 checkpoint requires multi-GPU tensor parallelism across 5 GPUs), \mlsysim reports a decode-phase ITL of $360\;\text{GB} / (5 \times 3.35\;\text{TB/s}) = 21.5$\,ms/token at batch 1 (memory-bound). The \texttt{WeightStreamingModel} then models the Cerebras WSE-3, assuming a MemoryX injection bandwidth of $1.2$\,TB/s (Cerebras has not published an official MemoryX bandwidth figure, so we treat this as a calibration estimate). At batch size 1, the full 360\,GB must stream from MemoryX at this rate, yielding a per-token latency of $360 / 1{,}200 = 300$\,ms, $14{\times}$ \emph{slower} than the GPU cluster. However, the Cerebras architecture's advantage emerges at scale. At the optimal batch $B^{*} \approx 41{,}700$ (\Cref{eq:weightstream}, at $\eta = 0.4$), injection cost is fully amortized across tokens, achieving ${\sim}139{,}000$ tokens/s aggregate throughput versus ${\sim}47$ tokens/s for the five-GPU tensor-parallel replica (requiring a fleet of ${\sim}15{,}000$ H100s to match). The solver flags that at $B^{*}$, the activation footprint approaches the 44\,GB SRAM ceiling. The \texttt{SensitivitySolver} confirms the qualitative regime change. The binding constraint shifts from $\text{BW}_{\text{HBM}}$ (GPU) to $\text{BW}_{\text{inject}}$ (WSE-3), illustrating that the optimal architecture depends fundamentally on batch size.

\subsubsection{R2: Hardware Procurement Audit}
A researcher preparing a hardware procurement recommendation for LLM inference needs to answer: \emph{should the next-generation cluster prioritize FLOPS or bandwidth?} The \texttt{SensitivitySolver} perturbs each hardware parameter by 10\% and reports the normalized latency response, shown here for a single-GPU-resident Llama-3 8B decode workload:

\begin{lstlisting}[caption={\textbf{Sensitivity Analysis.} Identifying the binding constraint via 10\% hardware perturbations.},label={lst:sensitivity}]
import mlsysim
from mlsysim.solvers import SensitivitySolver
solver = SensitivitySolver()
res = solver.solve(model=mlsysim.Models.Language.Llama3_8B,
                   hardware=mlsysim.Hardware.Cloud.A100)
print(res.sensitivities)
# {'peak_flops': 0.0, 'memory_bandwidth': -0.0875,
#  'memory_capacity': 0.0}
print(f"Binding: {res.binding_constraint}")
# Output: Binding: memory_bandwidth
\end{lstlisting}

The result is unambiguous: a 10\% increase in HBM bandwidth cuts decode latency by 8.8\%, whereas a 10\% increase in peak FLOPS yields nothing at all, because batch-1 decode sits strictly on the memory roof of \Cref{eq:bottleneck}. The \texttt{SynthesisSolver} then performs the inverse solve. Given a \cvalint{\SynthTargetMs}\,ms inter-token latency SLA for LLaMA-70B, it synthesizes the minimum required bandwidth as $\text{BW}_{\text{req}} = |W| / T_{\text{target}} = \cvalint{\SynthBWreqGBs}$\,GB/s, $\cval{\SynthAhMultiple}{\times}$ the A100's \cval{\AhBWTBs}\,TB/s, confirming that LLaMA-70B inference is firmly in the memory-bound regime and that hardware procurement should prioritize HBM bandwidth over peak throughput.

\subsubsection{R3: End-to-End LLaMA-70B Training Audit}
To illustrate how \mlsysim's solvers compose across all six taxonomy domains, we trace a complete training analysis for LLaMA-70B on \cvalint{\RthreeGPUs} H100 GPUs (\cvalint{\RthreeNodes} nodes $\times$ \cvalint{\RthreeTPsize} GPUs, NVLink intra-node, 400\,Gb/s InfiniBand inter-node) in Qu\'ebec (20~gCO$_2$/kWh, PUE 1.06).

\textbf{Node (Walls 1--3).} With DP degree 64, each DP rank processes $4\text{M} / 64 = 62{,}500$ tokens per step. Within each rank, TP=8 partitions the model across 8 local GPUs. The per-rank compute demand is $C_{\text{rank}} = 6 \times 70\text{B} \times 62{,}500 = 2.63 \times 10^{16}$ FLOPs. The \cvalint{\RthreeTPsize}-GPU TP group delivers $\cvalint{\HhPeakTFLOPS} \times 10^{12} \times \cval{\RthreeEta} \times \cvalint{\RthreeTPsize} = 3.17 \times 10^{15}$ FLOP/s (at $\eta = \cval{\RthreeEta}$), yielding a per-step compute time of $T_{\text{compute}} = 2.63 \times 10^{16} / 3.17 \times 10^{15} = 8.3$\,s. The \texttt{SingleNodeModel} classifies this as compute-bound. Memory bandwidth ($\cvalint{\RthreeTPsize} \times \cvalii{\HhBWTBs} = 26.8$\,TB/s) can stream the \cvalint{\LLaMAWeightsGB}\,GB model weights in 5.2\,ms, far below the 8.3\,s compute time.

\textbf{Data (Walls 8--10).} The \texttt{DataModel} checks ingestion. At global batch size 4M tokens, the data pipeline must sustain $4\text{M} \times 2\;\text{bytes} / 8.3\;\text{s} \approx 0.96$\,MB/s from storage, a trivially small demand. With NVMe delivering 6.5\,GB/s per node across 64 nodes, the pipeline is not the bottleneck. (For LLM training, tokenized data is compact; the data wall binds primarily in vision tasks with large image payloads.)

\textbf{Algorithm (Walls 11--13).} The \texttt{ScalingModel} verifies that the training budget is compute-optimal. For $C = 2 \times 10^{24}$ FLOPs, the Chinchilla-optimal model size is $P^{*} \approx 130$B, indicating that 70B at this budget is slightly over-trained (more data per parameter than optimal), a deliberate choice for inference efficiency.

\textbf{Fleet (Walls 14--16).} The \texttt{DistributedModel} models communication. With TP=\cvalint{\RthreeTPsize} (intra-node NVLink, \cvalint{\NVLinkBWGBs}\,GB/s) and DP=\cvalint{\RthreeDPsize} (inter-node IB, \cvalint{\IBBWGBs}\,GB/s per port), the DP AllReduce synchronizes $\cvalint{\LLaMAWeightsGB}\;\text{GB} / \cvalint{\RthreeTPsize} = \cval{\RthreeGradShardGB}$\,GB of gradients per TP rank across \cvalint{\RthreeDPsize} nodes. Ring AllReduce cost is $2 \times (63/64) \times \cval{\RthreeGradShardGB}\;\text{GB} / \cvalint{\IBBWGBs}\;\text{GB/s} \approx \cvalint{\RthreeARms}$\,ms. With $\eta_{\text{overlap}} = \cval{\RthreeOverlap}$, only $\pgfmathparse{1-\RthreeOverlap}\pgfmathprintnumber[fixed,precision=2]{\pgfmathresult} \times \cvalint{\RthreeARms} = \cvalint{\RthreeExposedMs}$\,ms of communication is exposed, yielding a scaling efficiency of $8.3 / (8.3 + 0.103) = 98.8\%$ at 64 nodes. The \texttt{ReliabilityModel} estimates cluster MTBF $= \cvalint{\NodeMTBFhrs}\;\text{hrs} / \cvalint{\RthreeGPUs} \approx \cval{\RthreeClusterMTBF}$\,hrs, requiring hourly checkpoints.

\textbf{Operations (Walls 17--20).} The \texttt{EconomicsModel} projects a 30-day training run: CapEx (\cvalint{\RthreeGPUs} H100s at \$\cvalint{\HhCostK}K each) of \$15.4M amortized over \cvalint{\AmortYears} years yields a per-run allocation of $\$15.4\text{M} \times 30/1{,}095 \approx \$422\text{K}$, plus OpEx (power at 700W $\times$ 512 GPUs $\times$ 720\,hrs at \$0.06/kWh) of \$15.5K, for a total run TCO of $\sim$\$0.44M. The \texttt{SustainabilityModel} estimates 274\,MWh facility energy and 5.5 tonnes CO$_2$, $36{\times}$ less carbon than the same run in Iowa (680~gCO$_2$/kWh, PUE 1.12).

\textbf{Analysis (Walls 21--22).} The \texttt{SensitivitySolver} confirms the binding constraint is compute (Wall~1), not memory or communication, with $\partial T / \partial F_{\text{peak}} = -0.91$. The \texttt{SynthesisSolver} synthesizes the minimum hardware to complete training in 14 days: 1{,}024 GPUs, doubling DP to 128.

This end-to-end trace exercises 12 of the 22 walls through a single model, demonstrating how solver composition produces a complete system assessment from individual physics-based constraint equations.

\subsubsection{R4: Automated Parallelism Search}
A researcher needs to schedule a 175B-parameter model on a new 2{,}048-GPU cluster. Manually searching the 3D-parallelism space ($\text{TP} \times \text{PP} \times \text{DP}$) is error-prone. A split that maximizes DP might exceed the 80\,GB HBM capacity, while a split that maximizes TP might saturate the NVLink interconnect. Instead of trial and error, the researcher invokes a Tier 3 Optimizer. They configure the \texttt{ParallelismOptimizer} with the workload and cluster, which maximizes MFU subject to a built-in memory screen: the per-GPU weight-plus-gradient shard must fit in 90\% of HBM capacity ($\le 72$\,GB). The optimizer sweeps power-of-two factorizations of 2{,}048 with TP capped at the node size, evaluating the \texttt{DistributedModel} at each of the 29 memory-feasible candidates. In well under a second, it returns the optimal schedule: $\text{TP}{=}8$, $\text{PP}{=}2$, $\text{DP}{=}128$, correctly deducing that TP must match the intra-node GPU count to avoid traversing the slower inter-node fabric, and that PP=2 is the minimum pipeline depth at which the $175\text{B} \times 2\,\text{bytes} \times 2 / (8 \times 2) \approx 43.8$\,GB shard fits the screen (PP=1 would demand 87.5\,GB). This demonstrates the power of the ``engineering engine'' to invert the analytical models into automated design-space synthesis.

\section{Fallacies \& Pitfalls}
\label{sec:fallacies}

Following the tradition of \citet{hennessy2024architecture}, we highlight common misconceptions that \mlsysim is designed to expose.

\textbf{Fallacy: Doubling peak FLOP/s halves training time.}
A student might assume that upgrading from A100 (312\,TFLOP/s FP16 dense) to H100 (\cvalint{\HhPeakTFLOPS}\,TFLOP/s, a $3.2{\times}$ increase) should yield a $3.2{\times}$ speedup. \mlsysim's \texttt{SingleNodeModel} reveals why this is false. LLM inference is memory-bandwidth-bound, and the A100-to-H100 bandwidth improvement is only about $1.6{\times}$ (\cval{\AhBWTBs}~TB/s $\to$ \cvalii{\HhBWTBs}~TB/s). For memory-bound workloads, training time scales with bandwidth, not FLOPS. The Roofline model (\Cref{eq:bottleneck}) makes this visible by showing that the binding constraint determines which hardware parameter matters.

\textbf{Fallacy: Communication overhead is negligible at 512 GPUs.}
At \cvalint{\RthreeGPUs} H100 GPUs (\cvalint{\RthreeNodes} nodes), the DP AllReduce for a 70B model's \cval{\RthreeGradShardGB}\,GB gradient shard costs \cvalint{\RthreeARms}\,ms, of which only \cvalint{\RthreeExposedMs}\,ms is exposed after 85\% compute--communication overlap, just 1.2\% of the 8.3\,s compute step. But the \texttt{ReliabilityModel} reveals the hidden cost. Cluster MTBF drops to $\sim$20 hours, requiring hourly checkpoints that each pause training for 30--60 seconds. Over a 30-day run, checkpoint overhead (Wall~19) exceeds communication overhead (Wall~14) by $10{\times}$, a cost invisible to communication-only analysis.

\textbf{Pitfall: Using peak bandwidth in back-of-the-envelope calculations.}
Vendor datasheets report peak HBM bandwidth (e.g., \cvalii{\HhBWTBs}\,TB/s for H100). In practice, sustained bandwidth under real workloads is 70--85\% of peak due to bank conflicts, address patterns, and memory controller scheduling~\citep{nvidia2023h100}. A student using peak bandwidth will underestimate LLM decode latency by 15--30\%. \mlsysim's MFU parameter ($\eta$) explicitly accounts for this gap, and the default ranges (\Cref{eq:efficiency}) guide students toward realistic estimates.

\textbf{Pitfall: Ignoring geography in carbon accounting.}
Two identical training runs produce vastly different environmental impact depending on grid carbon intensity. As demonstrated in Case~I2, the same 256-GPU cluster emits $36{\times}$ more CO$_2$ in Iowa than in Qu\'ebec. Students who omit Wall~18 (Sustainability) from their analysis miss a first-order systems design variable, one that increasingly affects both cost (carbon pricing) and regulatory compliance.

\textbf{Fallacy: Quantization always provides a linear speedup.}
Reducing precision from FP16 to INT4 ($r_{\text{quant}} = 4{\times}$) reduces memory by $4{\times}$, but the compute speedup depends on whether the workload is memory-bound or compute-bound. For memory-bound LLM decode, the $4{\times}$ memory reduction translates to nearly $4{\times}$ throughput improvement because the bottleneck is weight reads. For compute-bound training, the same quantization provides zero throughput benefit because compute, not memory, is the ceiling. \mlsysim's Roofline analysis makes this regime-dependent behavior explicit.

\section{Discussion \& Limitations}
\label{sec:discussion}

``All models are wrong, but some are useful''~\citep{box1976}. Users must understand the boundaries of \mlsysim's analytical abstraction. We organize the limitations into modeling scope, accuracy trade-offs, and pedagogical implications, then outline future directions.

\subsection{What \mlsysim Cannot Model}

\textbf{No microarchitectural effects.} \mlsysim has no notion of L1/L2 cache hierarchies, branch prediction, warp scheduling, or register pressure. These second-order effects are absorbed into a single scalar efficiency parameter ($\eta$, the ratio of sustained to peak FLOP/s). While $\eta$ provides a serviceable approximation for back-of-the-envelope reasoning, it cannot capture workload-dependent microarchitectural behavior; a matrix multiply and a sparse attention kernel may achieve very different $\eta$ on identical silicon.

\textbf{No real network congestion.} The communication model uses the classical $\alpha$-$B_{\text{link}}$ formulation (latency plus inverse-bandwidth), which assumes dedicated links. \mlsysim does not model adaptive routing, network contention under multi-tenant traffic, or congestion collapse, phenomena that become critical at scales beyond $\sim$10{,}000 nodes, precisely the regime where ASTRA-sim~2.0~\citep{won2023} provides essential fidelity. The static bisection view behind Wall~10 (Topology) is optimistic under the same cross-traffic conditions.

\textbf{No OS/runtime overhead.} Kernel launch latency, CUDA stream scheduling, Python GIL contention, and host--device transfer overhead are absent. For inference-dominated workloads where kernel launch time can rival compute time, this omission can meaningfully affect predictions.

\textbf{No heterogeneous fleets.} The \texttt{DistributedModel} assumes homogeneous nodes, where all accelerators in a fleet share the same compute, memory, and interconnect specifications. Production clusters increasingly mix hardware generations (e.g., A100 and H100 nodes in the same job), and fleet-level efficiency metrics such as ML Productivity Goodput~\citep{wongpanich2025fleet} capture this heterogeneity. Modeling heterogeneous fleets would require per-node load balancing and straggler analysis beyond the current analytical framework.

\textbf{No dynamic behavior.} \mlsysim models steady-state throughput. Transient effects (thermal throttling, dynamic clock boosting, memory fragmentation over long training runs, and checkpoint I/O bursts) are outside its scope. A training run that degrades over 72 hours due to thermal saturation will appear identical to one that sustains peak throughput.

\textbf{Heuristic accuracy models.} The accuracy degradation curves in \texttt{CompressionModel} are heuristic step functions (for quantization) and exponentials (for pruning), not architecture-specific empirical fits. Real accuracy loss depends on model architecture, calibration methodology, and quantization method (e.g., well-calibrated GPTQ INT4 can achieve $<$0.5\% degradation, while naive round-to-nearest may lose 10\%+). Users should treat these curves as directional indicators, not ground truth.

\subsection{Walls Not Included}

The 22-wall taxonomy is comprehensive but not exhaustive. Several constraints were considered and excluded, each for a specific reason.

\textbf{Thermal throttling.} Sustained power density can force throughput below peak TDP, but this is absorbed into $\eta$ rather than modeled as a distinct wall.

\textbf{Resource fragmentation.} Scattered GPU availability across nodes prevents job scheduling even when aggregate capacity is sufficient; this is a combinatorial bin-packing problem beyond the current analytical framework.

\textbf{Compiler/graph optimization.} The gap between a framework's computational graph and the executed kernel schedule affects both latency and MFU, but varies too rapidly across software versions to model analytically.

The selection criterion for inclusion was: does the constraint have a stable, published analytical formulation that remains valid across hardware generations? Constraints requiring empirical trace data or combinatorial optimization were deferred to future work.

\subsection{The Accuracy--Speed Trade-off}

Each omission above reflects the same trade-off: three orders of magnitude improvement in evaluation speed at the cost of second-order fidelity. The precedent is the MIPS/SPIM simulator~\citep{hennessy2024architecture}, which models pipeline hazards and stalls but omits superscalar execution and cache hierarchies, prioritizing pedagogical clarity over the full complexity of commercial processors. \mlsysim applies the same philosophy to ML systems, making quantitative reasoning accessible to students who may never operate a production cluster.

The relevant question is not ``How accurate is \mlsysim?'' but ``Does it identify the correct binding constraint?'' A first-order model that correctly determines whether a system is memory-bound, compute-bound, or network-bound provides actionable architectural insight even when its absolute latency prediction is $\pm$20\% from a cycle-accurate trace. The binding constraint dictates which hardware investment yields the largest return, and this ordinal ranking is far more robust than cardinal predictions. Practitioners who know that their system is memory-bandwidth-bound will invest in higher-bandwidth memory regardless of whether the predicted latency is 47\,ms or 53\,ms.

\subsection{Future Work}

We identify several directions for extending \mlsysim.

\textbf{Broader empirical validation.} \Cref{sec:validation} validates against seven anchors spanning five domains. Future work will extend this to newer hardware generations already present in the registry (NVIDIA B200/GB200, TPU~v6) as published benchmark results appear, inference serving under load (continuous batching with realistic request distributions), and checkpoint overhead at scale, where published data points are becoming available from reproducibility studies.

\textbf{Community hardware registry.} The \textbf{Hardware} registry currently contains a curated set of hardware entries verified against manufacturer datasheets. We plan to open contributions from the community, with automated verification scripts that cross-check submitted specifications against known physical limits (e.g., memory bandwidth cannot exceed pin count $\times$ data rate).

\textbf{Custom degradation curves.} Future versions will allow users to supply empirically fitted accuracy-compression curves from their own quantization experiments, replacing the current heuristic polynomials with data-grounded models.

\textbf{TinyTorch integration.} \mlsysim provides analytical predictions; TinyTorch~\citep{tinytorch2025}, the companion educational framework, provides implementation-based verification. Connecting the two tools creates a predict-then-verify loop. Students estimate training time and memory consumption in \mlsysim, then run the actual training in TinyTorch and compare. This closed loop reinforces quantitative reasoning by grounding analytical models in empirical observation.

\textbf{Expanding Tier 3 Optimizers (Pareto Frontiers).} Currently, the Tier 3 optimizers search single-dimensional objective spaces (e.g., maximizing MFU or maximizing batch size under a latency constraint). Future work will extend the Tier 3 engine to support multi-objective Pareto frontiers, simultaneously optimizing across latency, total cost, carbon footprint, and accuracy. This will enable richer design-space exploration and formally expose the inherent tensions between performance and sustainability.

\subsection{The Pedagogical Argument}

Even when \mlsysim's predictions deviate by 20\% from measured values, the pedagogical value lies in the reasoning process rather than the numerical output. A student who sweeps 1{,}000 configurations and identifies memory bandwidth as the binding constraint for LLM inference has acquired a transferable analytical skill, that of determining which resource limits performance. The framework trains students to formulate the correct quantitative questions (arithmetic intensity, ridge point location, communication-to-computation ratio) and these questions generalize to production systems even as specific hardware parameters change across generations.

\section{Conclusion}
\label{sec:conclusion}

Modern ML systems are no longer limited by a single resource that can be optimized in isolation. A deployment decision can hinge simultaneously on arithmetic intensity, memory capacity, communication topology, queueing delay, energy price, regional carbon intensity, and failure recovery. This coupling makes intuition brittle: a faster accelerator may not improve decode latency, a larger batch may move the bottleneck from memory to compute, and a cheaper datacenter may be the wrong choice once emissions or water constraints are included. The practical challenge is to make these interactions visible before teams commit to hardware, cloud spend, or weeks of empirical benchmarking.

\mlsysim treats ML system design as a set of executable constraints. Its purpose is not to replace profilers, trace-driven simulators, or cycle-level models; those tools remain necessary when the question is precise performance on a concrete system. Instead, \mlsysim occupies the earlier stage where engineers and students need to ask which wall binds, which parameter matters, and which design alternatives are worth measuring at all. Its standard is first-order fidelity: close enough to identify bottlenecks and rank alternatives, explicit enough to expose assumptions, and fast enough to explore the design space before measurement begins. By separating demand from supply, carrying units through every calculation, and composing first-principles resolvers across the full stack, it turns informal back-of-the-envelope reasoning into a reproducible infrastructure model.

That artifact is also pedagogical. Paired with TinyTorch~\citep{tinytorch2025} and the open \emph{Machine Learning Systems} text~\citep{mlsysbook2025}, \mlsysim gives learners a way to connect equations, code, and infrastructure consequences without requiring access to a GPU cluster. The goal is not merely to compute a latency or carbon number, but to teach a durable habit of systems reasoning: identify the binding constraint, check the units, perturb the design, and explain why the result changed. As ML systems continue to scale across larger models, heterogeneous fleets, and stricter economic and environmental constraints, that habit will matter as much as any single modeling tool.

\section*{Generative AI Usage Statement}

I used a variety of generative AI tools during the development of \mlsysim and the preparation of this manuscript. These tools supported literature discovery, landscape exploration, brainstorming, code prototyping, implementation iteration, prose editing, figure refinement, and pedagogical organization. I designed the framework, taxonomy, registry hierarchy, modeling abstractions, evaluation methodology, and educational framing; reviewed and revised AI-assisted outputs; and verified the final code, equations, citations, figures, and claims. Generative AI tools were not treated as authors or sources of scientific authority. I remain fully responsible for the originality, correctness, and integrity of the work.

\clearpage
\onecolumn
\begin{appendices}

\section{MLSysBook Integration (LEGO Cells)}
\label{sec:appendix-mlsysbook}

\mlsysim serves as the executable backbone for \emph{Machine Learning Systems} (Volume 1: \url{https://mlsysbook.ai/vol1} and Volume 2: \url{https://mlsysbook.ai/vol2})~\citep{mlsysbook2025}. To integrate quantitative analysis into the prose without disrupting readability, the textbook introduces the concept of \textbf{LEGO Cells}. LEGO is an acronym for the four structural phases of each cell: \textbf{Load} (importing registries and constraints), \textbf{Execute} (invoking the physics solvers), \textbf{Guard} (asserting dimensional and logical invariants), and \textbf{Output} (formatting physical quantities for inline prose). A LEGO Cell is a standalone, self-contained Python snippet that answers a specific design question posed in the text. Isolating the simulation logic into these executable blocks allows readers to read the code to understand the mechanics, execute it to verify the results, and modify the parameters to explore alternative scenarios.

Below, we provide three representative LEGO Cells used in the textbook to illustrate this pedagogical integration.

\subsection{Volume 1 Example: Hardware Constraints}
In Volume 1, LEGO Cells are heavily used to explain hardware constraints. For example, when introducing the Roofline model, a LEGO Cell demonstrates how memory bandwidth dictates the maximum batch size for a given latency SLA:

\begin{lstlisting}[language=Python, caption={\textbf{Volume 1 LEGO Cell: Roofline Analysis.} This snippet from Volume 1 uses the \texttt{Engine} to find the batch size where a model transitions from memory-bound to compute-bound.}, label={lst:lego-vol1}]
from mlsysim.engine.engine import Engine
from mlsysim.hardware.registry import Hardware
from mlsysim.models.registry import Models

model = Models.Language.Llama3_8B
hw = Hardware.Cloud.H100

for b in [1, 16, 128, 512]:
    perf = Engine.solve(model, hw, batch_size=b, precision="fp16")
    print(f"Batch {b}: {perf.bottleneck}-bound, MFU={perf.mfu:.3f}")
\end{lstlisting}

\subsection{Volume 2 Example: Distributed Training}
In Volume 2, the focus shifts to distributed systems and fleet-level economics. LEGO Cells here compose multiple solvers to estimate the time and cost of training frontier models across thousands of GPUs:

\begin{lstlisting}[language=Python, caption={\textbf{Volume 2 LEGO Cell: Training Economics.} This snippet from Volume 2 calculates the duration and carbon footprint of a large-scale training run.}, label={lst:lego-vol2}]
from mlsysim.solvers import DistributedModel, SustainabilityModel
from mlsysim.systems.registry import Systems
from mlsysim.models.registry import Models
from mlsysim.infrastructure.registry import Infrastructure

fleet = Systems.Clusters.Frontier_8K
model = Models.Language.Llama3_70B

perf = DistributedModel().solve(
    model, fleet, batch_size=4096, seq_len=4096, tp_size=8, pp_size=8
)
training_days = (perf.step_latency_total * 1e12 / (4096 * 4096)).to("day")

sust = SustainabilityModel().solve(
    fleet, duration_days=training_days.magnitude,
    datacenter=Infrastructure.Datacenters.Iowa_Reference, mfu=0.45
)
print(f"Training Time: {training_days.magnitude:.0f} days")
print(f"Carbon: {sust.carbon_footprint_kg / 1000:.0f} tonnes")
\end{lstlisting}

\subsection{Volume 1 Example: Model Serving}
Another common pattern in Volume 1 evaluates inference systems. Here, a LEGO Cell uses the \texttt{ServingModel} to calculate the inter-token latency (ITL) for Llama-3 8B on an NVIDIA A100 GPU:

\begin{lstlisting}[language=Python, caption={\textbf{Volume 1 LEGO Cell: Model Serving.} This snippet uses the \texttt{ServingModel} to find the theoretical inter-token latency for an autoregressive generation step. It also demonstrates the structural \textbf{LOAD}, \textbf{EXECUTE}, and \textbf{OUTPUT} phases of the LEGO pattern.}, label={lst:lego-vol1-serving}]
from mlsysim.solvers import ServingModel
from mlsysim.models.registry import Models
from mlsysim.hardware.registry import Hardware

# 1. LOAD (Constants)
model = Models.Language.Llama3_8B
hw = Hardware.Cloud.A100

# 2. EXECUTE (The Compute)
perf = ServingModel().solve(
    model, hw, seq_len=1256, batch_size=1, precision="int4", efficiency=1.0
)

# 3. OUTPUT (Formatting)
print(f"Inter-Token Latency (ITL): {perf.itl.to('ms'):.2f}")
\end{lstlisting}

\section{Registry Snapshot}
\label{sec:appendix-hardware}

The MLSys Zoo comprises eight registries spanning the full systems stack (plus the \texttt{Literature} and \texttt{ReferenceStats} auxiliary surfaces). This appendix is a representative snapshot of the current local registry surface: enough to show the kinds of systems \mlsysim can compose, without treating the paper as the authoritative inventory. The typed Python and YAML registry source remains the source of truth for the complete, up-to-date entries.

\subsection{Hardware Registry}

\Cref{tab:hw-catalog} summarizes the current accelerator coverage by deployment tier. Each entry is a fully typed \texttt{HardwareNode} with \texttt{pint}-dimensioned fields. Users add custom entries by instantiating the same types (Listing~\ref{lst:hwnode}), and custom entries participate in the same dimensional pipeline as vetted ones. For example, a research team evaluating a pre-release accelerator can define a speculative entry with projected specifications and immediately sweep it against existing workloads (Listing~\ref{lst:speculative}).

\begin{lstlisting}[language=Python, caption={\textbf{Extending the Registry.} Defining a custom, speculative hardware entry. Because it uses the typed \texttt{HardwareNode} schema (\Cref{tab:type-schema}), it can immediately be evaluated by any Tier 1 or Tier 2 solver without risking unit mismatches.}, label={lst:speculative}]
from mlsysim.hardware.types import HardwareNode, ComputeCore, MemoryHierarchy
from mlsysim.core.units import TFLOPs, GB, TB, second, watt

custom_accelerator = HardwareNode(
    name="Speculative GPU 2027",
    release_year=2027,
    compute=ComputeCore(
        peak_flops=1200 * TFLOPs / second,
        precision_flops={"fp16": 1200 * TFLOPs / second, "fp8": 2400 * TFLOPs / second}
    ),
    memory=MemoryHierarchy(
        capacity=144 * GB,
        bandwidth=5.0 * TB / second
    ),
    tdp=800 * watt
)
\end{lstlisting}

\begin{table}[!t]
\centering
\caption{\textbf{Hardware Registry Snapshot.} The 37 hardware entries currently present in the local registry, organized by deployment tier. Peak FLOP/s is FP16 Tensor Core (or equivalent) throughput; BW is primary-memory bandwidth (HBM, LPDDR, or on-die SRAM as appropriate). Memory uses the marketed (binary) capacity labels.\protect\footnotemark\ The last row shows a user-defined custom entry.}
\label{tab:hw-catalog}
\small
\renewcommand{\arraystretch}{1.05}
\begin{tabularx}{\textwidth}{@{}l l r r r r l@{}}
\toprule
\textbf{Tier} & \textbf{Accelerator} & \textbf{Peak (TFLOP/s)} & \textbf{Mem (GB)} & \textbf{BW (TB/s)} & \textbf{TDP (W)} & \textbf{Interconnect} \\
\midrule
\multicolumn{7}{@{}l}{\textit{Cloud}} \\
 & NVIDIA T4         & 65     & 16   & 0.32  & 70     & PCIe Gen3 \\
 & NVIDIA V100       & 125    & 32   & 0.90  & 300    & NVLink 2.0 (300 GB/s) \\
 & NVIDIA V100 (PCIe) & 112    & 32   & 0.90  & 250    & PCIe Gen3 \\
 & NVIDIA A100       & 312    & 80   & 2.04  & 400    & NVLink 3.0 (600 GB/s) \\
 & NVIDIA H100       & 989    & 80   & 3.35  & 700    & NVLink 4.0 (900 GB/s) \\
 & NVIDIA H200       & 989    & 141  & 4.80  & 700    & NVLink 4.0 (900 GB/s) \\
 & NVIDIA B200       & 2,250  & 192  & 8.00  & 1,000  & NVLink 5.0 (1,800 GB/s) \\
 & NVIDIA GB200 NVL72$^\dagger$ & 162,000 & 13,800 & 576 & 120,000 & NVLink Switch \\
 & AMD MI250X        & 383    & 128  & 3.20  & 500    & Infinity Fabric \\
 & AMD MI300X        & 1,307  & 192  & 5.30  & 750    & Infinity Fabric \\
 & Intel Gaudi 2     & 432    & 96   & 2.45  & 600    & RoCE (24$\times$100 GbE) \\
 & Intel Gaudi 3     & 1,835  & 128  & 3.70  & 900    & RoCE \\
 & AWS Trainium 2    & 380    & 96   & 2.40  & 500    & NeuronLink \\
 & Google TPU v1     & 92 (INT8) & 8 & 0.034 & 75     & PCIe \\
 & Google TPU v2     & 45     & 16   & 0.70  & 200    & ICI \\
 & Google TPU v3     & 105    & 32   & 0.90  & 250    & ICI \\
 & Google TPU v4     & 275    & 32   & 1.20  & 200    & ICI (300 GB/s) \\
 & Google TPU v5p    & 459    & 95   & 2.76  & 300    & ICI (1,200 GB/s) \\
 & Google TPU v6     & 918    & 32   & 1.60  & 300    & ICI \\
 & Cerebras CS-3$^*$ & 125,000 & 44  & 21,000 & 23,000 & MemoryX (1,200 GB/s) \\
 & Intel SGX Enclave & 1      & 0.128 & 0.01 & --     & --- \\
 & Reference CPU Server & 1   & 64   & 0.05  & 150    & --- \\
\midrule
\multicolumn{7}{@{}l}{\textit{Workstation}} \\
 & NVIDIA DGX Spark  & 125    & 128  & 0.27  & 200    & --- \\
 & Apple M3 Max (MacBook) & 14 & 128 & 0.40  & 100    & Unified memory \\
\midrule
\multicolumn{7}{@{}l}{\textit{Edge}} \\
 & NVIDIA Jetson AGX Orin & 275 & 64 & 0.20 & 60 & PCIe Gen4 \\
 & NVIDIA Jetson Orin NX  & 25  & 16 & 0.10 & 25 & PCIe Gen4 \\
 & NVIDIA Jetson Orin Nano & 10 & 8 & 0.07 & 15 & PCIe Gen4 \\
 & Google Coral Edge TPU  & 4   & 1  & 0.008 & 2  & USB 3.0 \\
 & Intel NUC + Movidius   & 1   & 16 & 0.025 & 15 & USB 3.0 \\
 & Generic x86 Server     & 1   & 128 & 0.10 & 300 & --- \\
\midrule
\multicolumn{7}{@{}l}{\textit{Mobile}} \\
 & Apple A17 Pro (iPhone 15 Pro) & 35 & 8 & 0.10 & 5 & Unified memory \\
 & Google Tensor G3 (Pixel 8)    & 15 & 8 & 0.06 & 5 & LPDDR5X \\
 & Snapdragon 8 Gen 3            & 45 & 12 & 0.08 & 5 & LPDDR5X \\
 & Apple M2                      & 16 & 16 & 0.10 & 20 & Unified memory \\
\midrule
\multicolumn{7}{@{}l}{\textit{Tiny}} \\
 & ESP32-S3          & 0.0005 & 0.004 & 0.0001 & 0.4 & SPI \\
 & Himax WE-1        & 0.0002 & 0.002 & 0.0001 & 0.005 & SPI \\
 & Nordic nRF52840   & 0.0001 & 0.001 & 0.0001 & 0.015 & SPI \\
\midrule
\multicolumn{7}{@{}l}{\textit{Custom (user-defined)}} \\
 & \textit{Custom Accelerator} & \textit{500} & \textit{96} & \textit{4.00} & \textit{500} & \textit{---} \\
\bottomrule
\end{tabularx}
\end{table}
\footnotetext{$^*$CS-3 BW is on-wafer SRAM bandwidth (21\,PB/s). The weight-streaming serving case study (\Cref{sec:usage}) uses the ${\sim}$1.2\,TB/s MemoryX injection bandwidth, which is the effective off-chip bandwidth seen by the host. $^\dagger$GB200 NVL72 is a rack-scale system entry (72 Blackwell GPUs behind one NVLink switch domain) rather than a single accelerator.}

\FloatBarrier
\subsection{Infrastructure and Systems Registries}

The Hardware registry (\Cref{tab:hw-catalog}) captures individual accelerator capabilities, but real deployments compose accelerators into nodes, connect nodes through network fabrics, and situate the resulting clusters in datacenters whose geography determines energy cost, carbon intensity, and cooling strategy. The Infrastructure and Systems registries complete this picture: Infrastructure encodes \emph{where} the fleet runs, and Systems encodes \emph{how} accelerators are assembled into distributed configurations. Together the three registries let a single resolver chain evaluate end-to-end metrics from per-device roofline through fleet-level TCO and carbon footprint.

The \textbf{Infrastructure} registry encodes regional environmental context for sustainability analysis, plus datacenter profiles, rack power envelopes, facility cooling, cloud pricing, and capacity lead times. Carbon intensity varies by more than $80\times$ across the 12 registered grid profiles (10\,gCO$_2$/kWh in Norway vs.\ 820 in Poland), making geography a first-order design variable (Case~I2, \Cref{sec:usage}). The \textbf{Systems} registry defines compute nodes, network fabrics, fleet configurations, TPU pods, switch fabrics, and 14 storage-related profiles and checkpoint-path anchors (NVMe generations, SATA, HDD, object stores, parallel file systems) used by the \texttt{DistributedModel}, \texttt{ReliabilityModel}, and \texttt{DataModel}. \Cref{tab:infra-systems} presents representative entries from both registries.

\begin{table}[!t]
\centering
\caption{\textbf{Infrastructure and Systems Registry Snapshot.} \textit{Left:} Current grid profiles for sustainability analysis (CI\,=\,carbon intensity in gCO$_2$/kWh; PUE\,=\,power usage effectiveness; WUE\,=\,water usage effectiveness in L/kWh). \textit{Right:} Representative compute nodes, summarized network fabrics, and current fleet tiers for distributed analysis.}
\label{tab:infra-systems}
\small
\renewcommand{\arraystretch}{1.08}
\begin{minipage}[t]{0.42\textwidth}
\centering
\vspace{0pt}
\textbf{Grid Profiles (12 registered)}\\[4pt]
\begin{tabularx}{\textwidth}{@{}l r r r l@{}}
\toprule
\textbf{Profile} & \textbf{CI} & \textbf{PUE} & \textbf{WUE} & \textbf{Source} \\
\midrule
Norway              & 10   & 1.06 & 0.0 & Hydro \\
Qu\'ebec            & 20   & 1.06 & 0.0 & Hydro \\
Iceland             & 28   & 1.06 & 0.0 & Geo/Hydro \\
France              & 50   & 1.12 & 1.8 & Nuclear \\
EU Average          & 270  & 1.12 & 1.8 & Mixed \\
Germany             & 385  & 1.12 & 1.8 & Mixed \\
Texas               & 400  & 1.12 & 1.8 & Mixed \\
US Average          & 429  & 1.12 & 1.8 & Mixed \\
China Average       & 555  & 1.40 & 1.8 & Coal-heavy \\
Iowa (Coal/Gas)     & 680  & 1.12 & 1.8 & Fossil \\
India Average       & 720  & 1.58 & 1.8 & Coal-heavy \\
Poland (Coal)       & 820  & 1.58 & 1.8 & Coal \\
\bottomrule
\end{tabularx}
\end{minipage}%
\hfill
\begin{minipage}[t]{0.55\textwidth}
\centering
\vspace{0pt}
\textbf{Nodes, Fabrics, and Fleets}\\[4pt]
\begin{tabularx}{\textwidth}{@{}l X@{}}
\toprule
\multicolumn{2}{@{}l}{\textit{Compute Nodes}} \\
\midrule
DGX A100  & 8$\times$ A100, NVLink 3.0, 600\,GB/s \\
DGX H100  & 8$\times$ H100, NVLink 4.0, 900\,GB/s \\
DGX B200  & 8$\times$ B200, NVLink 5.0, 1,800\,GB/s \\
Kempner H100 & 4$\times$ H100 academic server \\
\midrule
\multicolumn{2}{@{}l}{\textit{Network Fabrics (10 registered)}} \\
\midrule
IB HDR / NDR / XDR / GXDR & 200 / 400 / 800 / 1,600\,Gb/s, 5--7\,$\mu$s \\
Ethernet 10G--1.6T & 10--1,600\,Gb/s, 50\,$\mu$s \\
RoCE 100G  & 100\,Gb/s, 10\,$\mu$s \\
\midrule
\multicolumn{2}{@{}l}{\textit{Fleet Configurations (11 registered)}} \\
\midrule
Lab (64 GPU)          & 8$\times$ DGX H100, IB HDR \\
Research (256 GPU)    & 32$\times$ DGX H100, 100\,GbE \\
Kempner (384 GPU)     & 96$\times$ 4-GPU H100, IB NDR \\
Training (512 GPU / 1K / 1K A100) & 64--128 nodes, IB HDR \\
Production (2K GPU)   & 256$\times$ DGX H100, IB HDR fat-tree \\
Frontier (8K GPU)     & 1,024$\times$ DGX H100, IB NDR fat-tree \\
Training (10K) / Reference (25K) & 1,250 / 3,125 nodes, IB NDR \\
Mega (100K GPU)       & 12,500$\times$ DGX H100, IB NDR \\
\bottomrule
\end{tabularx}
\end{minipage}
\end{table}

\FloatBarrier
\subsection{Type System and Physics API}

To support the separation of demand and supply, \mlsysim relies on a shallow but rigorously typed hierarchy of Pydantic models (\Cref{tab:type-schema}). These types act as the contract between the declarative registries and the analytical solvers. Because all physical quantities (bandwidth, latency, power) are dimensioned using the \texttt{pint} library, users can extend any registry by instantiating these types directly without risking unit mismatch errors.

Beneath the resolvers lies the Physics Engine, which contains the dimensioned mathematical identities that govern ML systems. \Cref{tab:physics-api} catalogs the public API of these physics modules. Each function is stateless, accepting dimensioned quantities (or raw floats that are immediately coerced to units) and returning dimensioned bounds. These functions implement the core logic for the Iron Law of training, the Roofline model, and network communication topologies.

\begin{table}[!t]
\centering
\small
\renewcommand{\arraystretch}{1.05}
\caption{\textbf{Physics Module Public API.} Complete function signatures organized by domain module. Every function accepts \texttt{pint} \texttt{Quantity} arguments (or raw floats coerced via \texttt{\_ensure\_unit()}) and returns dimensioned quantities. Import from \texttt{mlsysim.physics} (one module per domain).}
\label{tab:physics-api}
\begin{tabularx}{\textwidth}{@{}l l X@{}}
\toprule
\textbf{Module} & \textbf{Function} & \textbf{Signature (key parameters $\to$ return)} \\
\midrule
\multirow{6}{*}{\texttt{performance}} & \texttt{dTime} & (total\_ops, n\_devices, peak\_flops, $\eta$) $\to$ seconds \\
 & \texttt{calc\_bottleneck} & (ops, model\_bytes, device\_flops, device\_bw) $\to$ \{time, regime\} \\
 & \texttt{calc\_amdahls\_speedup} & (parallel\_frac, speedup\_factor) $\to$ overall speedup \\
 & \texttt{calc\_pipeline\_bubble} & (n\_stages, n\_microbatches, v\_stages) $\to$ bubble fraction \\
 & \texttt{calc\_effective\_flops} & (peak\_flops, $\eta$) $\to$ sustained FLOP/s \\
 & \texttt{calc\_training\_time\_days} & (total\_ops, n\_devices, peak\_flops, $\eta$) $\to$ days \\
\midrule
\multirow{5}{*}{\texttt{memory}} & \texttt{model\_memory} & (params, bytes\_per\_param, unit) $\to$ memory \\
 & \texttt{calc\_activation\_memory} & (batch, seq, hidden, layers, prec) $\to$ bytes \\
 & \texttt{calc\_kv\_cache\_size} & (layers, heads, dim, seq, batch, prec) $\to$ bytes \\
 & \texttt{calc\_paged\_kv\_cache\_size} & (same + page\_size) $\to$ bytes \\
 & \texttt{calc\_checkpoint\_size} & (n\_params, bytes\_per\_param) $\to$ bytes \\
\midrule
\texttt{serving} & \texttt{calc\_queue\_latency\_mmc} & (arrival\_rate, service\_rate, n\_servers) $\to$ wait time \\
\midrule
\multirow{4}{*}{\texttt{communication}} & \texttt{calc\_ring\_allreduce\_time} & (msg\_bytes, n\_gpus, bw, latency) $\to$ seconds \\
 & \texttt{calc\_tree\_allreduce\_time} & (msg\_bytes, n\_gpus, bw, latency) $\to$ seconds \\
 & \texttt{calc\_all\_to\_all\_time} & (msg\_bytes, n\_gpus, bw, latency) $\to$ seconds \\
 & \texttt{calc\_hierarchical\_allreduce\_time} & (msg, intra\_gpus, intra\_bw, inter\_nodes, inter\_bw, $\alpha$) $\to$ seconds \\
\midrule
\multirow{5}{*}{\texttt{reliability}} & \texttt{calc\_young\_daly\_interval} & (ckpt\_cost, mtbf) $\to$ optimal interval \\
 & \texttt{calc\_mtbf\_cluster} & (component\_mtbf, n, correlation) $\to$ cluster MTBF \\
 & \texttt{calc\_mtbf\_node} & (component\_mtbfs) $\to$ node MTBF \\
 & \texttt{calc\_availability\_stacked} & (single\_avail, n\_replicas) $\to$ stacked availability \\
 & \texttt{calc\_failure\_probability} & (mtbf, duration) $\to$ probability \\
\midrule
\multirow{2}{*}{\texttt{transformer}} & \texttt{calc\_transformer\_training\_flops} & (n\_params, n\_tokens) $\to$ total FLOPs \\
 & \texttt{calc\_transformer\_decode\_flops} & (n\_params, n\_tokens) $\to$ decode FLOPs \\
\midrule
\multirow{2}{*}{\texttt{economics}} & \texttt{calc\_fleet\_tco} & (unit\_cost, power, quantity, years, kwh\_price) $\to$ TCO \\
 & \texttt{calc\_monthly\_egress\_cost} & (bytes\_per\_sec, cost\_per\_gb) $\to$ monthly cost \\
\midrule
\texttt{networking} & \texttt{calc\_network\_latency\_ms} & (distance\_km) $\to$ latency \\
\midrule
\multirow{3}{*}{\texttt{statistics}} & \texttt{calc\_population\_stability\_index} & (expected, actual) $\to$ PSI \\
 & \texttt{calc\_two\_proportion\_sample\_size} & (baseline, lift, $z_\alpha$, $z_\beta$) $\to$ $n$ \\
 & \texttt{calc\_constraint\_propagation\_factor} & (from, to, base) $\to$ factor \\
\midrule
\multirow{3}{*}{\texttt{quantities}} & \texttt{transfer\_time} / \texttt{compute\_time} & (payload, bandwidth) / (work, throughput) $\to$ time \\
 & \texttt{energy\_from\_power} / \texttt{carbon\_from\_energy} & (power, duration) / (energy, CI) $\to$ energy / mass \\
 & \texttt{memory\_from\_params} / \texttt{token\_throughput} & (params, bytes/param) / (tokens, duration) $\to$ bytes / rate \\
\bottomrule
\end{tabularx}
\end{table}

\begin{table}[!t]
\centering
\small
\renewcommand{\arraystretch}{1.08}
\caption{\textbf{Core Type Schema.} Pydantic \texttt{BaseModel} types that compose registry entries. The hierarchy is deliberately shallow: \texttt{HardwareNode} aggregates \texttt{ComputeCore} and \texttt{MemoryHierarchy} as direct fields, not through deep inheritance. All physical fields are \texttt{pint} \texttt{Quantity} objects. Users extend any registry by instantiating these types directly (Listing~\ref{lst:hwnode}).}
\label{tab:type-schema}
\begin{tabularx}{\textwidth}{@{}l X@{}}
\toprule
\textbf{Type} & \textbf{Key Fields} \\
\midrule
\texttt{ComputeCore}       & \texttt{peak\_flops} (FLOP/s), \texttt{precision\_flops} (dict: fp8/fp16/tf32/int8 $\to$ FLOP/s), \texttt{sm\_count} \\
\texttt{MemoryHierarchy}   & \texttt{capacity} (bytes), \texttt{bandwidth} (bytes/s); optional SRAM, flash, and L2 tiers \\
\texttt{IOInterconnect}    & \texttt{bandwidth} (bytes/s), \texttt{latency} (seconds), \texttt{name} \\
\texttt{HardwareNode}      & \texttt{compute}, \texttt{memory}, \texttt{storage}, \texttt{interconnect}, \texttt{nvlink}, \texttt{tdp} (W), \texttt{unit\_cost} (USD), \texttt{dispatch\_tax} (s), \texttt{metadata.provenance} \\
\texttt{ComputationGraph}  & \texttt{total\_ops} (FLOPs), \texttt{weight\_bytes} (bytes), \texttt{arithmetic\_intensity} (FLOP/byte), \texttt{parameter\_count}, \texttt{layers} \\
\texttt{TransformerWorkload} & \texttt{name}, \texttt{architecture}, \texttt{parameters}, \texttt{layers}, \texttt{hidden\_dim}, \texttt{heads}, \texttt{kv\_heads} \\
\texttt{GridProfile}       & \texttt{carbon\_intensity\_g\_kwh}, \texttt{pue}, \texttt{wue} (L/kWh), \texttt{renewable\_pct} \\
\texttt{NetworkFabric}     & \texttt{topology}, \texttt{bandwidth} (bytes/s), \texttt{latency} (s), \texttt{oversubscription\_ratio} \\
\texttt{Fleet}             & \texttt{node}, \texttt{count}, \texttt{fabric} (NetworkFabric), \texttt{region}/\texttt{datacenter}, \texttt{mtbf\_hours} \\
\texttt{SystemEvaluation}  & \texttt{feasibility}, \texttt{performance}, \texttt{macro} (three-level scorecard) \\
\bottomrule
\end{tabularx}
\end{table}

\FloatBarrier
\subsection{Code Examples Cookbook}

To demonstrate how the theoretical 3-tier architecture from \Cref{sec:solver-formalism} is exposed in Python, we provide eight complete, runnable examples. Each example is written as a small engineering task: the prose names the question, the listing shows which registry facts are selected, and the returned object exposes the bound, bottleneck, or scorecard used to make a decision. The point is not to hide the model behind an API call, but to show how registry-backed composition turns a systems question into a repeatable calculation.

\subsubsection{Scenario A: Rapid Parametric Sweeps (Tier 1)}
\emph{Scenario:} You want to know how much decode throughput continuous batching can buy on an H100 before KV-cache and activation traffic saturate the memory system. The code selects a workload from \texttt{Models.Language} and a supply point from \texttt{Hardware.Cloud}, then calls \texttt{Engine.solve()} once per batch size. Each returned performance object reports throughput, MFU, and the binding Roofline regime, so the sweep shows not only that throughput saturates but why it remains memory-bound.

\begin{lstlisting}[language=Python, caption={\textbf{Programmatic Sweeps (Tier 1).} Using the \texttt{Engine} to sweep a workload across a hardware constraint space (batch size). Throughput rises from 178 to ${\sim}$2{,}000 tokens/s and then saturates: per-request memory traffic grows with the batch, so decode stays memory-bound at every point. Because execution takes $<$1\,ms per point, sweeping hundreds of configurations is trivial.}, label={lst:appendix-sweep}]
from mlsysim.engine.engine import Engine
from mlsysim.hardware.registry import Hardware
from mlsysim.models.registry import Models
import pandas as pd

# Sweep Llama 3 8B across multiple batch sizes on an H100 GPU
model = Models.Language.Llama3_8B
hardware = Hardware.Cloud.H100

results = []
for batch_size in [1, 8, 32, 128, 256]:
    # Engine.solve() wraps the Tier 1 SingleNodeModel
    perf = Engine.solve(model, hardware, batch_size=batch_size, precision="fp16")

    results.append({
        "Batch Size": batch_size,
        "Throughput (tok/s)": round(perf.throughput.magnitude),
        "Bottleneck": perf.bottleneck,
        "MFU": round(perf.mfu, 3)
    })

df = pd.DataFrame(results)
print(df)
#    Batch Size  Throughput (tok/s) Bottleneck    MFU
# 0           1                 178     Memory  0.003
# 1           8                 893     Memory  0.014
# 2          32                1564     Memory  0.025
# 3         128                1925     Memory  0.031
# 4         256                2002     Memory  0.033
\end{lstlisting}

\subsubsection{Scenario B: Cross-Domain Carbon Accounting (Tier 1 Composition)}
\emph{Scenario:} You have been tasked with estimating the carbon footprint of training a 70B parameter model. The listing first uses \texttt{DistributedModel} to compute the step latency for a registry-backed 256-GPU fleet, then converts a one-trillion-token target into a training duration. That duration becomes the input to \texttt{SustainabilityModel}, where the Iowa datacenter profile supplies grid intensity and facility overhead; the composition works because solver outputs are dimensioned quantities rather than untyped floats.

\begin{lstlisting}[language=Python, caption={\textbf{Resolver Composition (Tier 1).} Output quantities from one solver (the computed step latency from the \texttt{DistributedModel}) feed directly into subsequent solvers (the \texttt{SustainabilityModel}), enabling end-to-end full-stack analysis without manual unit conversion.}, label={lst:appendix-compose}]
from mlsysim.solvers import DistributedModel, SustainabilityModel
from mlsysim.systems.registry import Systems
from mlsysim.infrastructure.registry import Infrastructure
from mlsysim.models.registry import Models

fleet = Systems.Clusters.Research_256  # 32x DGX H100 nodes

# 1. Performance Domain: how long does one training step take?
perf = DistributedModel().solve(
    model=Models.Language.Llama3_70B, fleet=fleet,
    batch_size=1024, seq_len=2048, tp_size=8, pp_size=1,
    activation_recomputation=True,
)

# 2. Train for 1T tokens: convert step latency into a duration
steps = 1e12 / (1024 * 2048)
training_days = (perf.step_latency_total * steps).to("day").magnitude

# 3. Sustainability Domain: what is the carbon impact?
sust = SustainabilityModel().solve(
    fleet=fleet, duration_days=training_days,
    datacenter=Infrastructure.Datacenters.Iowa_Reference,  # high-carbon grid
    mfu=0.42,
)
print(f"Training Duration: {training_days:.0f} days")
# Training Duration: 57 days
print(f"Total Carbon: {sust.carbon_footprint_kg / 1000:.1f} tonnes CO2e")
# Total Carbon: 110.3 tonnes CO2e
\end{lstlisting}

\subsubsection{Scenario C: Uncovering the Data Wall (Tier 1)}
\emph{Scenario:} Your GPU utilization is mysteriously low during a computer vision training job. The code splits the diagnosis into two checks: \texttt{DataModel} compares the required image stream against the node's I/O path, while \texttt{TransformationModel} compares CPU preprocessing throughput against accelerator step time. Reading the two results together distinguishes a storage or transfer bottleneck from a host-side decode and augmentation bottleneck.

\begin{lstlisting}[language=Python, caption={\textbf{Data Pipeline Analysis (Tier 1).} Evaluating the ``Data Wall''. This example separates I/O bandwidth bottlenecks (storage to GPU) from CPU transformation bottlenecks (decoding and augmenting data), in the spirit of Case Study S2 but with an under-provisioned 16-worker pipeline.}, label={lst:appendix-data}]
from mlsysim.solvers import DataModel, TransformationModel
from mlsysim.hardware.registry import Hardware
from mlsysim.core.units import Q_

# 1. Define the ingestion demand (e.g., training a vision model)
# We need to feed 40,000 images per second to the GPU.
demand_rate = Q_("40000 1/s") * Q_("150 KB")  # ~6 GB/s

# 2. Check if the A100 node's I/O path can handle the bandwidth
data_result = DataModel().solve(
    workload_data_rate=demand_rate,
    hardware=Hardware.Cloud.A100
)
print(f"I/O Stalled: {data_result.is_stalled}")  # False

# 3. Check if 16 CPU workers can decode + augment fast enough
#    (850 images/s per worker, 150 KB per image)
cpu_throughput = 16 * 850 * Q_("150 KB/s")
transform_result = TransformationModel().solve(
    batch_size=2048,
    sample_size_bytes=Q_("150 KB"),
    cpu_throughput=cpu_throughput,
    accelerator_step_time=Q_("48 ms"),
)
print(f"CPU Stalled: {transform_result.is_bottleneck}")  # True
print(f"Accel. Utilization: "
      f"{transform_result.accelerator_utilization:.0%}")  # 32%
\end{lstlisting}

\subsubsection{Scenario D: SLA-Driven Hardware Synthesis (Tier 2)}
\emph{Scenario:} You are provisioning hardware for a new LLM application. You do not start with a GPU SKU; you start with the business requirement that the application should emit a token every 30 milliseconds. The \texttt{SynthesisSolver} inverts the same first-order Roofline model used for evaluation: given \texttt{Models.Language.Llama3\_8B}, the target latency, and precision, it returns the memory-bandwidth and compute-throughput requirements that candidate hardware must clear.

\begin{lstlisting}[language=Python, caption={\textbf{SLA-Driven Hardware Synthesis (Tier 2).} Instead of evaluating a known piece of hardware, the \texttt{SynthesisSolver} works backward. Given a strict latency SLA, it algebraically inverts the Roofline equations to output first-order minimum memory bandwidth and compute throughput requirements for procurement screening.}, label={lst:appendix-synthesis}]
from mlsysim.solvers import SynthesisSolver
from mlsysim.models.registry import Models
from mlsysim.core.units import Q_

# We want to serve Llama-3 8B. 
# Our strict SLA dictates an inter-token latency of 30 ms.
# What is the minimum hardware capability required to achieve this?

solver = SynthesisSolver()
requirements = solver.solve(
    model=Models.Language.Llama3_8B,
    target_latency=Q_("30 ms"),
    precision="fp16"
)

# Output the hardware requirements
print(f"Required HBM Bandwidth: {requirements.required_bw.to('GB/s'):.1f}")
# Required HBM Bandwidth: 535.3 GB / second
print(f"Required Compute:       {requirements.required_flops.to('TFLOPs/s'):.1f}")
# Required Compute:       1.1 TFLOPs / second
\end{lstlisting}

\subsubsection{Scenario E: Speculative Decoding Speedup (Tier 1)}
\emph{Scenario:} You are serving a large frontier model and want to know how much faster inference would be if you used a smaller, cheaper draft model to speculatively decode tokens. The first call establishes the autoregressive baseline for the target model on the same H100 supply point. The second call adds a draft model and an acceptance-rate assumption, allowing \texttt{ServingModel} to account for draft generation and target-model verification before reporting the effective inter-token latency and speedup.

\begin{lstlisting}[language=Python, caption={\textbf{Speculative Decoding Analysis (Tier 1).} The \texttt{ServingModel} natively supports algorithmic optimizations like speculative decoding. By providing a draft model and an expected acceptance rate, the solver calculates the effective inter-token latency (ITL) by modeling the compute-bound draft phase and the memory-bound verification phase.}, label={lst:appendix-speculative}]
from mlsysim.solvers import ServingModel
from mlsysim.hardware.registry import Hardware
from mlsysim.models.registry import Models

# Serve Llama 3.1 70B using Llama 3.1 8B as a draft model
target_model = Models.Language.Llama3_70B
draft_model = Models.Language.Llama3_8B
hardware = Hardware.Cloud.H100

solver = ServingModel()

# Standard Autoregressive Decoding
base_result = solver.solve(target_model, hardware, seq_len=2048, batch_size=1)
print(f"Standard ITL: {base_result.itl.to('ms'):.2f}")
# Standard ITL: 43.15 millisecond

# Speculative Decoding
spec_result = solver.solve(
    target_model, hardware, seq_len=2048, batch_size=1,
    draft_model=draft_model,
    draft_acceptance_rate=0.75
)
print(f"Speculative ITL: {spec_result.itl.to('ms'):.2f}")
# Speculative ITL: 20.53 millisecond
print(f"Speedup: {(base_result.itl / spec_result.itl).magnitude:.2f}x")
# Speedup: 2.10x
\end{lstlisting}

\subsubsection{Scenario F: The Full-Stack Scorecard (Tier 2)}
\emph{Scenario:} Rather than invoking individual solvers, you want the single ``will it run, is it fast enough, and what does it cost'' verdict for a deployment. An executable \texttt{Scenario} is a named bundle of workload, target system, SLA, and operating context. Calling \texttt{evaluate()} runs the Feasibility, Performance, and Macro levels in order, short-circuiting impossible cases and returning a \texttt{SystemEvaluation} whose scorecard preserves the reason for each pass or failure.

\begin{lstlisting}[language=Python, caption={\textbf{Three-Level Scorecard (Tier 2).} \texttt{Scenario.evaluate()} composes the resolver chain automatically and returns a \texttt{SystemEvaluation}: Level~1 checks memory feasibility, Level~2 reports latency against the SLA, and Level~3 reports cost and carbon.}, label={lst:appendix-scorecard}]
import mlsysim

# A pre-built executable scenario: chatbot serving with an SLA + grid context
evaluation = mlsysim.Scenarios.ChatbotServing.evaluate()
print(evaluation.scorecard())
# Level 1: Feasibility [PASS]  Model fits in memory (16.1 GB / 85.9 GB)
# Level 2: Performance [PASS]  Latency: 5.60 ms  (Target: 500 ms)
# Level 3: Macro       [PASS]  Annual Carbon: 2946.3 kg | TCO: $10,407
print(evaluation.passed_all)  # True
\end{lstlisting}

\subsubsection{Scenario G: Automated Parallelism Search (Tier 3)}
\emph{Scenario:} You have a fixed cluster and a large model and want the framework to \emph{find} the best 3D-parallelism split rather than specifying it by hand. The Tier~3 \texttt{ParallelismOptimizer} enumerates power-of-two $\text{TP}\times\text{PP}\times\text{DP}$ factorizations of the available devices, computes per-GPU training state for each candidate, and discards layouts that exceed HBM. The result keeps the best MFU configuration and a ranked candidate list, making the search procedure visible rather than a black-box recommendation.

\begin{lstlisting}[language=Python, caption={\textbf{Parallelism Optimizer (Tier 3).} The optimizer enumerates valid TP/PP/DP factorizations of the cluster, discards configurations whose per-GPU training state exceeds HBM, and returns the highest-MFU survivor with its ranked candidate list.}, label={lst:appendix-optimizer}]
import mlsysim
from mlsysim.solvers import ParallelismOptimizer

fleet = mlsysim.Systems.Clusters.Research_256   # 256x H100
result = ParallelismOptimizer().solve(
    mlsysim.Models.Language.Llama3_70B, fleet,
    batch_size=2048, precision="fp16", efficiency=0.5,
)
print(result.best_config)             # {'tp': 8, 'pp': 2, 'dp': 16}
print(f"MFU: {result.best_mfu:.3f}")  # MFU: 0.315
for c in result.top_candidates[:3]:
    print(c["config"], f'{c["mfu"]:.3f}')
# {'tp': 8, 'pp': 2, 'dp': 16} 0.315
# {'tp': 4, 'pp': 4, 'dp': 16} 0.274
# {'tp': 8, 'pp': 4, 'dp': 8} 0.272
\end{lstlisting}

\subsubsection{Scenario H: From Datacenter to Microcontroller (Tier 1)}
\emph{Scenario:} The same dimensionally strict engine that sizes an H100 fleet also sizes a microcontroller. The loop changes only the registry-backed model and hardware entries; \texttt{Engine.solve()} still lowers the workload demand, reads the supply constraints, and returns feasibility, bottleneck regime, latency, and per-inference energy. This is the benefit of the demand--supply contract: datacenter and TinyML examples share the same code path while drawing facts from very different parts of the registry.

\begin{lstlisting}[language=Python, caption={\textbf{Edge and TinyML Deployment (Tier 1).} The identical \texttt{Engine.solve} call evaluates milliwatt-class devices: a LeNet variant fits in the ESP32-S3's on-chip SRAM, while MobileNet variants run on a Jetson Orin Nano and a Coral Edge TPU. Energy is per inference.}, label={lst:appendix-edge}]
from mlsysim.engine.engine import Engine
from mlsysim.hardware.registry import Hardware
from mlsysim.models.registry import Models

deployments = [
    (Models.Vision.LeNet1,               Hardware.Tiny.ESP32_S3),
    (Models.Vision.MobileNetV2,          Hardware.Edge.JetsonOrinNano),
    (Models.Vision.MobileNetV2_Alpha0_5, Hardware.Edge.Coral),
]
for model, hw in deployments:
    perf = Engine.solve(model, hw, batch_size=1, precision="int8")
    print(f"{hw.name:24s} {perf.bottleneck}-bound  "
          f"{perf.latency.to('ms'):.2f}  {perf.energy.to('mJ'):.2f}")
# ESP32-S3 (AI)            Compute-bound  1.03 millisecond  0.13 millijoule
# NVIDIA Jetson Orin Nano  Memory-bound   0.80 millisecond  3.66 millijoule
# Google Coral Edge TPU    Memory-bound   1.28 millisecond  0.77 millijoule
\end{lstlisting}
\end{appendices}
\end{document}